# A Case Study on Quality Attribute Measurement using MARF and GIPSY

Masoud Bozorgi, Rohan Nayak, Arslan Zaffar, Mohammad Iftekharul Hoque,
Saad Anwer Ghouri, Harmeet Singh, Parminder Singh Kalshan

Concordia University
Montreal, QC, Canada

{m_bozo, r_nayak, a_zaffa, mo_hoqu, s_ghouri, harme_s, p_kalsha}@encs.concordia.ca

**Abstract**

*This literature focuses on doing a comparative analysis between Modular Audio Recognition Framework (MARF) and the General Intentional Programming System (GIPSY) with the help of different software metrics. At first, we understand the general principles, architecture and working of MARF and GIPSY by looking at their frameworks and running them in the Eclipse environment. Then, we study some of the important metrics including a few state of the art metrics and rank them in terms of their usefulness and their influence on the different quality attributes of a software. The quality attributes are viewed and computed with the help of the Logiscope and McCabe IQ tools. These tools perform a comprehensive analysis on the case studies and generate a quality report at the factor level, criteria level and metrics level. In next step, we identify the worst code at each of these levels, extract the worst code and provide recommendations to improve the quality. We implement and test some of the metrics which are ranked as the most useful metrics with a set of test cases in JDeodorant. Finally, we perform an analysis on both MARF and GIPSY by doing a fuzzy code scan using MARFCAT to find the list of weak and vulnerable classes.*

## 1. INTRODUCTION

In this survey we study about the MARF and GIPSY frameworks as well as do a comparative study on different software metrics to understand the working and applications of these case studies along with a suite of software metrics. The main motivation behind this research is to gain a deeper insight in the working of an object-oriented software and it's relationships with General Intentional Programming language. We survey the relationships between classes, methods and attributes in the MARF and GIPSY frameworks by using several tools and eclipse plugins which will be described in this report. We will also be performing a comparative study on the different metric tools and techniques on MARF and GIPSY. These techniques can be used by measurement toolkits to compare the complexities of the case studies. Moreover in this survey, JDeodorant is used to test the metric suites with corresponding test cases and then we compare the case studies for vulnerabilities by using MARFCAT [72].

## 2. BACKGROUND

MARF and GIPSY are the two Open Source Software (OSS) case studies used in this study where we attempt to gain an understanding about them along with their working and applications.

### 2.1. MARF

Modular Audio Recognition Framework (MARF) is an open source framework implemented in Java which is a collection of various algorithms for Audio recognition, Text Processing and Natural Language Processing. Most of the program is written and compiled in Java [1, 2, 5, 8, 9, 10, 11, and 12].

#### 2.1.1. Purpose

MARF has an extensible framework, which facilitates addition of newer algorithms and can run on a distributed network. It can act as a library for audio processing applications and perform all the pattern recognition tasks. It provides a platform for developers to test and compare new and existing algorithms among each other with the help of various performance metrics [1, 8].

#### 2.1.2. Design and Architecture

The architecture of MARF can be seen in figure 1 in which "MARF" class is the central server and the "configuration" class is the placeholder, which contains the important functions that are carried out in the pattern recognition process. The major pattern recognition methods which form the "core pipeline" of the entire process is represented in figure 2. In this figure the training module is responsible for the storage of the feature vectors or the clusters of a given subject. The classification module ranks a list of possible subject identifiers from least likely to the most likely [1].

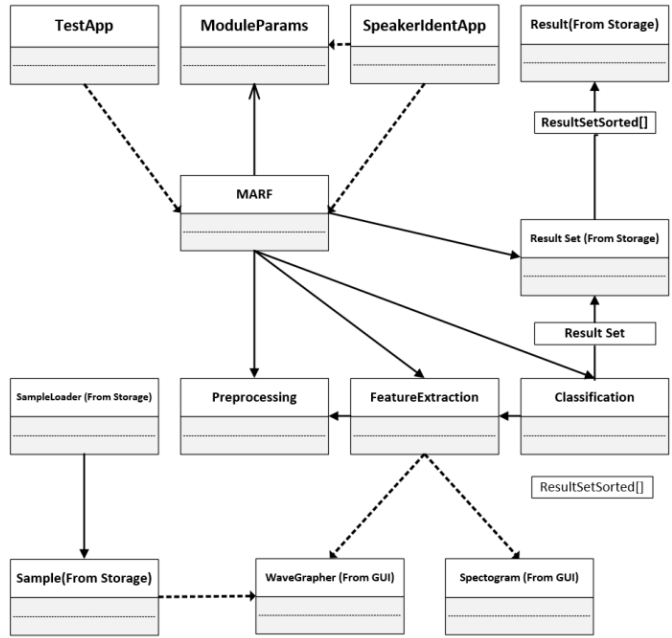

Figure 1: Overall Architecture of MARF System [2]

### 2.1.3. Methodology and Working of MARF

There are several MARF approaches that use several voice samples for audio recognition. MARF has three main functions, which are listed below. Moreover, MARF pattern recognition pipeline flow is shown in figure 2.

1. *Pre-processing*: At first, voice is loaded into the system in the form of wave, text or image file. Then, noisy characters and other irrelevant things remove from that and system uses median or mean clusters to recognize the voice and displays the result. The mean cluster is the cluster of the audio samples that have values lying close to the mean of the all the audio samples. This tells us that the audio samples lying in the cluster are similar [3] [8].

2. *Feature Extraction*: MARF is responsible for extracting fine details such as type of voice, gender, etc. The techniques used are Fast Fourier Transformation (FFT) and Linear Predictive Coding Algorithms (LPC) [1, 8].

3. *Classification*: This module corresponds to the training and classification of MARF sub-framework, which is the last stage of the pipeline. The implementations include comparison of multiple algorithms together including Supervised and Unsupervised, statistical, etc. [1,9]. Supervised algorithms use "tagged" data where the data can be audio, speech samples, etc. to identify and cluster samples. Unsupervised algorithms find correlations between samples without any external input [73].

### 2.1.4. Experimental Study

An experiment was studied in which SpeakerIdentApp application performs various pre-processing tasks [8]. The sample test cases consisted of 319 voice samples from 28 different speakers. To obtain noticeable characteristics, gender and accent identification, it has an options to use LPC, FFT [8,12], Min/Max Amplitudes, random extraction of features, aggregated feature extraction. SpeakerIdentApp can use different algorithms like Chebyshev Distance, Euclidean Distance, Minkowski Distance, etc. for classification [8]. The statistics collected by the system are based on the successful versus unsuccessful guesses as well as the second-best approach. The testing of all audio samples are based on the scripts of all available configurations to achieve the best possible accuracy. The reliability of the system depends on the accuracy of the system in processing audio signals as well as good software practices during implementation [8].

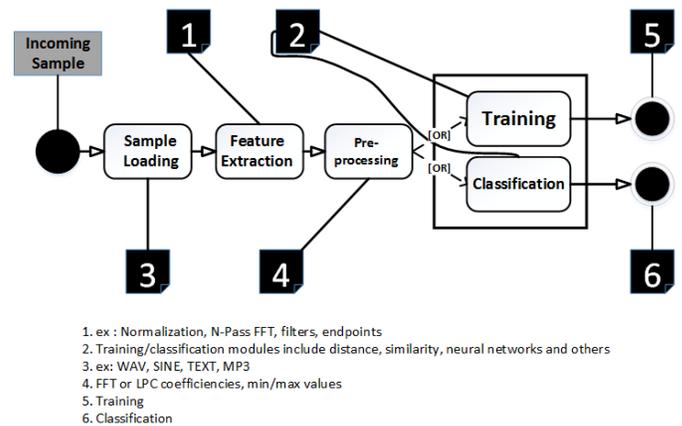

Figure 2: MARF Pattern Recognition Pipeline [1]

### 2.1.5. Applications

There are a large number of applications that can be written using MARF however there are four major applications that include Text-Independent Speaker Identification, Language Identification [8], Probabilistic Parsing, and Zipf's Law analysis [1]. MARF is not necessarily used in audio processing [8]. It can be used to perform a lot of other NLP tasks [5] as well as network-related tasks shown in the pipeline. It is also used in research and implementation and testing of security framework and data security [1]. MARF was also applied for the DEFT (DÉfiFouille de Textes) 2010 in which the challenge was to acquire the date of publication and the geographical location, where the journal was published [9]. Vulnerabilities in computer security mean a weakness that allows an attacker to extract information from software [4]. To discover these vulnerabilities MARFCAT application of MARF's NLP framework is used [5].

## 2.2. GIPSY

GIPSY is a General Programming Intentional System, which provides a common platform for the compilation and execution of intentional programming languages where other systems were not successful because the change in syntax and semantics is often. It also allows the creation of hybrid languages. [13, 14, 15, 16, 17, 18, 19, 20].

GIPSY has a system, which is a collection of rules that assign a property or type system. This system binds the static and dynamic typing between intentional and imperative languages at runtime in its own compiler to support expression by intentional evaluation [14, 16, 17].

Intentional means computation in terms of expressions or senses, and imperative means computation in terms of statements. Lucid is dataflow programming language. Different types of segments are introduced in GIPSY programs along with matching the Lucid and Java data types and some implementation of type system. It focuses on specific features without pointing out specific language. Here a type system is designed to form a bridge between intentional and imperative programming paradigms [14, 16, 17].

### 2.2.1. Design and Implementation Goals

The main purpose of GIPSY is to set contexts as first-class values and usage of context calculus operators, which will lead into the construction and manipulation of contexts [16].

### 2.2.2. Basic Requirements

There should be a bridge between the defined, used and shared data types, which will be shared by hybrid counterparts. Other one, the tag, which is associated with dimension, has to be in valid range. Generic Intentional Programming Language (GIPL) plays a vital role in the GIPSY environment; GIPL is helpful for the execution of programs using primitive operators associated with target machine [16].

### 2.2.3 Architecture

GIPSY has a multi-tiered architecture as seen in figure 3.

It is divided into four tasks assigned to four layers shown in table 1:

|    | Layer Type | Purpose |
|----|-----------|---------|
| **1.** | Demand Generator Tier (DGT) | Encapsulate demands [13]. |
| **2.** | . Demand Store Tier(DST) | Middleware for other tier [13]. |
| **3.** | Demand Worker Tier (DWT) | Process Demands [13]. |
| **4.** | GIPSY Manager Tier (GMT) | Register GIPSY nodes [13]. |

**Table 1: Summary of Code Analysis of MARF and GIPSY**

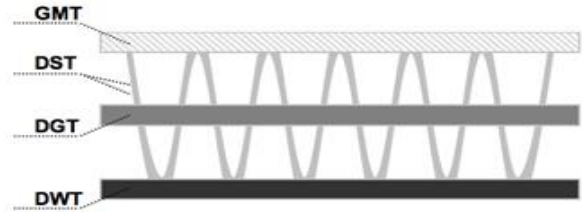

**Figure 3: GIPSY Multi-Tier Architecture [13, 20]**

The architecture follows Demand Migration System (DMS), which consist of two parts: Demand Dispature (DD) and Transport Agent (TA) [13].

GIPSY tiers can interact with each other using demands. Here, based on initial demand, the Demand Generator Tier (DGT) creates intentional and procedural demands. For migration of the demands between various tiers the Demand Store Tier (DST) is used. It is designed like a peer-to-peer architecture. Here, the Demand Worker Tier (DWT) is responsible to process the procedural demands. The GIPSY Instance Manager (GIM) communicates with DGT, DST, DWT controllers to decide if any new tiers or nodes are required and based on the situation it registers them [13, 19, 20].

*General Eduction Engine (GEE)*

GIPSY uses eduction, which is a demand-driven computational technique, which works with a value cache called a warehouse. A demand generates a procedure call and the computed values are stored in the warehouse. An already computed value in demand will be extracted from the warehouse, which results in reduced overhead. The GIPSY has generator-worker execution architecture as shown in figure 4.

The system consists of two systems the Intentional Demand Propagator (IDP), which generates and propagates demands, and the Intentional Value Warehouse (IVW), which is built to store and retrieve values that have already been computed [18].

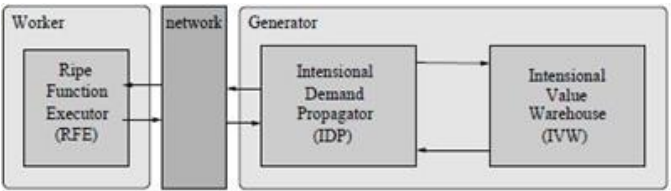

**Figure 4: Generator-Worker Execution Architecture [18]**

*Run-Time Interactive Programming Environment (RIPE)*

RIPE is used to provide visualization to the Lucid program. It helps in visualizing the data flow in the program. The user is given more control over the program at runtime with the aid of RIPE [18].

### 2.2.4. Purpose and Quality Requirements

To design a run-time system, which will execute procedural hybrid programs, the design should have quality like no language dependency, which will be able to run any program. It also should be effective on distributed computing, allows modification in run time without consideration and an observable system which can be studied for further improvements [19].

In figure 5 we can clearly see there are 6 GIPSY Nodes. Here, Node 1 is an independent GIPSY instance, which can run hybrid programs. Node 2 is made of DGT and DST instances. Node 3 hosts GIPSY Instance 2 with multiple DWT Instances. Node 4 hosts multiple DWT and one DGT Instance. Node 5 is composed of DST and DWT Instances. Node 6 is the main host of GIPSY Instance 3 [19].

### 2.2.5. Jini and JMS

In order to have the point of: usability, stability, programmability and deployment in the GIPSY multi-tier environment, implementation of Jini DMS and the JMS DMS under GEE multi-tier architecture has been unified. Jini is a service-oriented middleware technology in java for creating distributed systems that is based on Demand Generators and Demand Workers. On the other hand, JMS is a Java-based and message-oriented middleware technology based on Open Message Queue and JBoss Messaging [15].

### 2.2.6. AGIPSY Architecture

As can be seen in figure 6 a network composition of GNs, act autonomously without a direct interposition of external GIPSY entities like human user or other software, and control by its internal state. This internal control is presented by node manager (NM) [13].

*1. AGIPSY architecture specification*

- Multi-agent
- Loosely coupled
- Distributed
- Decentralized control and data allocation.

GNs are GIPSY Nodes, and GMs are GIPSY Managers. The coordination of GNs in this architecture is very important. The GNs communicate with each other if their tasks are beyond their knowledge [13].

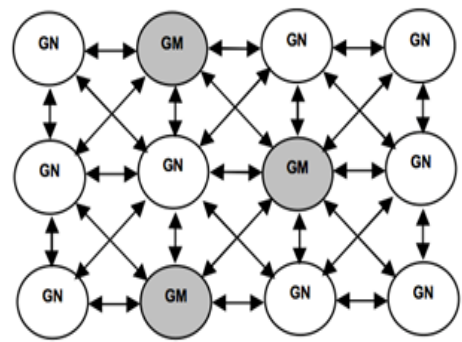

**Figure 6: AGIPSY Architecture [13]**

*2. AGIPSY Behavior*

Fault Tolerance & Recovery: GNs save the state of the GMs after every issue to be able to recover in case of any possible fault include DG and DW [13].

*Self-Maintenance*: NMs are assigned to all GNs which able them and consequently the whole model to follow self-management policies [13].

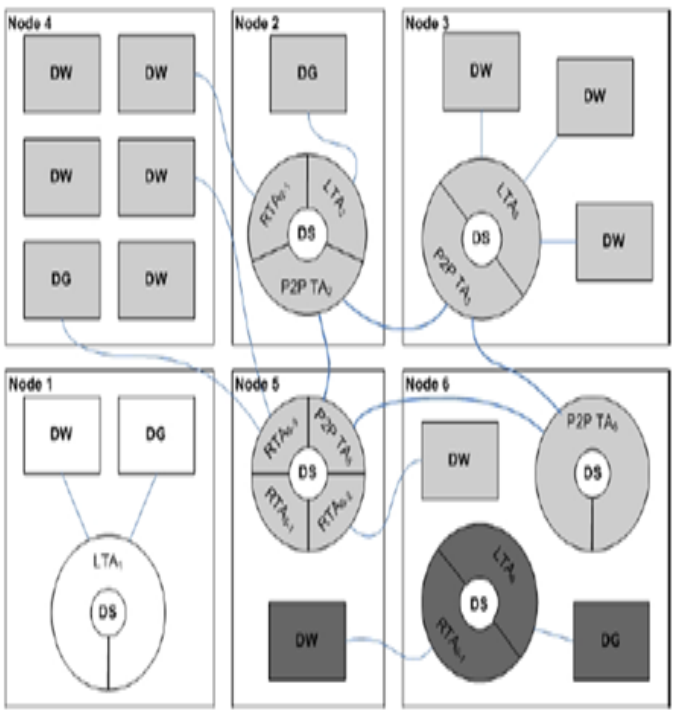

**Figure 5: Example of a GIPSY Nodes Network [19]**

*Self-Configuration*: By registering each user computer to a web portal on GM, each node has access to other nodes computational resource [13].

*Self-Optimization*: A user can advertise new performance optimum to all other nodes. AGIPSY should tackle any resource request conflict between GNs [13]

*Self-Healing*: All GNs iterate their essential state. NMs monitor GNs state and inform their GMs. GMs ask high load GNs to transfer some of their tasks to other GNs [13].

*Self-Protection*: By using well-defined message pattern and two-level communicating mechanism AGIPSY secure itself. At first level GNs communicate by following Demand Migration System (DMS) standard. It means registered in DMS. In second level NMs talk on behalf of their GNs and GMs which have to be registered first. So communication is only possible between registered entities [13].

## 2.3. SUMMARY

Quality Attributes are absolutely necessary project measurements that are required to understand and further improve the source code in a project. The important quality attributes are Understandability, extensibility, adaptability, reliability, accuracy and Maintainability [7]. The purpose of this study is to understand the usage of MARF and GIPSY if they can help deliver these quality attributes.

### 2.3.1. MARF

We summarize the different aspects of MARF, its working and its architecture and how it allows research in audio recognition by working as a platform for research to improve existing algorithms. We also look at the different quality attributes that MARF helps to achieve in different fields and projects. It increases the reusability and maintainability of the application from the engineering perspective [11]. MARF is adaptable as its approach is used for DEFT challenges [9]. MARF has also quality of extensibility as it allows to add new algorithms in its framework [1, 8]. As shown in the results in Figure 3 the system is reliable with high accuracy [8].

### 2.3.2. GIPSY

GIPSY and AGIPSY bind the static and dynamic typing between intensional and imperative languages, which present more Flexibility and Interoperability. In addition, AGIPSY architecture and its behavior like Self-Configuration, Self-Optimization, Self-Healing and Self-Protection present quality attributes like Fault-Tolerability, Recoverability and Stability [13].

Node structure in GIPSY can present Replaceability; in case of facing any serious problem in a node, that node can be replaced instead of whole system [13].

By GIPSY architecture, the difficulty of inter-process communication in distributed system is solved because of scalable DST, which distributes the load. Moreover, when programs are running, the multi-tier architecture allows new nodes and tiers introduction or elimination dynamically for better output [19].

Furthermore, various execution topologies can be chosen for the same source code, which can be decided at run time or before execution. Besides, components interact by exchanging demands, which are kept in the DST. It also saves lifetime result, which can be studied for optimization in run time [19].

### 2.3.3. Code Analysis

To calculate metrics on the given projects we use different tools and plugins such as McCabe, Logiscope, Plugins CodePro [21] and Metrics 1.3.6 [22]. These plugins are installed in eclipse. The results of the metrics are achieved after compiling the project. SLOCC [23] is also used to achieve the number of programming languages and is used in Linux. To compute the number of lines of text, first the projects were compiled using the eclipse plugins. The projects were then run in McCabe, Logiscope and SLOC Count. The results are compared in the tables. Linux is used to compute some of the metrics of the project which are described as follows-

For computing number of files in project, following commands are used on the terminal of Linux -

find marf –type f | wc –l (Files in whole project)

find marf/src – type f | wc –l (Files in source folder)

find gipsy –type f | wc –l (Files in whole project)

find gipsy/src – type f | wc –l (Files in source folder)

For finding the number of classes first project is compiled with "make" command and then following commands are executed on Linux

find marf/src –type f – name "*.class"| wc –l

For finding the number of languages following command is executed that provides all the distinct type of files from which numbers of languages are inferred - find marf –type f | rev | cut –d marf –f1 | rev | sort | uniq –ic | sort – rn

find gipsy –type f | rev | cut –d gipsy –f1 | rev | sort | uniq –ic | sort – rn

The number of programming languages used for each project was deduced by calculating the number of files with extensions that are used for each programming language. For example, a .java file corresponds to the Java language, a.pl file corresponds to the Perl language. SLOC Count was also used to calculate the number of programming languages. The limitation of this tool is it unable to identify every programming language [23]. The results are computed as follows.

|  | MARF | GIPSY | |
|---|---|---|---|
|  | Value Achieved | Value Achieved | Support Used |
| Files | 963 (Whole Project) | 2498 (Whole Project) | LINUX |
|  | 391 (SRC folder) | 1824 (SRC folder) | LINUX |
| Classes | 199 | ` | Metrics 1.3.6 [22] |
|  | 200; 197 (public), 2 (protected), 1 (private) | 642 | Code Pro [21] |
|  | 201 | ~ | LINUX |
|  | 181 | ~ | McCabe |
| Lines of Text | 52633 (Total Number of Lines) [MARF Package] | 139677 (Total Number of Lines) | Code Pro, Metrics 1.3.6, Logiscope |
|  | 24596 (Lines of Code) [MARF Package] | 104073 (Lines of Code) |  |
|  | 5115 (Comments) [MARF Package] | 7966 (Comments) |  |
|  | 113804 | 100605 | SLOCCount |
| Programming Languages Used | Java, TeX/LaTeX, Make, Perl, Shell Script, DOS Batch Script | C, ShellScript, Java, HaskellScript, UML, JSP, DOS Batch Script | ohloh.net [24], LINUX |

**Table 2: Summary of Code Analysis of MARF and GIPSY**

**Graphical Representation of Programming Languages**

| MARF | | |
|---|---|---|
| Language | Lines of Code | Tool Used |
| Java | 109236 (95.99%) | SLOCCount |
| Shell Script | 2573 (2.26%) |  |
| Perl | 1540 (1.35%) |  |
| Python | 256 (0.22%) |  |
| C Shell Script | 199 (0.17%) |  |

**Table 3: Programming Language Analysis of MARF**

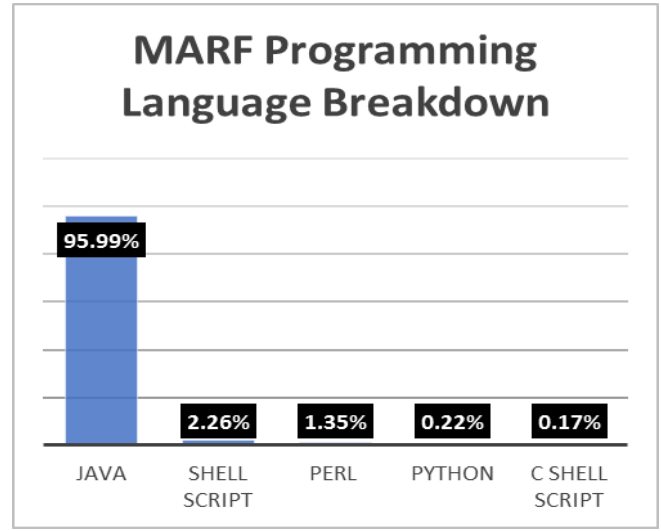

**Figure 7: Summary of Programming Languages used in MARF**

| GIPSY | | |
|---|---|---|
| Language | Lines of Code | Tool Used |
| Java | 98478 (97.89%) | SLOCCount |
| ANSIC | 1631 (1.62%) |  |
| Shell Script | 452 (0.45%) |  |
| Haskell | 44 (0.04%) |  |

**Table 4: Programming Language Analysis of GIPSY**

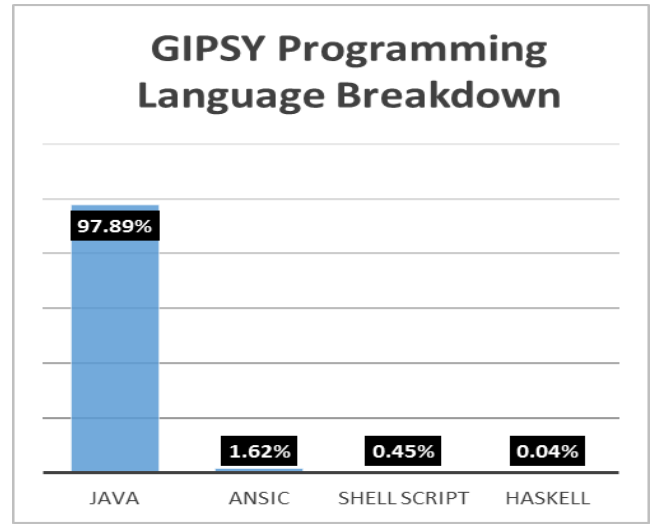

**Figure 8: Summary of Programming Languages used in GIPSY**

## 3. METRICS

Software metrics is defined as a measure of some property of a piece of software or its specifications. In other terms, a software metric measures (or quantifies) a characteristic of the software [25]. Some of the common software metrics are lines of code, complexity, coupling and cohesion. Metrics play a very important role in developing quality software. The demand for quality software has greatly increased over time and to keep up with that demand traditional metrics were insufficient. To keep up with the demand of the industry Object Oriented (OO) metrics are introduced. Most metric suites are only capable of performing analysis on software after it is completed or in final stages of development. OO metrics help conduct analysis in early stages of development as well.

Before we get into the study of the different software metric suites first we need to understand the different quality attributes and design property attributes.

*Quality Attributes*

1. *Reusability:* The implementation should have the capability to be reapplied in a new problem without the application of significant effort.
2. *Flexibility*: A design should allow the incorporation of change.
3. *Understandability*: The properties of design should have less complexity for easy comprehensibility.
4. *Functionality*: The responsibilities assigned to the class in the requirement analysis phase of the project should be satisfied by the class interfaces.
5. *Extendibility*: Ability of the implementation to incorporate new requirements.
6. *Effectiveness*: Ability of the implementation to achieve the desired functionality.

*Design Property Attributes*

The design property attributes comprise of the features of the system and the relationships between the different components of the system. These include *design size, hierarchies, abstraction, encapsulation, coupling, cohesion, composition, inheritance, polymorphism, messaging, and complexity* [26].

*Software Measurement Metrics*

In this section we study different software measurement metrics and understand their features along and establish pros and cons of each metric so as to understand which measurement metric is the best choice to satisfy the different quality attributes and also to understand which design property attributes and object oriented principles are utilized by the measurement metrics.

**3.1. Coupling Measurement and Object Oriented Programming**

Coupling is defined as: "The *measure of the strength of association established by a connection from one module to another"* [27].

When the coupling is strong between two modules then the higher the similarities between them. This may or may not increase the comprehensibility of the classes making it easier or harder to correct and use elsewhere. In Object Oriented (OO) systems the coupling measurement focuses on the usage dependencies between classes from static and dynamic analysis of the source code.

Object oriented coupling has been divided into three frameworks. First, Eder et al, classified coupling in terms of relationships which include interaction relationships between methods, component relationships between classes and inheritance between classes. Second, Hitz and Montazeri derived class level coupling and object level coupling based on the state of the object and its implementation and finally, Briand et al define coupling as the relationship between classes.

*Formalism*

Coupling helps in remedying the lack of standardized terminology and help define a formalism to express software measure. This will help in expressing a project in a more consistent, understandable and meaningful manner. This formalism can be expressed in an OO system in terms of the System with a set of classes *C,* Methods, Method Invocations, Attributes, Attribute References, Types and Predicates. Well defined relationships contribute to the concept of formalism. The various coupling measures are explained as follows.

*Coupling Measures*

Coupling Measures have been proposed by Chidamber and Kemerer , Li and Henry , Martin, Abreu et al, Lee at al and Briand et al .

*Definition of Measures*

1. *Coupling between objects (CBO)* – "CBO for a class is a count of the number of classes coupled with other classes" [28].

Def – $CBO(c) = |\{d \in C - \{c\}| \text{ uses } (c,d) \vee \text{ uses } (d,c)\}|$ where: C = set of all classes, c and d are the classes coupled to each other.

2. *Response for Class (RFC)* – RFC = |RS| where RS is the response set for the class where RS can be expressed as a union of the methods and responses for the class [28].

3. *Message passing Coupling (MPC)* – Defined as number of send statements present in a class.

$$MPC(c) = \sum_{m \in M_I(c)} \sum_{\acute{m} \in SIM(m) - M_I(c)} NSI(m, \acute{m})$$

Where SIM (m) -Statically Invoked Methods- is the set of m and NSI (m, m') -responses [29].

*4. Data Abstraction Coupling (DAC)* – Defined as the number of abstract data types defined in a class [29].

*5. Coupling Factor Briand et al. (CF)* – Defined as the ratio between the total number of client-server relationships that are not related by inheritance to the total number of client-server relationships in the system [30].

$$CF = \frac{\sum_{i=1}^{TC} \sum_{j=1}^{TC} client(C_i, C_j)}{TC^2 - TC - 2 \times \sum_{i=1}^{TC} |Descendants(c_i)|}$$

Where $(C_i, C_j)$ = client-server relationship

TC = Total number of classes

*6. Weighted Methods per Class (WMC)* – Measures overall complexity of class by summing up the cyclomatic complexity of each method in the class [33].

*7. Depth Inheritance Tree (DIT) and Number of Children (NOC)* - DIT measures the maximum path length from root to node and NOC is the count of the immediate descendants [33].

*Advantages of Coupling Measurement*

Coupling measurement imbues the classes with formalism which helps improve comprehensibility of the project. It helps in extracting multiple features such as granularity [27], stability [27], Criticality, Testability, Direct and Indirect connections between attributes and methods and most importantly gives insight into external quality attributes such as Understandability, Maintainability and Reusability. It also helps in error detection and fault detection in specific areas of the project [33].

*Disadvantages of Coupling Measurement*

Coupling Measurement extremely relies on relationships between classes and methods in the system. These relationships should be clear and comprehensible which is not always the case which causes this measurement to provide improper data. It also highly relies on formalism in the absence of which the project cannot be comprehended. There are other kinds of measurements that have a simpler approach and deliver more insight into the project specifics.

**3.2. Cohesion Metric Suites**

Cohesion is defined as the degree of belongingness of module components or elements together and measures how strongly modules are related by elements, and if the system has high cohesion it is better but if it has low then we should split it so that it can have greater cohesion individually [51].

*1. LCOM (Chidamber & Kemerer)*

It measures how similar the methods are by taking into consideration the instance variable or attributes [45]. As we are considering Chidamber & Kemerer so the formula of LCOM is

**P=Ii, | Ii∩Ij=∅** the set of method pairs that do not share attributes

**Q=Ii, | Ii∩Ij≠∅** the set of method pairs that share at least one attribute

**LCOM= P−Q, if P>Q**

*2. LCOM (Li & Henry)*

It counts the number of disjoint components in the graph in which it considers the methods as nodes and the sharing of attributes between these methods as an edge [46].

*3. LCOM (Hitz & Montazeri)*

It also counts the number of disjoint components in the graph but her the difference from Li and Henry method is that we are also taking into consideration the calls between methods as an edge [47].

*4. LCOM (Henderson-Sellers)*

$$LCOM = \frac{m - \frac{\sum_{i=1}^{a} p(A_i)}{a}}{m-1}$$

Here **m** is the number of methods, **a** is the number of instance variables and **p ($A_i$)** is the number of methods that access instance variable $A_i$ [49].

*5. COH (Briand et al.)*

$$Coh = \frac{\sum_{i=1}^{a} p(A_i)}{m * a} \quad [32]$$

*Tight and Loose Class Cohesion (Bieman & Kang)*

In tight class cohesion it counts the number of directly connected pairs of methods. Here if the methods are connected by an attribute then it is called directly connected but in loose class cohesion we also count the transitively connected pairs of methods indirectly [48].

*6. Class cohesion (Bonja and Kidanmariam)*

Here we count the ratio of the total number of similarities between all pairs of methods to the total number of pair of methods [50].

$$Similarity\ (m_i, m_j) = \frac{I_i \cap I_j}{I_i \cup I_j}$$

Here $I_i$ and $I_j$ are the sets of attributes and $m_i$ and $m_j$ are methods [48].

*Advantages of Cohesion Metric Suites*

Modules have increased understanding (They become simpler with fewer operations). Maintenance of the system becomes easier, because fewer modules are changed logically in the domain affect, and because changes in one module require fewer changes in other modules. Reuse becomes easier because application developers will find the component they require more easily among the cohesive set of operations given by module.

*Disadvantages of Cohesion Metric Suites*

If the cohesion is low then relationship between classes decreases and reusability increases.

### 3.3. MOOD Set of Object-Oriented Software Metric

Evaluating the effectiveness of MOOD metrics: encapsulation, inheritance, coupling and polymorphism is an essential question in software quality. Could they be used for providing overall realistic and effective assessment in software systems?

We have to distinguish between different types of attributes types in evaluation:

1) Direct measurement, which measures attributes that does not depend on any other attribute.
2) Indirect measurement, which measures one attribute is related to one or more than one other attributes.
3) Internal attributes of product or process, which can be measured purely by product or process.
4) External attributes of product or process depends on relation between product or process and its environment.

### 3.3.1. Mood Quality Properties

*1. Encapsulation*

The Method Hiding Factor (MHF) and Attribute Hiding Factor (AHF) are metrics we use for measuring encapsulation. Encapsulation means that which methods and attributes of one class are visible to other classes. In language like C++ protected methods and attribute make this calculation problematic. For protected method and attributes we use below formula, and protected methods is counted between 0 and 1.
(Number of classes not inheriting the method) / (Total number of classes – 1)
Third direct criteria in Kitchenham framework ask us if all methods or attributes for MHF or AHF are equivalent. At first it looks the answer is no, because the need to hide methods or attributes depends on functionality and their condition. But we have to consider that MHF and AHF just care about amount of information hiding not the quality of this hiding, and so they meet direct criteria 3.

*2. Inheritance*

For Method Inheritance Factor (MIF) and Attribute Inheritance Factor (AIF) metrics:

1. Two systems with two different amount of inheritance definitely will have two different MIF.
2. A system with more inheritance will show larger MIF.
3. MIF and AIF are normalized so adding an inherited method to a large system will increase MIF less than adding an inherited method in small system.
4. It is obvious that different system and programs can have the same MIF.

*3. Coupling*

Two approaches can be considered here -

Direct measure of inter-class coupling or indirect measure of related characteristics like:

• Lack of encapsulation
• Lack of reuse potential
• Lack of understandability
• Lack of maintainability

For direct measure it is clear that system with more interclass coupling will have larger Coupling Factor (CF). Moreover, different systems can have same CF and we can consider it as a valid measure of coupling.

In indirect approach we cannot consider CF as a valid measurement for external attributes, because: Higher CF does not mean more complexity. We can design a complicated system with low coupling or a small simple system with high level of coupling. In addition, relationship between encapsulation and coupling is not clear. Moreover, for relation between complexity and coupling, we can find a class which is not coupled but it is difficult to understand.

*4. Polymorphism*

Polymorphism Factor (PF) is an indirect metric. PF for system without inheritance is undefined and we have an unexpected discontinuity, so we have to consider PF as an invalid metric. If we remove metric discontinuity we can consider PF as a valid metric.

### 3.3.2. Theoretical Validation

According to the framework of Kitchenham et al. metrics should exhibit below properties to be acceptable:

*Direct Metrics:*

1. Measureable attribute should allow different entities to be distinguished.
2. Valid metric should preserve all intuitive notions about the attribute, and the way metric distinguishes between entities.
3. Each unit contributing to a valid metric is equivalent.
4. It is possible that different entities have the same attribute value.

*Indirect Metrics*:

1. The metric should be based on precisely defined model of relationship between attributes.
2. The model has to be consistent in different dimensions.
3. The model should not have any unexpected discontinuities.
4. The metric have to use units and scales in an accurate manner.

### 3.3.2. Empirical Validation

Practical measure in three releases (R1,R2,R3) of a laboratory electronic retail system (ERS) and the second release of a suite of image processing programs (EFOOP2) confirms above theatrical approach and we can consider the six MOOD metrics as valid measurement metric base on Kitchenham framework. The results of this validation are shown in Table 5.

|     | R1   | R2        | R3        | EFOOP2    |
|-----|------|-----------|-----------|-----------|
| AHF | 100  | 100       | 100       | 100       |
| MHF | 0    | 20.4      | 20.4      | 20.4      |
| AIF | 12.5 | 0         | 0         | 0         |
| MIF | 9.1  | 0         | 0         | 0         |
| CF  | 0    | 5.8       | 5.8       | 3.0       |
| PF  | 60   | Undefined | Undefined | Undefined |

**Table 5: The MOOD Metric s (ERS) [40]**

*Advantages of the MOOD Metric Suite*

MOOD Metrics operate at system level and they offer different assessments of systems which help in performing a comparative study of the projects in different conditions and environments. It is very useful for project managers as they provide an overall assessment of the system.

*Disadvantages of the Mood Metric Suite*

The validity and authenticity of the results of MOOD are not always considered as there are insufficient empirical validations and studies performed with this metric suite. The relationships between the metrics and the quality attributes are generally not considered.

### 3.4. Quality Model for Object Oriented Design (QMOOD)

QMOOD (Quality model for Object Oriented Design) is a hierarchical model which is used for assessment of high level design quality attributes in object oriented design. This model identify and analyze the metric of project at the early stages of development so that developer can fix problems, remove irregularities and eliminated unwanted complexity at the initial of the development cycle, which leads to the quality product. The QMOOD identify and analyze the metrics for quality product in following steps [34].

*1. Identifying Design Quality Attributes*

Initially the suite identifies the quality attributes of design, which are functionality, effectiveness, understandability, extendibility, reusability and flexibility [34].

*2. Identifying Object-Oriented Design Properties*

In the next step Object Oriented Design properties are identified, which are abstraction, encapsulation, coupling, cohesion, complexity, design size, messaging, composition, inheritance, polymorphism, and class hierarchies, where last five are new design concepts which plays an important role in the quality of an object-oriented design [34].

*3. Identifying object oriented design metrics*

Metrics are defined according to the design properties.

*4. Identifying Object Oriented Design Component*

A set of components which can help in analyzing, and implementing an object-oriented design are attributes, methods, objects (classes), relationships, and class hierarchies [34].

Table 6 gives the descriptions of each metric and their relation to the design properties [34].

| METRIC | NAME | DESCRIPTION |
|---|---|---|
| DSC | Design Size in Classes | This metric is a count of the total number of classes in the design. |
| NOH | Number of Hierarchies | This metric is a count of the number of class hierarchies in the design. |
| ANA | Average Number of Ancestors | This metric value signifies the average number of classes from which a class inherits information. It is computed by determining the number of classes along all paths from the "root" class(es) to all classes in an inheritance structure. |
| DAM | Data Access Metric | This metric is the ratio of the number of private (protected) attributes to the total number of attributes declared in the class. A high value for DAM is desired. (Range 0 to 1) |
| DCC | Direct Class Coupling | This metric is a count of the different number of classes that a class is directly related to. The metric includes classes that are directly related by attribute declarations and message passing (parameters) in methods. |
| CAM | Cohesion Among Methods of Class | This metric computes the relatedness among methods of a class based upon the parameter list of the methods [3]. The metric is computed using the summation of the intersection of parameters of a method with the maximum independent set of all parameter types in the class. A metric value close to 1.0 is preferred. (Range 0 to 1) |
| MOA | Measure of Aggregation | This metric measures the extent of the part-whole relationship, realized by using attributes. The metric is a count of the number of data declarations whose types are user defined classes. |
| MFA | Measure of Functional Abstraction | This metric is the ratio of the number of methods inherited by a class to the total number of methods accessible by member methods of the class. (Range 0 to 1) |
| NOP | Number of Polymorphic Methods | This metric is a count of the methods that can exhibit polymorphic behavior. Such methods in C++ are marked as virtual. |
| CIS | Class Interface Size | This metric is a count of the number of public methods in a class |
| NOM | Number of Methods | This metric is a count of all the methods defined in a class. |

Table 6: Definition of QMOOD Design Properties [34]

5. *Mapping Quality-Carrying Component Properties to Design Properties*

The properties of design component are then classified based on the design properties. For example, *name* is a quality-carrying property of attributes, methods and class. When *names* are self-descriptive, this means that they provide better understandability and reduces the design property complexity. Where the quality is carrying property, *encapsulation* of attributes, methods, and classes, it is the same as the general design property, encapsulation. Similarly, the remaining quality-carrying properties can be grouped into design properties described above [34].

| Design Property | Derived Design Metric |
|---|---|
| Design Size | Design Size in Classes (DSC) |
| Hierarchies | Number of Hierarchies (NOH) |
| Abstraction | Average Number of Ancestors (ANA) |
| Encapsulation | Data Access Metric (DAM) |
| Coupling | Direct Class Coupling (DCC) |
| Cohesion | Cohesion Among Methods in Class (CAM) |
| Composition | Measure of Aggregation (MOA) |
| Inheritance | Measure of Functional Abstraction (MFA) |
| Polymorphism | Number of Polymorphic Methods (NOP) |
| Messaging | Class Interface Size (CIS) |
| Complexity | Number of Methods (NOM) |

Table 7: Design Metrics for Design Properties [34]

6. *Assigning Design Metrics to Design Properties*

The definition of each design property is referring at least one design metrics so here we map the design metrics to design property which are shown IN Table 7.

7. *Linking Design Properties to Quality Attributes*

There are many development books [35], [36] and publications [37], [38] that describe a basis for relating product characteristics to quality attributes. Here we link the design properties to quality attributes. For instance if we take abstraction design property that has a great influence on the quality attributes like flexibility, effectiveness, functionality, and extendibility quality attributes of a design. Similarly other design properties also have an effect on quality attributes which are described in Table 8 [34].

| | Reusability | Flexibility | Understandability | Functionality | Extendibility | Effectiveness |
|---|---|---|---|---|---|---|
| Design Size | ↑ | | | ↑ | | |
| Hierarchies | | | | ↑ | | |
| Abstraction | | | | | ↑ | ↑ |
| Encapsulation | | ↑ | ↑ | | | ↑ |
| Coupling | | | | | | |
| Cohesion | ↑ | | ↑ | ↑ | | |
| Composition | | ↑ | | | | ↑ |
| Inheritance | | | | | ↑ | ↑ |
| Polymorphism | | ↑ | | ↑ | ↑ | ↑ |
| Messaging | ↑ | | | ↑ | | |
| Complexity | | | | | | |

Table 8: Quality Attributes—Design Property Relationships [34]

Furthermore based upon the relationship shown in above table, quality attributes are weighted proportionally according to the design properties. Table 8 described the weighted relation between design properties and design attributes [34].

3.4.1. Validation of QMOOD

To gain a further insight into the working of QMOOD we study two experiments. In the first experiment several versions of two popular WINDOWS application frameworks, Microsoft Foundation Classes (MFC), and Borland Object Windows Library (OWL) are analyzed by QMOOD model. In the result it was found that the quality attributes increase from one release to the next. This is because early structure of the framework may be unstable but after several initial releases, a framework can be expected to improve their usability, reduce complexity and can be easily understood [34].

In the second experiment a set of 14 projects are evaluated by QMOOD and they are ranked by total quality index (TQI) measure [39]. On the other hand group of 13 independent evaluators study the quality of the 14 projects. Each evaluator develops a set of design and implementation heuristic. The evaluator then assigns these heuristics to the quality attributes and then measures the quality of each project according to these heuristics. Results of the evaluations show that there was considerable disagreement in the rankings of one of the project between the evaluators. This disagreement in the rank is due to the number of classes in that project. The numbers of classes are significantly less than the other projects.

Evaluators who did not include a heuristic for the number of classes in their evaluations, ranked the project higher than evaluators who included a in their assessments [34].

| Quality Attribute | Index Computation Equation |
|---|---|
| Reusability | -0.25 * Coupling + 0.25 * Cohesion + 0.5 * Messaging + 0.5 * Design Size |
| Flexibility | 0.25 * Encapsulation - 0.25 * Coupling + 0.5 * Composition + 0.5 * Polymorphism |
| Understandability | -0.33 * Abstraction + 0.33 * Encapsulation - 0.33 * Coupling + 0.33 * Cohesion - 0.33 * Polymorphism - 0.33 * Complexity - 0.33 * Design Size |
| Functionality | 0.12 * Cohesion + 0.22 * Polymorphism + 0.22 Messaging + 0.22 * Design Size + 0.22 * Hierarchies |
| Extendibility | 0.5 * Abstraction - 0.5 * Coupling + 0.5 * Inheritance + 0.5 * Polymorphism |
| Effectiveness | 0.2 * Abstraction + 0.2 * Encapsulation + 0.2 * Composition + 0.2 * Inheritance + 0.2 * Polymorphism |

**Table 9: Computation Formulas for Quality Attributes [34]**

While comparing a results of QMOOD and evaluators it was also found that evaluators, who did not include heuristics of the number of classes on the quality of the design shows significant difference in result [34].

*Advantages of QMOOD Suite*

The QMOOD model has the ability to estimate overall design quality and give comprehensive design information. The use of this tool is mainly to collect data or design metrics and compute quality attributes. It is found to be non-intrusive and easy to apply. It is the most useful software metric suite as it is easy to use and also gives comprehensive data [34].

*Disadvantages of QMOOD Suite*

QMOOD metrics have not been very effective in detecting fault proneness and they have practical limitations to the effectiveness of the metrics over the course of many software updates and releases. With many changes to the software or in other words the dynamic nature of software reduces the effectiveness of the QMOOD metric suite [34].

### 3.5. State of the Art Metric Suites

The field of software measurement has become a huge field of interest for developers and managers and has led to the development of many new state of the art metric suites. In some cases the previously mentioned metric suites might not give all the required information for the success of the project. Developers, researchers and managers have come up with their own metric suites to derive new metric suites to suit their requirements. We study a few of them.

### 3.5.1. Code Churn

The evolution of code travels constantly in the overall development process. Every time a modification in the code generates a new risk that introduces susceptibility. This vulnerability is tried to be dealt working replacement has been introduced to make the measurements precise enough for the development outcomes. Due to the software changes overtime, the understandability and prediction of the effect that the changes are imposing becomes difficult. Study made it possible to relate complexity and faults in software [41].

*Purpose*

Now, we measure the software evolution process accurately when the measurement relates to specifically fault injections in the software [42]. Under the explanation of relative complexity [44] the discussion of a baseline is done which tells us that there should be a scale to compare the evolving software.

*Relative Complexity ($\rho$)*:
Can be obtained in 2 steps.
(a) Scoring program module in baseline for each metric.
(b) Each program module should be characterized by a single value (cumulative measure of complexity).

$$\rho_i = \sum_j \lambda_i d_{ij}$$

Where $\lambda_i$ is the Eigen value associated with the $j^{th}$ factor $d_{ij}$ is the $j^{th}$ domain metric of the $i^{th}$ program module. For getting the relative complexity: statements and cycles were being considered and observed and all other metrics related to faults. For getting the relative complexity: statements and cycles were being considered and observed and all other metrics related to faults. For this purpose as many relative complexities will be present so average relative complexity can be obtained where $N_b$ is the cardinality of the set of modules on build $b$ which the baseline build and $i$ is starting build .

$$\rho^{-b} = \frac{1}{N^b} \sum_{i=1}^{N^b} \rho_i^b \rightarrow \text{Relative Complexity for the baseline system}$$

$$\rho^{-b} = \frac{1}{N^b} \sum_{i=1}^{N^b} \rho_i^b \rightarrow \text{Relative Complexity for the baseline system}$$

$$R = \sum_{i=1}^{N^b} \rho_i^b \rightarrow \text{Relative Complexity of each module}$$

*Experiment*

Modules of software were categorized into MA (Set of modules that were in the early build and were removed prior to the later build), MB (Set of modules that have been added to the system since the earlier build) and MC (Set of modules present in both builds) [39].

$$R^1 = \sum_{c \in Mc} \rho_c^i + \sum_{a \in Ma} \rho_a^i \rightarrow \text{Relative Complexity at early project}$$

$$R^1 = \sum_{c \in Mc} \rho_c^i + \sum_{a \in Ma} \rho_a^i \rightarrow \text{Relative Complexity at later project}$$

Where $\rho_a$ is the relative complexity of later builds, $\rho_b$ is relative complexity of earlier build and $\rho_c$ is the relative complexity of both builds. Complexity increases as the size of the software increases and as the changes are made, and the system developed later is considered more complex if $Ri > Rj$. After that the difference between the 2 builds is judged based on the relative complexities as shown in figure 9.

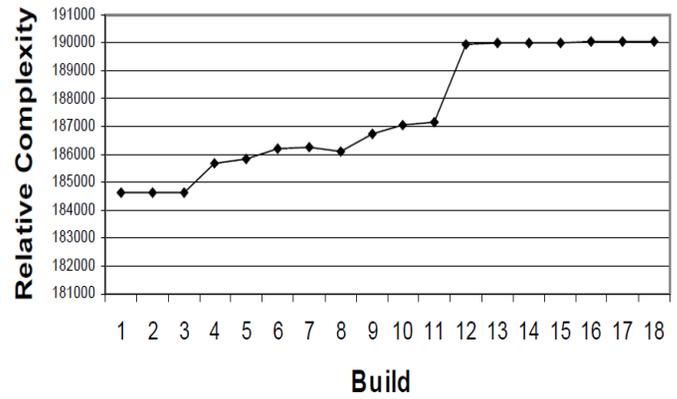

**Figure 9: Trend w.r.t Relative Complexity**

This is the demonstration of a real-time embedded system which shows comparison between 19 successive builds [44]. Keeping in mind the stability factor or measure it has been concluded that software complexity metric along with the relative complexity metric [43] with code churn and deltas are very important for the quality of the program as they are used as a help in different kinds of testing.

*Advantage of Code Churn*

It lets us know the measure of code churn as study made it possible to relate complexity and faults in software and it has the greatest correlation with the criterion measure of trouble reports.

*Disadvantage of Code Churn*

It does not tell us that how much change the system has undergone and it also has a low correlation with the people measure as number of software developers involved in changes is apparently not related to the magnitude of the change.

### 3.5.2. Maintainability Index (MI)

Maintainability Index (MI) [53] [54] is a software metric which measures how maintainable (easy to support and change) the source code is. The maintainability index is calculated as a factored formula consisting of Lines of Code, Cyclomatic Complexity and Halstead volume. To calculate first we need to measure the following metrics from the source code:

- V = Halstead Volume
- G = Cyclomatic Complexity
- LOC = count of Source Lines of Code (SLOC)
- CM = percent of lines of Comment (optional)

And then apply this formula:

**MI = 171 - 5.2 * log2 (V) - 0.23 * G - 16.2 * log2 (LOC) + 50 * sin (sqrt (2.4 * CM))**

There are some limitations of MI as follows:

• As it is a composite number it is hard to understand what causes a particular value.
• Halstead Volume is difficult to compute in some cases.
• Generally counting the number of lines have no relation with maintainability.
• There is no logical argument why this formula contains particular constants, variables and symbols [55].

### 3.5.3. SIG Maintainability Model

Based on limitations of MI minimal requirements for new model are that it should have a straightforward definition that is easy to compute, technology independent, simple to understand and enable root-cause analysis. According to these requirements SIG defines an intrinsic model which follows ISO 9126 [52] guidelines. It uses following metrics for measuring maintainability in existing software systems:

• *Volume*: volume of source code determines the analyzability of system.
• *Complexity*: complexity of source code influences changeability and testability of system
• *Size per unit*: influences analyzability and testability.
• *Unit testing*: influences analyzability, testability and stability
• *Duplication*: influences analyzability and changeability

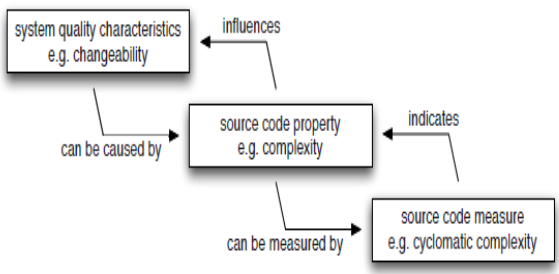

**Figure 10: SIG Maintainability Model [55]**

*Benefits of this model*:
• Uses simple code metrics (LOC, Cyclomatic Complexity, etc.), easier to measure.
• Allowing root-cause analysis as maintainability is broken down.
• It is independent from the implementation language and environment of technology.
• Volume influences complexity, but also intrinsic characteristics ("how is the software") are considered, so mere volume does not dominate the results.

*Limitations*:
• Ranking systems by volume is not extremely accurate.
• This metric does not give a report on attributes that affect stability, testability and changeability

• Do not provide specific recommendations useful for better maintainability, reducing future maintenance costs.

### 3.5.4. Yesterday's Weather

This metric is used illustrate the changes in the evolution of object-oriented software systems in order to predict changes in the class in future. Yesterday's weather (YW), is the approach which emphasizes on selecting those classes which changes more recently and this is helpful to start reverse engineering. This approach uses the historical data which tells the information about change in the classes, system versions etc. [56].

There are 3 measurements that are necessary to compute YW and they are:
1. Evolution of Number of Methods (ENOM)
2. Latest Evolution of Number of Methods (LENOM)
3. Earliest Evolution of Number of Methods (EENOM)

*Evolution of Number of Methods*
We use ENOMi to represent number of methods in versions i-1 and i of class C:

$(i > 1)$ $ENOM_i(C) = |NOM_i(C) - NOM_{i-1}(C)|$ (NOM- Number of Methods)
$ENOM_{j..k}$ is the sum of methods added or removed from the subsequent version of j to k out of n versions of class C: [56]

$(1 \le j < k \le n)$ $ENOM_{j..k}(C) = \sum_{i=j+1}^{k} ENOM_i(C)$
We use this measurement to show overall changes occurred during the lifetime of class.

*Latest Evolution of Number of Methods*
In this we consider only the recent changes over the past changes by using weighting function $2^{i-k}$ which decrease the importance of change of version i, where in previous method all changes has same importance [56].

$(1 \le j < k \le n)$ $LENOM_{j..k}(C) = \sum_{i=j+1}^{k} ENOM_i(C) 2^{i-k}$

*Earliest Evolution of Number of Methods*
In this we consider the changes closer to the first version of the history over the changes appear closer to latest version [56].
$(1 \le j < k \le n)$ $EENOM_{j..k}(C) = \sum_{i=j+1}^{k} ENOM_i(C) 2^{k-i+1}$

*Benefits and Limitations of Yesterday's Weather*
Yesterday's weather is useful in detecting the candidates for reengineering which are not detectable if we focus only on the last version of the system by analysing the size of the class [56]. Reengineering is the process of creating design or other documentation from code [57].

The drawback of this approach is, they compress the large amounts of data into numbers which emerges the inherent noise [56], i.e., the loss of predictability [58] in terms of predicting changes in the system.

*Comparative Study*

While we have discussed a variety of software metrics, we need to look at the performance of these metrics under different circumstances. Empirical Validation is a necessary factor for demonstrating the usefulness of a metric. We perform a study from two different experiments and compare the metric suites against each other to further gain an understanding about them and to look at how they affect the quality attributes that are important for the success of a project.

*Comparison of CK metric suite, MOOD and QMOOD in terms of detecting Fault Proneness [59]*

In this experiment, an empirical validation is conducted on the mentioned OO metrics to answer if these metrics can detect fault-prone classes and if these metrics can be reused in multiple future releases of software. The experiment was conducted with the help of the open-source software, Rhino [60]. To answer the second question, we need to have a bottom-up approach and Rhino does exactly that. We perform a comparison on a project with respect to fault detection by utilizing CK metrics, MOOD and QMOOD.

*Experiment*

A project with a considerable number of faulty classes to test the hypothesis of detection of fault-prone classes by CK metrics, MOOD and QMOOD. To get a better understanding of the result set the experiment was conducted with different versions of Rhino namely 14R3, 15R1, 15R2 15R3 and 15R4. The result of the experiment is shown in the graph in figure 11. From the figure we can answer the first question and conclude that the tested OO software metrics are very well capable of detecting fault-prone classes in its iterative phase. The accuracy of the metrics was ranged between 69%-85.9% [60]. Figure 11 also shows that the OO software metric suites can identify fault prone classes in multiple, sequential releases of a software.

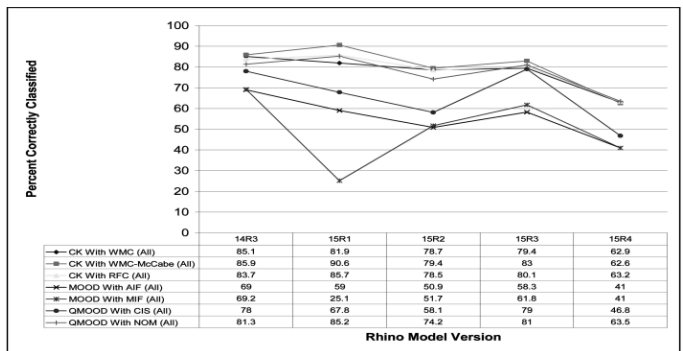

**Figure 11: Rhino Model Validation of all models [60]**

*Result Discussion*

The results of this experiment indicate CK and QMOOD OO class metrics are useful to develop quality classification models and predict defects. On the other hand MOOD metrics were not that good at detecting faults. Further analysis into the experiment indicated that the CK metrics suite was the best at detecting fault-proneness. Hence, we can conclude that CK metric suites are more reliable when compared to other metric suites.

*Comparison of QMOOD with System Design Instability (SDI) and System Design Instability with Entropy ($SDI_e$)*

The goal of this paper is to examine Bansiya and Davis quality model [62] to find connection with the stability studies of Olague et al. [63] and Li et al. [64] using same data [65].

*System Design Instability (SDI) and System Design Instability with Entropy (SDIe)*

A method is usually categorized as agile method based on the twelve principles declared by the Agile Alliance [61]. Extreme programming is one way to implement these twelve principles which makes the design of the software straightforward with better functionalities. Stability is a quality factor which helps to estimate the simplicity of software development protecting the design. One of the metrics for stability is proposed by Li *et al.* is System Design Instability (SDI) which detects design changes by tracking the percentage of classes added, deleted and whose name changed. SDI metric is also able to detect development progress which can help to make decisions during software development. Olague *et al.* proposed $SDI_e$ metric where he joined the idea of entropy. Because of partial automation $SDI_e$ metric is much easier to calculate than SDI metric. Based on previous iteration this metric calculates. Bansiya and Davis proposed QMOOD Quality Model based on their study of the ISO 9126. The six quality factors of this model are reusability, flexibility, understandability, functionality, extendibility and effectiveness. They also proposed eleven design properties which have good and bad relation with mentioned quality factors. Based on the weights of positive and negative values they calculate the quality factors and then sum them to calculate the Total Quality Index (TQI). So,

TQI = Reusability + Flexibility + Understandability + Functionality + Extendibility + Effectiveness [65].

*Experiment*

We have involves six projects which were developed using extreme programming. Here, first the normalized SDI, SDI$_e$ and TQI values has been computed and then graphed. From figure 12 we can see that in Project A, SDI metric graph follows the graph of TQI value but SDI$_e$ metric graph does not. In Project B, both SDI metric and SDI$_e$ metric graphs follow the TQI graph but SDI$_e$ metric graph is closer than SDI metric graph. But in Project C, both stability metrics do not follow the TQI graph. In Project D, SDI$_e$ metric graph follows the TQI graph. In Project E, we again see that both SDI metric and SDI$_e$ metric graphs follow the TQI graph. In Project F, because of not using extreme programming both stability metrics graph do not follow the TQI graph [65].

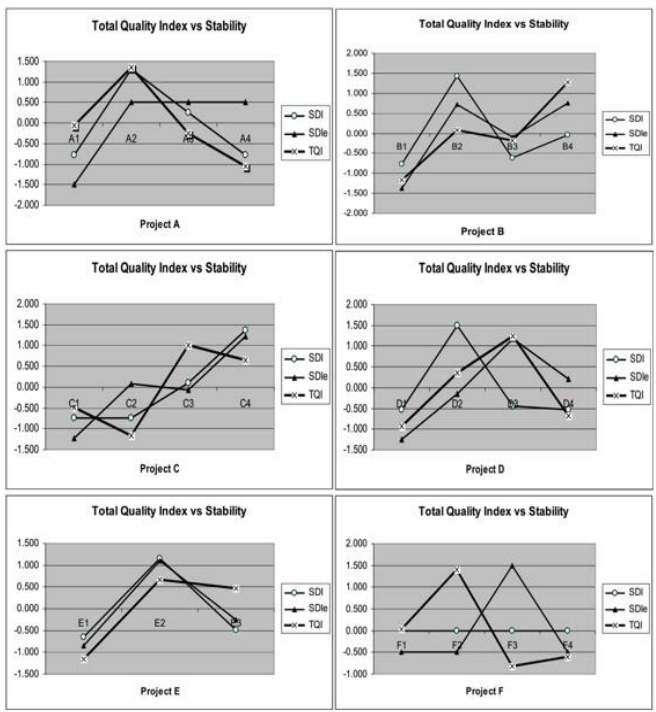

**Figure 12: TQI vs Stability [65]**

*Result and Discussion*

There is a definite association among stability metrics and TQI value especially when extreme programming has been used. Here, SDI metric and TQI value have more powerful connection than SDI$_e$ metric and the TQI value. But to estimate stability it is better to use TQI rather than SDI metric because of the human involvement in SDI in analysis phase [65].

**3.6. Summary**

We have studied a variety of metrics which include OO metrics and state of the art metrics and have also performed studies on various experiments conducted by different researchers to understand the authenticity and comprehensiveness of the result sets achieved by each metric. The main conclusions we drew from the study were as follows –

1. Class size impacts metric performance.
2. Metrics have limitations due to the dynamic nature of software and its iterations.
3. In terms of the amount of information provided by the metrics suite, QMOOD provides a comprehensive list of metrics which helps understand the project best.
4. In terms of fault-detection CK metrics suite was the best with QMOOD not far behind.
5. In terms of quality, CK metrics is the best at predicting OO class quality, followed by QMOOD model.
6. The results given by MOOD are always validated because of insufficient empirical validations. It is also not very good at fault-detection.
7. The state of the art metrics only provide information in specific conditions for specific projects and might not give valid results in every scenario.
8. In terms of stability, in the experiment conducted with QMOOD, SDI and SDI$_e$ we conclude that QMOOD is better.
9. In terms of common characteristics, we can say that classes with low cohesion have less in common and have higher reusability and vice versa.

With regard to all the observations made we can rank the metrics as follows in table 10.

| Rank | Metric |
|---|---|
| 1 | CK Metrics Suite (Coupling) |
| 2 | QMOOD |
| 3 | MOOD |
| 4 | Cohesion |
| 5 | State of the Art Metrics (YF, SIG, Code Churn, MI) |

**Table 10: Ranking of Software Measurement Metrics**

## 4. METHODOLOGY

In the first part of this section consists of an overview of the Logiscope and McCabe Tools which are used to understand and preview the quality of the case studies. The second section consists of the design, interpretation and analysis of the top ranked metrics discussed in the previous section by using JDeodorant and the McCabe tool.

**4.1. Metrics with Tools: Logiscope and McCabe IQ**

Object-oriented programming is probably the most widely used in the development of software. As can be seen in the code-analysis, the case studies MARF and GIPSY are written in Java. The most preferred tool sets to perform code-analysis and get a further insight into the metrics we have studied earlier are LOGISCOPE [66] and the McCabe IQ tool [67].

**I. Logiscope**

We use the LOGISCOPE tool to perform quality analysis on the MARF and GIPSY case studies. To understand the results generated, first we have to understand the definitions and the formulas that calculate these different quality metrics.

*Description*

The capability of the product is to be modified. Modification can include correction, improvements or adaptation of the software due to change in environment and functional requirements [68]. The formula to calculate Maintainability is

Maintainability = Analyzability + Changeability + Stability + Testability

The Maintainability is calculated for the case studies for the four factors in the class criteria level.

*Analyzability*

Analyzability is defined as the capability of the software product to be diagnosed for deficiencies or causes of failures in the software, or for the parts to be modified. [68].

Analyzability = cl_wmc + cl_comf + in_bases + cu_cdused

*Changeability*

Changeability is defined as the capability of the software product to enable a specified modification to be implemented [68].

Changeability = cl_stat + cl_func + cl_data

*Stability*

Stability is defined as the capability of the software product to avoid unexpected effects from modifications of the software [68].

Stability = cl_data_publ + cu_cdusers + in_noc + cl_func_publ

*Testability*

Testability is defined as the capability of the software product to enable modified software to be validated [68].

Testability = cl_wmc + cl_func + cu_cdused

The operands being used to calculate these factors are explained as follows –

- cl_wmc: Weighted Methods per Class:
  Sum of the static complexities of the class methods. Static complexity is represented by the cyclomatic number of the functions.
  cl_wmc = SUM (ct_vg)
- cl_comf: Class comment rate
  Ratio between the number of lines of comments in the module and the total number of lines:
  cl_comf = cl_comm / cl_line
  Where:
  *cl_comm* is the number of lines of comments in the package,
  *cl_line* is the total number of lines in the package.
- in_bases: Number of base classes
  Number of classes from which the class inherits directly or not.
- cu_cdused: Number of direct used classes
  Number of classes used directly by the current class.
- cl_stat: Number of statements
  Number of executable statements in all methods and initialization code of a class.
- cl_func: Total number of methods
  Total number of methods declared inside the class declaration.
- cl_data: Total number of attributes
  Total number of attributes declared inside the class declaration.
- cl_data_publ: Number of public attributes
  Number of attributes declared in the public section or in the public interface of a Java class.
- cu_cdusers: Number of direct users classes
  Number of classes which use directly the current class.
- in_noc: Number of children
  Number of classes which directly inherit from the current class.
- cl_func_publ: Number of public methods
  Number of methods declared in the public section.

*Acceptable Values for Metric Criteria*

The metrics that are used to calculate the maintainability of the system need to have an acceptable range to understand in which module the quality of the software can be improved. We take two values *M1* and *M2* which lie between the minimum and maximum values of the acceptable range. We assume *M2* is greater than *M1*.

⇨ Acceptable Range = {Min < M1 < M2 < Max}

Now we compare the values M1 and M2 with the different software metrics as follows –

- Class Comment Rate:
The class comment rate is the ratio of comments in the class to the total number of lines. The total number of lines includes blank lines, comments and every other statement. For the code to be understandable, a minimum of 20% of the total number of lines should be comments.

⇨ cl_comm / cl_line = 0.20 (minimum) = 20%

Therefore, the higher the value, the better the comprehensibility, which increases fault detection as well.

- Number of Lines of Comment:
The larger number of comments, the better the understandability, although relevance of comments is important. Compactness of the code is as important as readability. Good programming style reduces the requirement of comments.

- Total Number of attributes:
Larger the number of attributes, larger the complexity, hence a lower value is preferred.

- Number of public attributes:
The presence of public attributes means that there are many methods that will be utilizing them, which increases inheritance, relationships with objects and method calls. This increases the complexity of the code and the testing time.

- Total Number of methods:
More the methods, more the complexity, hence lesser value is preferred.

- Number of public Methods:
The number of public methods is a measure of the total amount of effort required for the class development and writing functionalities for the class. It is a count of the number of public methods. It is useful to estimate the effort but the larger the value, more the testing and more the complexity.

- Number of Lines:
Lesser this value does not necessarily mean that the code is less complex. It can go either way depending on the programming style and understandability of the code. Hence, the number of lines is an inconclusive quality metric.

- Number of Statements:
The numbers of statements are defined by the lines of executable code. More the executable code, increase complexity and decrease maintainability. The number of statements should be less to improve maintainability.

- Weighted Methods for Class (WMC)
The Weighted Methods for Class measure the static complexity. It is defined by a cyclomatic number. Lower that number, lower the complexity of the code. This means that the code requires less amount of testing. Also if the number of methods are more reusability decreases.

- Number of direct used classes
The more the number of classes used by a class, the more the dependability and coupling will be.

- Number of direct users classes
The more the number of classes being dependent on a class again increases coupling.

- Number of base classes
This metric shows the count of parent classes. The lesser the value the lower the coupling is.

- Number of children
This metric shows the count of child classes. The lesser the value the lower the coupling is.

| | | Comparison of Metrics (*M1<M2*) | | | |
|---|---|---|---|---|---|
| S.NO | Mnemonic | Metric | Min | Max | Value |
| 1 | cl_comf | Class comment rate | 0.2 | ∞ | M2 |
| 2 | cl_comm | Number of lines of comment | −∞ | ∞ | M2 |
| 3 | cl_data | Total number of attributes | 0 | 7 | M1 |
| 4 | cl_data_publ | Number of public attributes | 0 | 0 | M1 |
| 5 | cl_func | Total number of methods | 0 | 25 | M1 |
| 6 | cl_func_publ | Number of public methods | 0 | 15 | M1 |
| 7 | cl_line | Number of lines | −∞ | ∞ | M1 |
| 8 | cl_stat | Number of statements | 0 | 100 | M1 |
| 9 | cl_wmc | Weighted Methods per Class | 0 | 60 | M1 |
| 10 | cu_cdused | Number of direct used classes | 0 | 10 | M1 |
| 11 | cu_cdusers | Number of direct users classes | 0 | 5 | M1 |
| 12 | in_bases | Number of base classes | 0 | 3 | M1 |
| 13 | in_noc | Number of children | 0 | 3 | M1 |

**Table 11: Summary of acceptable values for *M1* and *M2***

## A. Quality Analysis of MARF and GIPSY

The Quality analysis of the case studies is performed in three phases – *Extract, Evaluate and Execute,* which form the feedback phase.

*Extraction Phase*

With the help of LOGISCOPE we analyze all the classes of MARF and GIPSY and then calculate the maintainability to compare which case study is better in terms of quality. The first step of the quality analysis phase is the extraction phase where we extract information from the classes and rank them under four categories namely *Excellent, Good, Fair, Poor.*

We analyze the case studies at the class factor level where analyze the maintainability of the system, the class criteria level, where we look at the four factors that make up for maintainability of the system at the class metrics level. The results of the experiment are shown in the following diagrams, pie charts and Kyviat diagrams.

*Class Factor Level*

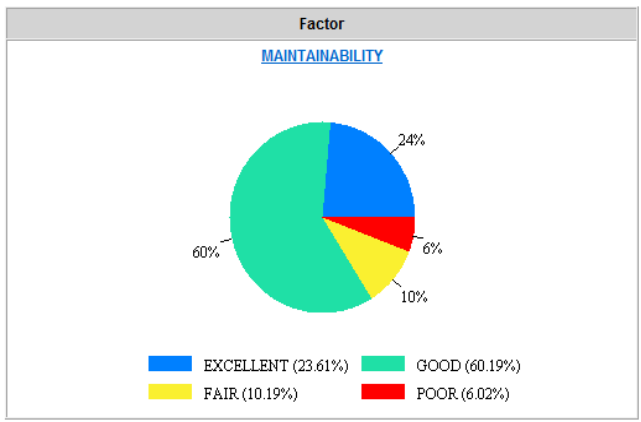

**Figure 13: Maintainability of MARF**

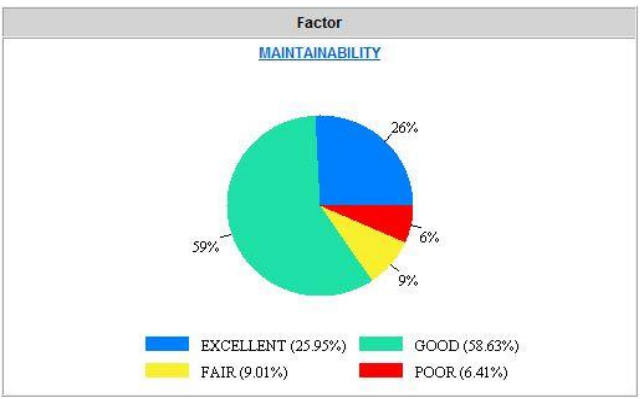

**Figure 14: Maintainability of GIPSY**

From the graphs we can conclude that GIPSY has a slightly higher maintainability as the percentage of fair and poor classes is slightly lesser when compared to MARF.

*Class Criteria Level*

At the class criteria level we measure the Analyzability, Changeability, Stability and Testability of the project.

*a. Analyzability*

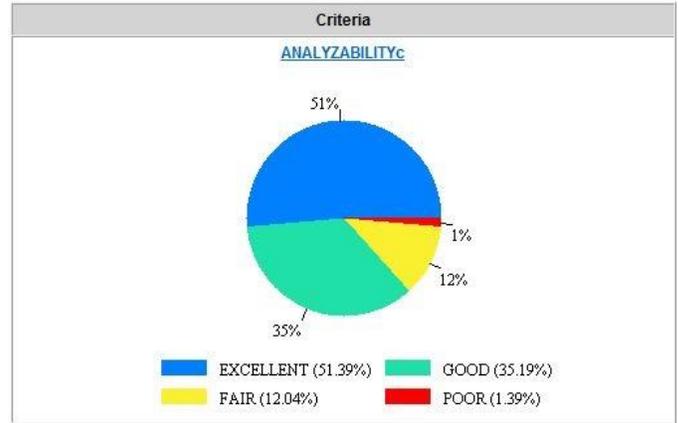

**Figure 15: Analyzability of MARF**

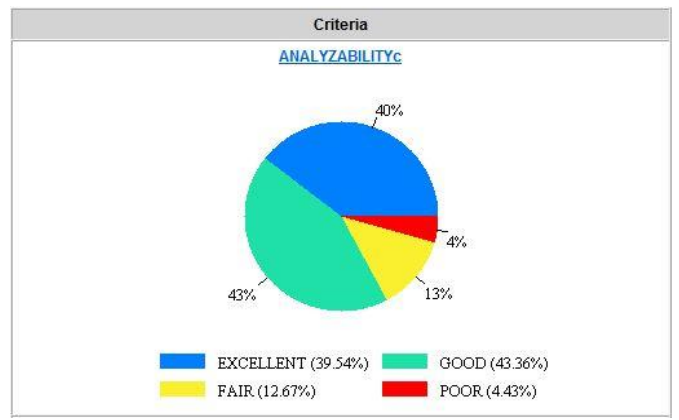

**Figure 16: Analyzability of GIPSY**

MARF has a better Analyzability when compared to GIPSY.

*Changeability*

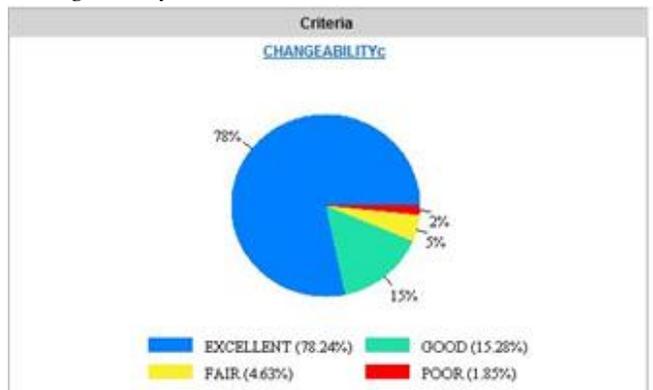

**Figure 17: Changeability of MARF**

MARF has a better Analyzability when compared to GIPSY.

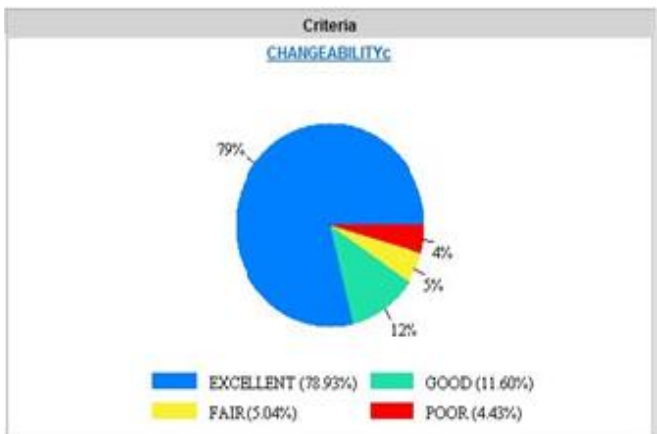

**Figure 18: Changeability of GIPSY**

*Stability*

GIPSY has a better Stability when compared to MARF.

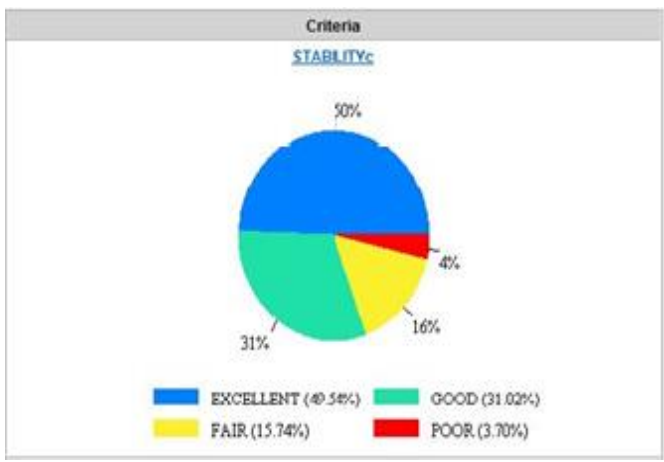

**Figure 19: Stability of MARF**

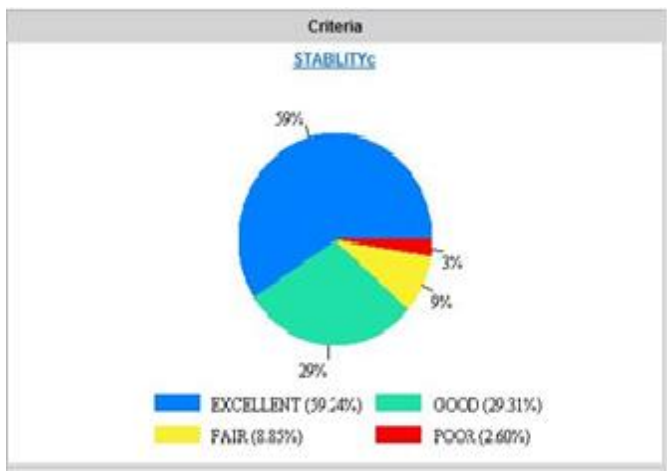

**Figure 20: Stability of GIPSY**

*Testability*

MARF has a better Testability when compared to GIPSY.

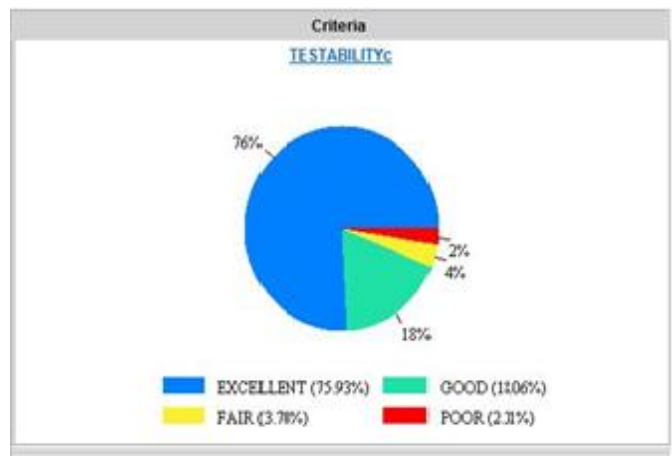

**Figure 21: Testability of MARF**

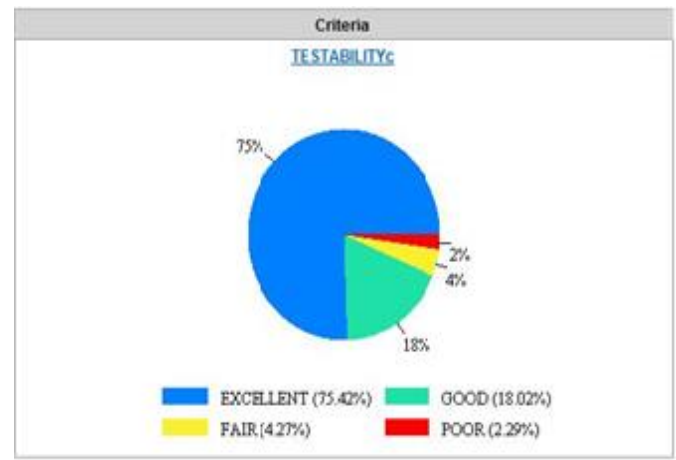

**Figure 22: Testability of GIPSY**

*Class Metric Level*

The class criteria level analysis is used to analyze the quality of each individual class in terms of each individual metric. The metrics that are used to analyze the quality of the class are shown in table 10. We consider two classes from both MARF and GIPSY; one class which has excellent analyzability and the other with poor analyzability. We compare the classes based on the Kiviat Diagram LOGISCOPE generates for each and every class that have been ranked under different criteria that make up the scope of maintainability.

*MARF*

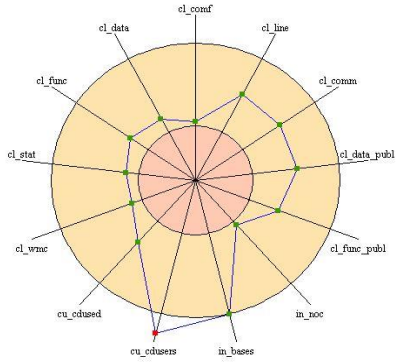

**Figure 23: Example MARF class with excellent Analyzability**

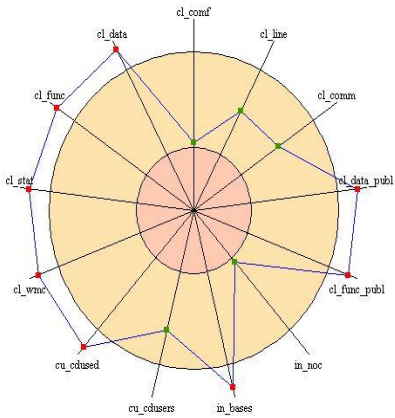

**Figure 24: Example MARF class with poor Analyzability**

*b. GIPSY*

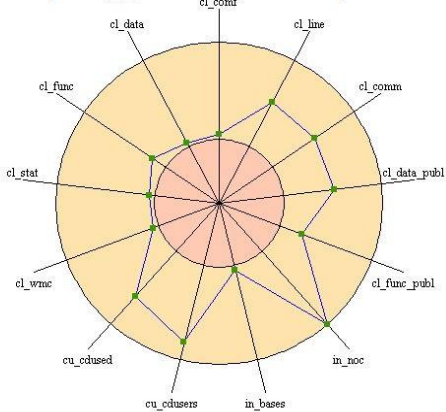

**Figure 25: Example GIPSY class with excellent Analyzability**

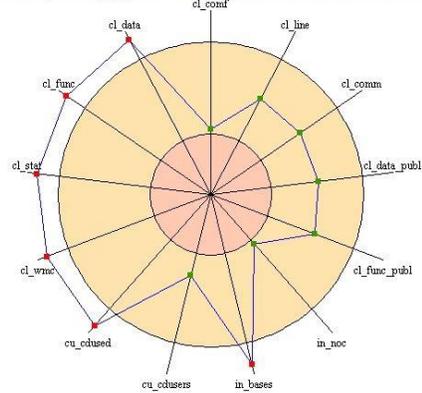

**Figure 26: Example GIPSY class with poor Analyzability**

From the Kiviat Charts we observe the variance of the different class metrics for an excellent class and a poor class for both the case studies. The inner most circle shows the boundary for the minimum of the range and the second circle shows the acceptable range for each metric. If the values of the metric fall inside the small circle or outside the second circle they are marked with red flags. These flags help us to understand the areas in which the quality of the code can be improved. For example, in figure 24 in class:

marf.Classification.NeuralNetwork.NeuralNetwork only the values of cl_comf, cl_line, cl_comm, and cd_users fall under the normal range. All the other metrics lie outside the maximum range and are marked as red flags. This tells us that we'll need to reduce the values of those metrics to improve the analyzability of the class. We discuss these recommendations in detail in the Execution phase of the quality analysis.

*Evaluation Phase*

In the evaluation phase, we compare the maintainability of the code of both case studies. The lesser the number of classes ranked under the fair and the poor categories, the better the maintainability of the code.

*Class Factor Level*

First, we rank the case studies based on the percentage of fair and poor classes under the class factor level. Figures 13 and 14 show the percentage of poor and fair classes for the case studies.

| S.No | Case Study | Fair Code % | Poor Code % |
|------|------------|-------------|-------------|
| 1    | MARF       | 10.19%      | 6.02%       |
| 2    | GIPSY      | 9.01%       | 6.41%       |

**Table 12: Comparison of MARF and GIPSY based on Factor Level**

As can be seen in the table 12, GIPSY has a slightly less percentage of classes ranked under the fair and poor categories. Therefore, we can conclude that in terms of maintainability GIPSY is better, and MARF is identified as the worst quality code.

*Class Criteria Level*

Next, we rank the case studies based on the percentage of fair and poor classes under the class criteria levels.

*Analyzability*

Figures 15 and 16 show the percentage of poor and fair classes for the case studies based on analyzability.

| S.No | Case Study | Fair Code % | Poor Code % |
|---|---|---|---|
| 1 | MARF | 12.04% | 1.39% |
| 2 | GIPSY | 12.67% | 4.43% |

**Table 13: Comparison of the Analyzability of MARF and GIPSY**

The table 13 tells us that MARF has a better analyzability when compared to GIPSY as it has a fairly less percentage of classes ranked under the fair and poor categories. In terms of analyzability, GIPSY is identified as the worst code.

*Changeability*

Figures 17 and 18 describe the changeability of the code and rank the classes in terms of metrics that compute changeability.

| S.No | Case Study | Fair Code % | Poor Code % |
|---|---|---|---|
| 1 | MARF | 4.63% | 1.85% |
| 2 | GIPSY | 5.04% | 4.43% |

**Table 14: Comparison of the Changeability of MARF and GIPSY**

In table 13 MARF has better changeability when compared to GIPSY as it has a fairly less percentage of classes ranked under the fair and poor categories. In terms of changeability, GIPSY is identified as the worst code.

*Stability*

Figures 19 and 20 describe the stability of the code and rank the classes in terms of metrics that compute stability.

| S.No | Case Study | Fair Code % | Poor Code % |
|---|---|---|---|
| 1 | MARF | 15.74% | 3.70% |
| 2 | GIPSY | 8.85% | 2.60% |

**Table 15: Comparison of the Stability of MARF and GIPSY**

In table 14 MARF has 19.44% of combined fair and poor classes which means that almost one fifth of the entire project poor stability. GIPSY on the other hand has only 9.47% of combined fair and poor classes which means that MARF is identified as the worst quality code in terms of stability.

*Testability*

Table 16 show the percentage of poor and fair classes for the case studies based on testability.

| S.No | Case Study | Fair Code % | Poor Code % |
|---|---|---|---|
| 1 | MARF | 3.7% | 2.3% |
| 2 | GIPSY | 4.27% | 2.29% |

**Table 16: Comparison of the Testability of MARF and GIPSY**

*Classification of Fair and Poor Classes in MARF and GIPSY at Class Factor Level*

The classes ranked under the fair and poor categories are listed in this section.

| Class Names for MARF with fair Maintainability |
|---|
| **marf.Classification.Classification** |
| **marf.FeatureExtraction.FFT.FFT** |
| **marf.FeatureExtraction.FeatureExtractionAggregator** |
| **marf.FeatureExtraction.LPC.LPC** |
| **marf.Preprocessing.CFEFilters.CFEFilter** |
| **marf.Preprocessing.FFTFilter.FFTFilter** |
| **marf.Preprocessing.Preprocessing** |
| **marf.Stats.ProbabilityTable** |
| **marf.Storage.Loaders.AudioSampleLoader** |
| **marf.Storage.Loaders.TextLoader** |
| **marf.Storage.Sample** |
| **marf.Storage.TrainingSet** |
| **marf.nlp.Parsing.GenericLexicalAnalyzer** |
| **marf.nlp.Parsing.GrammarCompiler.GrammarAnalyzer** |
| **marf.nlp.Parsing.LexicalAnalyzer** |
| **marf.nlp.Parsing.LexicalError** |
| **marf.nlp.Parsing.Parser** |
| **marf.nlp.Parsing.ProbabilisticParser** |
| **marf.nlp.Parsing.SyntaxError** |
| **marf.nlp.Parsing.TransitionTable** |
| **marf.util.OptionProcessor** |

**Table 17: MARF Classes with Fair Maintainability**

| Class Names for MARF with poor Maintainability |
|---|
| **marf.Classification.NeuralNetwork.NeuralNetwork** |
| marf.Classification.Stochastic.ZipfLaw |
| marf.Configuration |
| marf.MARF |
| marf.Stats.StatisticalEstimators.StatisticalEstimator |
| marf.Storage.ResultSet |
| marf.Storage.StorageManager |
| marf.math.ComplexMatrix |
| marf.math.Matrix |
| marf.nlp.Parsing.GrammarCompiler.Grammar |
| marf.nlp.Parsing.GrammarCompiler.GrammarCompiler |
| marf.nlp.Storage.Corpus |
| marf.util.Arrays |

**Table 18: MARF Classes with Poor Maintainability**

| Class Names in GIPSY with Fair Maintainability |
|---|
| gipsy.GEE.GEE |
| gipsy.GEE.IDP.DemandGenerator.DemandGenerator |
| gipsy.GEE.IDP.DemandGenerator.LegacyEductiveInterpreter |
| gipsy.GEE.IDP.DemandGenerator.LegacyInterpreter |
| gipsy.GEE.IDP.DemandGenerator.jini.rmi.JINITransportAgent |
| gipsy.GEE.IDP.DemandGenerator.jms.DemandController |
| gipsy.GEE.IVW.Warehouse.NetCDFFileManager |
| gipsy.GEE.multitier.DGT.DGTWrapper |
| gipsy.GEE.multitier.DST.jini.JiniDSTWrapper |
| gipsy.GEE.multitier.DST.jini.JiniERIDSTWrapper |
| gipsy.GEE.multitier.DST.jms.JMSDSTWrapper |
| gipsy.GEE.multitier.DWT.DWTWrapper |
| gipsy.GEE.multitier.TAExceptionHandler |
| gipsy.GIPC.DFG.DFGAnalyzer.LucidCodeGenerator |
| gipsy.GIPC.imperative.ImperativeCompiler |
| gipsy.GIPC.imperative.Java.JavaCompiler |
| gipsy.GIPC.intensional.GIPL.GIPLParserTreeConstants |
| gipsy.GIPC.intensional.GenericTranslator.TranslationLexer |
| gipsy.GIPC.intensional.IntensionalCompiler |
| gipsy.GIPC.intensional.SIPL.IndexicalLucid.IndexicalLucidParserTreeConstants |
| gipsy.GIPC.intensional.SIPL.JOOIP.JOOIPCompiler |
| gipsy.GIPC.intensional.SIPL.JOOIP.ast.Node |
| gipsy.GIPC.intensional.SIPL.JOOIP.ast.body.TypeDeclaration |
| gipsy.GIPC.intensional.SIPL.JOOIP.ast.expr.StringLiteralExpr |
| gipsy.GIPC.intensional.SIPL.JOOIP.ast.visitor.GenericVisitor |
| gipsy.GIPC.intensional.SIPL.JOOIP.ast.visitor.VoidVisitor |
| gipsy.RIPE.RIPE |
| gipsy.RIPE.editors.RunTimeGraphEditor.core.GIPSYTier |
| gipsy.RIPE.editors.RunTimeGraphEditor.core.GraphDataManager |
| gipsy.RIPE.editors.RunTimeGraphEditor.ui.ActionsLog |
| gipsy.RIPE.editors.RunTimeGraphEditor.ui.InstancesNodesPanel |
| gipsy.RIPE.editors.RunTimeGraphEditor.ui.MapEditor |
| gipsy.RIPE.editors.RunTimeGraphEditor.ui.dialogs.GIPSYNodeDialog |
| gipsy.RIPE.editors.RunTimeGraphEditor.ui.dialogs.TierPropertyDialog |
| gipsy.RIPE.editors.WebEditor.WebEditor |
| gipsy.apps.marfcat.MARFCATDWT |
| gipsy.apps.marfcat.MARFPCATDWT |
| gipsy.apps.memocode.genome.AlignDGT |
| gipsy.apps.memocode.genome.AlignDWT |
| gipsy.interfaces.GIPSYProgram |
| gipsy.lang.GIPSYFunction |
| gipsy.lang.GIPSYInteger |
| gipsy.lang.GIPSYType |
| gipsy.lang.context.TagSet |
| gipsy.tests.GEE.IDP.demands.DemandTest |
| gipsy.tests.GEE.multitier.GMT.GMTTestConsole |
| gipsy.tests.GEE.simulator.DGTDialog |
| gipsy.tests.GEE.simulator.DGTSimulator |
| gipsy.tests.GEE.simulator.DSTSpaceScalabilityTester |
| gipsy.tests.GEE.simulator.GlobalDef |
| gipsy.tests.GEE.simulator.ProfileDialog |
| gipsy.tests.GEE.simulator.demands.WorkResultPi |
| gipsy.tests.GEE.simulator.jini.WorkerJTA |
| gipsy.tests.Regression |
| gipsy.tests.jooip.CopyOfGIPLtestVerbose |
| gipsy.tests.junit.GEE.multitier.DGT.DGTWrapperTest |
| gipsy.tests.junit.GEE.multitier.DWT.DWTWrapperTest |
| gipsy.tests.junit.lang.GIPSYContextTest |
| gipsy.tests.junit.lang.context.GIPSYContextTest |

**Table 19: GIPSY Classes with Fair Maintainability**

| Class Names of GIPSY with poor Maintainability |
|---|
| gipsy.Configuration |
| gipsy.GEE.IDP.DemandGenerator.jini.rmi.JINITA |
| gipsy.GEE.IDP.DemandGenerator.jini.rmi.JiniDemandDispatcher |
| gipsy.GEE.IDP.DemandGenerator.jms.JMSTransportAgent |
| gipsy.GEE.IDP.demands.Demand |
| gipsy.GEE.multitier.GIPSYNode |
| gipsy.GEE.multitier.GMT.GMTWrapper |
| gipsy.GIPC.DFG.DFGAnalyzer.DFGParser |
| gipsy.GIPC.DFG.DFGAnalyzer.DFGParserTokenManager |
| gipsy.GIPC.DFG.DFGGenerator.DFGCodeGenerator |
| gipsy.GIPC.DFG.DFGGenerator.DFGTranCodeGenerator |
| gipsy.GIPC.GIPC |
| gipsy.GIPC.Preprocessing.PreprocessorParser |
| gipsy.GIPC.Preprocessing.PreprocessorParserTokenManager |
| gipsy.GIPC.SemanticAnalyzer |

| gipsy.GIPC.intensional.GIPL.GIPLParser |
|---|
| gipsy.GIPC.intensional.GIPL.GIPLParserTokenManager |
| gipsy.GIPC.intensional.GenericTranslator.TranslationParser |
| gipsy.GIPC.intensional.SIPL.ForensicLucid.ForensicLucidParser |
| gipsy.GIPC.intensional.SIPL.ForensicLucid.ForensicLucidParserTokenManager |
| gipsy.GIPC.intensional.SIPL.ForensicLucid.ForensicLucidSemanticAnalyzer |
| gipsy.GIPC.intensional.SIPL.IndexicalLucid.IndexicalLucidParser |
| gipsy.GIPC.intensional.SIPL.IndexicalLucid.IndexicalLucidParserTokenManager |
| gipsy.GIPC.intensional.SIPL.JLucid.JGIPLParser |
| gipsy.GIPC.intensional.SIPL.JLucid.JGIPLParserTokenManager |
| gipsy.GIPC.intensional.SIPL.JLucid.JIndexicalLucidParser |
| gipsy.GIPC.intensional.SIPL.JLucid.JIndexicalLucidParserTokenManager |
| gipsy.GIPC.intensional.SIPL.JOOIP.JavaCharStream |
| gipsy.GIPC.intensional.SIPL.JOOIP.JavaParser |
| gipsy.GIPC.intensional.SIPL.JOOIP.JavaParserTokenManager |
| gipsy.GIPC.intensional.SIPL.JOOIP.ast.visitor.DumpVisitor |
| gipsy.GIPC.intensional.SIPL.Lucx.LucxParser |
| gipsy.GIPC.intensional.SIPL.Lucx.LucxParserTokenManager |
| gipsy.GIPC.intensional.SIPL.ObjectiveLucid.ObjectiveGIPLParser |
| gipsy.GIPC.intensional.SIPL.ObjectiveLucid.ObjectiveGIPLParserTokenManager |
| gipsy.GIPC.intensional.SIPL.ObjectiveLucid.ObjectiveIndexicalLucidParser |
| gipsy.GIPC.intensional.SIPL.ObjectiveLucid.ObjectiveIndexicalLucidParserTokenManager |
| gipsy.GIPC.util.SimpleCharStream |
| gipsy.RIPE.editors.RunTimeGraphEditor.core.GlobalInstance |
| gipsy.RIPE.editors.RunTimeGraphEditor.ui.GIPSYGMTOperator |
| gipsy.lang.GIPSYContext |
| gipsy.tests.GIPC.intensional.SIPL.Lucx.SemanticTest.LucxSemanticAnalyzer |

**Table 20: GIPSY Classes with Poor Maintainability**

Please refer to the Appendix to view the classes which are categorized as fair and poor in the class criteria level.

*Class Metric Level*

At the class metric level, we visualize the data in terms of Kiviat Graphs for two classes. We compare the quality of both the classes by looking at the values of the metrics shown in the Kiviat graphs to the acceptable range provided by the tool. We then look at the values that fall out of bounds and provide a quality report for the chosen classes. The classes we chose are both from MARF:marf.nlp.Parsing.LexicalAnalyzer, marf.Classification.NeuralNetwork.NeuralNetwork.

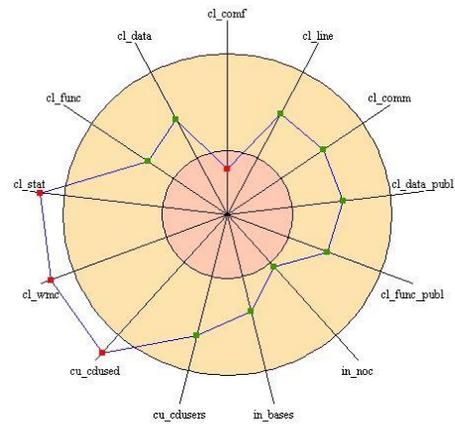

**Figure 27: Kiviat Graph for LexicalAnalyzer Class**

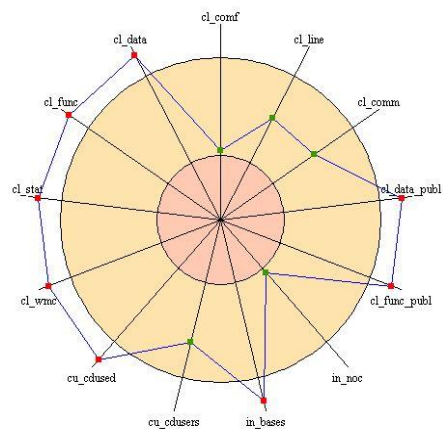

**Figure 28: Kiviat Graph for a NeuralNetwork Class**

| S.No | Mnemonic | Metric | Min | Max | Lexical Analyzer | Neural Network |
|---|---|---|---|---|---|---|
| 1 | cl_comf | Class comment rate | 0.2 | ∞ | 0.19 | 0.26 |
| 2 | cl_comm | Number of lines of comment | −∞ | ∞ | 147 | 354 |
| 3 | cl_data | Total number of attributes | 0 | 7 | 3 | 17 |
| 4 | cl_data_pulb | Number of public attributes | 0 | 0 | 0 | 8 |
| 5 | cl_func | Total number of methods | 0 | 25 | 7 | 27 |
| 6 | cl_func_pulb | Number of public methods | 0 | 15 | 6 | 21 |
| 7 | cl_line | Number of lines | −∞ | ∞ | 788 | 1348 |
| 8 | cl_stat | Number of statements | 0 | 100 | 268 | 372 |
| 9 | cl_wmc | Weighted Methods per Class | 0 | 60 | 65 | 115 |
| 10 | cu_cdused | Number of direct used classes | 0 | 10 | 17 | 33 |
| 11 | cu_cdusers | Number of direct users classes | −∞ | 5 | 3 | 3 |
| 12 | in_bases | Number of base classes | 0 | 3 | 1 | 6 |
| 13 | in_noc | Number of children | 0 | 3 | 0 | 0 |

**Table 21: Comparison of classes at Class Metric Level**

The highlighted rows signify the metrics for the classes that are out of bounds. In terms of maintainability we can conclude that the quality of the code for the LexicalAnalyzer class is better than the NeuralNetwork class as it has only four metrics that are out of bounds whereas the second class has 8 metrics that are out of bounds. However, these results do not necessarily mean that the first class is better than the second class in every case. Here it just so happens that the NeuralNetwork is a really bad class and has a lot of metrics that are out of bounds and therefore we could come to a conclusion. Earlier, we had discussed the formulae for calculating maintainability and other criteria levels. These values cannot be calculated without normalization. In addition, the weightage or in other words the importance that is given to each metric that is used to calculate maintainability is unknown. The impact these metrics generate for the class cannot be measured either.

*Execution Phase*

In this phase, we use the results of the extraction and evaluation phases to come up with recommendations on how to improve the quality of the code for the case studies at the class factor, class criteria and class metric levels.

*Class Factor Level*

The overall maintainability of the code depends on the ranking of classes. The classes ranked in excellent should be were most or all of the classes are supposed to be. We focus more on the classes that are ranked under the fair and poor categories as they are responsible for bringing down the overall quality of the project. An optional step to maximize the maintainability of the code is to try to analyse the Kiviat graphs of the classes present in the good category as well. For both MARF and GIPSY the maintainability of the code is almost the same, GIPSY having a slightly higher maintainability. In both the case studies most of the classes are ranked as good. 60.19% of the MARF classes and 58.63% of the GIPSY classes are marked under the good category. To improve the quality of code, work needs to be done on the fair and poor ranked classes. The percentage of poor and fair classes are relatively high and around 15-16% for both the case studies. The more the presence of fair\poor code, the lesser the quality of the code. The list of classes ranked fair and poor can be referred in the evaluation phase of the report. We list the recommendations to improve the quality of the code in detail in the class criteria level.

*Class Criteria Level*

In this section, we list the recommendations under each maintainability criteria for both the case studies.

MARF

*Analyzability*

The analyzability of the code is decent. 51.39% of the classes are ranked excellent. However, we focus our attention on the fair and poor classes. The percentage of these classes is relatively high. We have studied and analysed the Kiviat diagrams and the class metrics for all the classes ranked under the fair and poor categories and have noticed a trend for the metrics that have gone out of bounds on most occasions. The metrics that have mostly shown problems are cl_stat, cl_wmc, cu_cdused and cl_comf. The values of the first three values are more than the maximum values of the acceptable range whereas cl_comf falls below the minimum range. To improve the analyzability of the code we need to reduce the Number of Statements, Weighted Methods per Class and Number of Directly Used Classes. We need to increase the number of comments as well.

| Class Names | Recommendations |
|---|---|
| marf.Classification.NeuralNetwork.NeuralNetwork | Decrease number of Directly Used Classes (Reduce Inheritance Relationships), Number of Weighted Methods per Class, Number of Statements (Reduce Complexity), Increase Comment Rate (Improves Understandibility) |
| marf.nlp.Parsing.GrammarCompiler.GrammarCompiler | |
| marf.nlp.Parsing.LexicalAnalyzer | |

**Table 22: Example Test Classes and Recommendations to improve Analyzability (MARF)**

*Changeability*

The code has very good changeability. Only 4.63% of the classes are ranked under fair and a meagre 1.85% under poor. We can still improve changeability by performing analysis. The attributes that go out of bounds are cl_data, cl_func, cl_stat, cu_cdused and cl_func_publ. The values of these metrics should be reduced to fit into the acceptable range.

| Class Names | Recommendations |
|---|---|
| marf.Classification.NeuralNetwork.NeuralNetwork | Fit the values of Total number of attributes, Total number of methods, Number of public methods, Number of statements and Number of direct used classes under the acceptable range to reduce complexity and try to inculcate the habit of coding in the same standard |
| marf.Configuration | |
| marf.MARF | |
| marf.nlp.Storage.Corpus | |

**Table 23: Example Test Classes and Recommendations to improve Changeability (MARF)**

*Stability*

The code does not have good stability as almost 20% of the classes are ranked in the fair and poor categories. The reason for increased instability was found out by analyzing problematic classes. The number of directly used classes go way out of bounds cause increased coupling and complexity. The metrics that have mostly shown problems are cl_stat, cl_wmc, cu_cdused and cl_comf. The values of the first three values are more than the maximum values of the acceptable range whereas cl_comf falls below the minimum range. These value should be adjusted to improve the stability.

| Class Names | Recommendations |
|---|---|
| marf.MARF | Reduce number of directly used classes to reduce inheritance relationships and complexities. This improves overall stability of the project. |
| marf.Preprocessing.FFTFilter.FFTFilter | |
| marf.Stats.StatisticalEstimators.StatisticalEstimator | |
| marf.Storage.Sample | |
| marf.Storage.StorageManager | |
| marf.math.Matrix | |
| marf.math.Vector | |
| marf.nlp.Parsing.SyntaxError | |

**Table 24 Example Test Classes and Recommendations to improve Stability (MARF)**

*Testability*

The code has a good testability. Only 3.70% of the classes are ranked under fair and 2.31% of the classes are ranked under poor.

| Class Names | Recommendations |
|---|---|
| marf.Classification.NeuralNetwork.NeuralNetwork | Reduce Total number of methods, Number of public methods, Number of statements, Weighted Methods per Class and Number of direct used classes and reduce the complexity of the system. |
| marf.MARF | |
| marf.Storage.StorageManager | |
| marf.math.Matrix | |
| marf.util.Arrays | |

**Table 25: Example Test Classes and Recommendations to improve Testability (MARF)**

GIPSY

*Analyzability*

The amount of fair and poor code present is fairly high when compared to MARF. Almost 17% of the code is problematic. The metrics that show a trend for being out of bounds are in_bases, cu_cdusers, cl_comf, cu_cdused and cl_data. To improve analyzability of the code we need to increase the comment rate to improve understandability of the code, reduce the number of user classes, base classes and used classes to reduce inheritance relationships as well as complexity and coupling.

| Class name | Recommendations |
|---|---|
| gipsy.GIPC.SemanticAnalyzer | Increase Comment Rate to improve understandability, reduce number of statements to reduce complexity, Reduce the number of public methods, number of used methods and number of weighted methods per class to reduce inheritance relationships and coupling. |
| gipsy.GIPC.intensional.GIPL.GIPLParser | |
| gipsy.GIPC.Preprocessing.PreprocessorParser | |

**Table 26: Example Test Classes and Recommendations to improve Analyzability (GIPSY)**

*Changeability*

The code has around 10% of classes that are ranked under the fair and poor categories. This is fairly high and the reason for this is that the metrics cl_comf, cl_data, cu_cdusers, cu_cdused, cl_data, in_noc and cl_stat go out of bounds. We should bring the values into the acceptable range to improve the stability of the code.

| Class name | Recommendations |
|---|---|
| gipsy.GIPC.util.SimpleCharStream | Reduce Complexity, Coupling and Inheritance relationships by reducing the number of children classes, number of public methods, number of used methods and number of weighted methods per class. Increase Comment Rate to improve understandability, reduce number of statements to reduce complexity. |
| gipsy.GIPC.intensional.SIPL.Lucx.LucxParser | |
| gipsy.GIPC.intensional.SIPL.Lucx.LucxParserTokenManager | |

**Table 27: Example Test Classes and Recommendations to improve Changeability (GIPSY)**

*Stability*

The code has around 10% of classes that are ranked under the fair and poor categories.

| Class name | Recommendations |
|---|---|
| gipsy.GEE.IDP.DemandGenerator.jini.rmi.JiniDemandDispatcher | Increase Comment Rate to improve understandability, reduce number of statements to reduce complexity. Reduce Complexity, Coupling and Inheritance relationships by reducing the number of children classes, number of public methods, number of used methods and number of weighted methods per class. |
| gipsy.RIPE.editors.RunTimeGraphEditor.ui.GIPSYGMTOperator | |
| gipsy.GIPC.DFG.DFGAnalyzer.DFGParser | |

**Table 28: Example Test Classes and Recommendations to improve Stability (GIPSY)**

*Testability*

The code has good testability and only around 6.5% of the classes are ranked under the fair and poor categories.

| Class name | Recommendations |
|---|---|
| gipsy.GIPC.intensional.GIPL.GIPLParser | Reduce the Number of Statements, Weighted Methods per Class and Number of Directly Used Classes to reduce coupling and complexity. Increase the number of comments to improve understandability. |
| gipsy.GIPC.DFG.DFGAnalyzer.DFGParser | |
| gipsy.GIPC.intensional.SIPL.Lucx.LucxParser | |

Table 29: Example Test Classes and Recommendations to improve Testability (GIPSY)

*Class Metric Level*

The quality of the code can be increased in numerous ways by looking at the Kiviat graphs and the corresponding values of the metrics listed in the tables. Since the quality of the class is measured in terms of the class metrics, the metrics that go out of bounds with respect to the acceptable range are responsible for the reduction in code quality. We consider the same two classes that we considered in the evaluation phase and give recommendations at the class metric level. The classes being used are:

marf.nlp.Parsing.LexicalAnalyzer
marf.Classification.NeuralNetwork.NeuralNetwork.

**Recommendations to improve Quality of Code:**

**cl_comf, cl_comm:** More the number of comments, better the understandability and readability of the code.

**cl_data:** The number of attributes is directly proportional to the amount of functionalities the system has to offer. According to the table the number of attributes should be limited to seven. The attributes should be the responsibility of one class only.

**cu_cdused:** The lesser the number of used classes, the lesser the coupling. The value should lie between 0 and 10. Higher the value, the more the class is sensitive to changes. Keep this value within the range.

Figure 29: Table displayed by LOGISCOPE for LexicalAnalyzer class

Figure 30: Table displayed by LOGISCOPE for NeuralNetwork class

**cu_cdusers:** If the class is being used by many other classes, it becomes the center of multiple functionalities which results in increased complexity. The value should lie between 0 and 5. Higher the value, the more the class is sensitive to changes. Keep this value within the range.

**cl_func:** The number of classes should be limited to 25. If the value exceeds this, the class will require more amount of effort to test. Keep the amount of functions within the acceptable range.

**cl_func_publ:** If the number of public methods are more, one class will have more to do and many other classes will try to access it. This will increase the effort required during testing. The class should be broken to reduce the complexity.

**cl_wmc:** The weighted methods per class is used to calculate the cyclomatic complexity of a class. The value of this metric should be less.

**cl_stat:** Limit the number of executable statements of a class to less than 100. This will reduce the complexity and increase understandability.

**II. McCabe IQ Measurement Tool**

We use the McCabe tool to perform quality analysis in terms of different complexities and Object Oriented principles on the MARF and GIPSY case studies. The quality trends are analyzed with respect to the methods and classes of the case studies.

*Quality Trends of Methods of the Case Studies*

The Average values of Cyclomatic Complexity, Essential complexity and Module Design Complexity for MARF and GIPSY are taken from the System Complexity report in McCabe. The thresholds discussed are according to McCabe IQ metrics provided by the McCabe IQ tool.

The Thresholds for the different types of complexities are:

● Average Cyclomatic Complexity - V(G) = 10.00
● Average Essential Complexity - ev(G) = 4.00

● Average Module Design complexity - Iv(G) = 7.00 [70]

*Cyclomatic Complexity*

Cyclomatic complexity (abbreviated as v(G)) is a measure of the complexity of a module's decision structure. It is the number of linearly independent paths and, therefore, the minimum number of paths that should be tested to reasonably guard against errors. A high cyclomatic complexity indicates that the code may be of low quality and difficult to test and maintain. In addition, empirical studies have established a correlation between high cyclomatic complexity and error-prone software [69]. The result of Cyclomatic complexity for MARF and GIPSY is shown in table 30 and figure 31.

| CASE STUDY | Essential Complexity [EV(G)] | | Essential Complexity Range | |
|---|---|---|---|---|
| | Total | Avg | Min | Max |
| MARF | 2546 | 1.2 | 1 | 37 |
| GIPSY | 11166 | 1.84 | 1 | 96 |

**Table 30: Cyclomatic Complexities for the Case Studies**

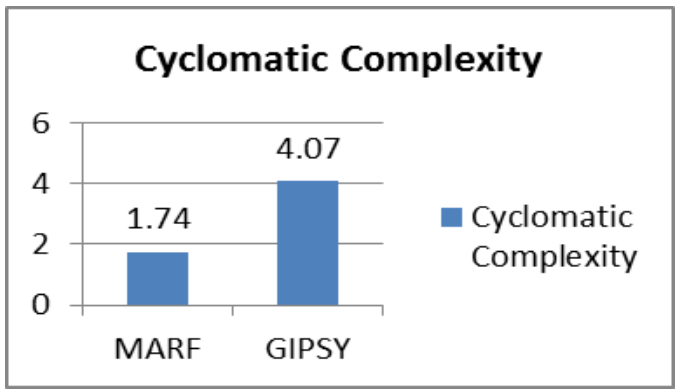

**Figure 31: Graphical Comparison of Cyclomatic Complexities for the Case Studies**

*KEY OBSERVATIONS:*

In Marf minimum Cyclomatic Complexity is 1 and maximum Cyclomatic Complexity 56 is found in module LexicalAnalyzer.getNextToken(). Where in GISPY minimum Cyclomatic Complexity is also 1 and maximum 294 is found in modules:
GIPLParserTokenManager.jjMoveNfa_0(int,int), JGIPLParserTokenManager.jjMoveNfa_0(int,int), ObjectiveGIPLParserTokenManager.jjMoveNfa_0(int,int).

*Essential Complexity*

Essential complexity (abbreviated as ev(G)) is a measure of the degree to which a module contains unstructured constructs. Unstructured constructs decrease the quality of the code and increase the effort required to maintain the code and break it into separate modules. Therefore, when essential complexity is high, there is a high number of unstructured constructs, so modularization and maintenance is difficult. In fact, during maintenance, fixing a bug in one section often introduces an error elsewhere in the code [69]. The results of the Essential Complexities of the case studies are described in table 31 and figure 32.

| CASE STUDY | Cyclomatic Complexity [V(G)] | | Cyclomatic Complexity Range | |
|---|---|---|---|---|
| | Total | Avg | Min | Max |
| MARF | 3968 | 1.74 | 1 | 56 |
| GIPSY | 24718 | 4.07 | 1 | 294 |

**Table 31: Essential Cyclomatic Complexities for the Case Studies**

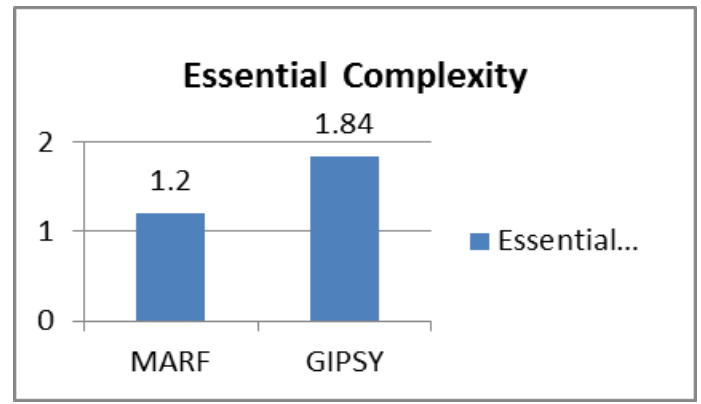

**Figure 32: Graphical Comparison of Essential Cyclomatic Complexities for the Case Studies**

*KEY OBSERVATIONS*

In Marf minimum Essential Complexity is 1 and maximum Essential Complexity 37 is found in module LexicalAnalyzer.getNextToken(). Where in GISPY minimum Essential Complexity is also 1 and maximum 96 is found in modules:

ForensicLucidSemanticAnalyzer.check(gipsy.GIPC.intensional.SimpleNode,gipsy.GIPC.intensional.SimpleNode),

LucxSemanticAnalyzer.check(gipsy.GIPC.intensional.SimpleNode,gipsy.GIPC.intensional.SimpleNode).

*Module Design Complexity*

Module Design Complexity is also known as Integration Complexity (abbreviated as iv(G)) is a measure of a module's decision structure as it relates to calls to other modules. This quantifies the testing effort of a module with respect to integration with subordinate modules. Software with high integration complexity tends to have a high degree of control coupling, which makes it difficult to isolate, maintain, and reuse individual software components [69]. The results of the Essential Complexities of the case studies are described in table 31 and figure 33.

| CASE STUDY | Module Design Complexity [IV(G)] | | Module Design Complexity Range | |
|---|---|---|---|---|
|  | **Total** | **Avg** | **Min** | **Max** |
| MARF | 3310 | 1.56 | 1 | 50 |
| GIPSY | 18272 | 1.84 | 1 | 160 |

**Table 32: Module Design Complexities for the Case Studies**

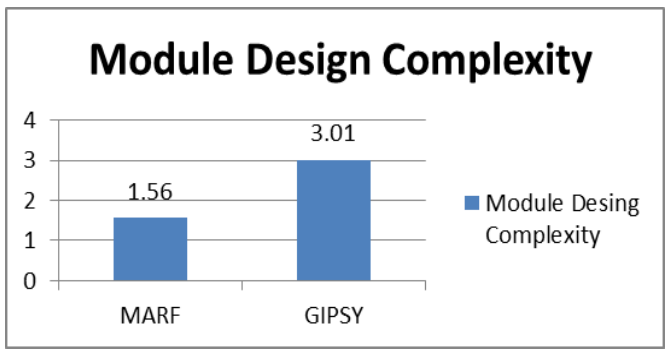

**Figure 33: Graphical Comparison of Module Design Complexities for the Case Studies**

*KEY OBSERVATIONS*

In Marf minimum Essential Complexity is 1 and maximum Module Design Complexity 50 is found in module LexicalAnalyzer.getNextToken(). Where in GISPY minimum Essential Complexity is also 1 and maximum 160 is found in modules:

SemanticAnalyzer.check(gipsy.GIPC.intensional.SimpleNode,gipsy.GIPC.intensional.SimpleNode).

*SCATTER PLOTS*

The scatterplots provided here pertain to the Modular View of each milestone from M1 through M5.

McCabe IQ defines the thresholds and three categories based on the threshold for analyzing code complexity.

● Low Complexity: V(G) < = 10 and ev(G) < =4

● Moderate Complexity: V(G) > 10 and ev(G) <=4

● High Complexity: ev(G) > 4 irrespective of V(G)

The Scatterplots have Cyclomatic Complexity along the X axis and the essential Complexity along the Y axis. Number of classes falling under each quadrant is a measure of corresponding number of classes falling into the four categories.

• I Quadrant - Unreliable/Unmaintainable

   High V(G) and High ev(G)

   Values with V(G) > 10 and ev(G) >4

• II Quadrant - Reliable/Unmaintainable

   Low v(G) and High ev(G)

   Values with V(G) < 10 and ev(G)> 4

• III Quadrant - Reliable/Maintainable

   Low v(G) and Low ev(G)

   Values with V(G) < 10 and ev(G) < 4

• IV Quadrant - Unreliable/Maintainable

   High v(G) and Low ev(G)

   Values with V(G) > 10 and ev(G)< 4

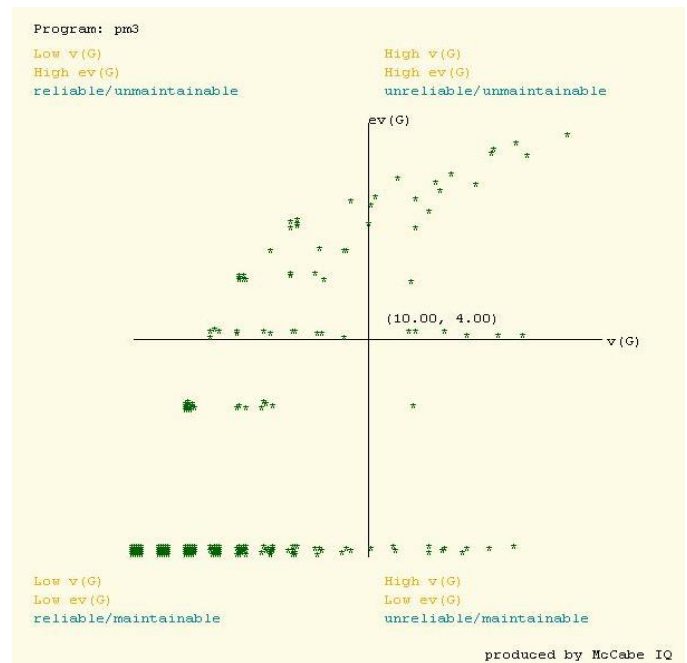

**Figure 34: Scatter Plot showing Complexities for MARF**

*KEY OBSERVATIONS*

From the Scatter plot of MARF we can see that most of the modules lies on the third quadrant that means they are more reliable and maintainable and very few modules are in unreliable and maintainable quadrant.

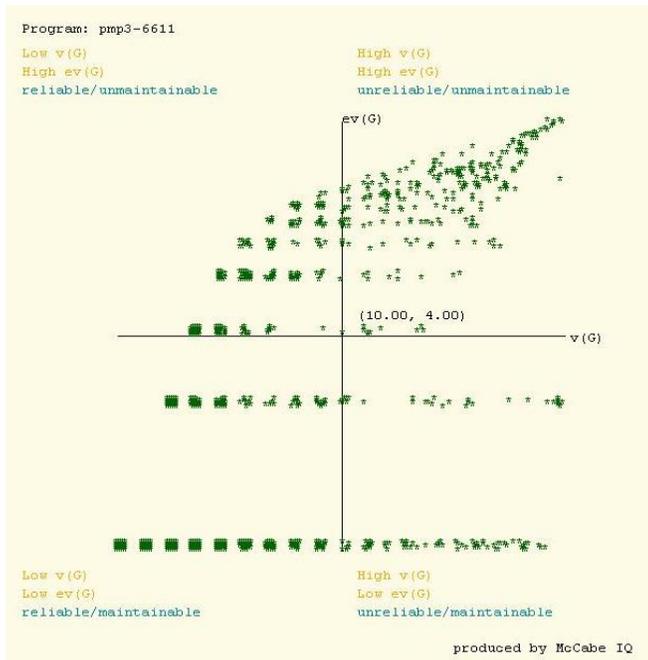

**Figure 35: Scatter Plot showing Complexities for GIPSY**

*KEY OBSERVATIONS*

From the Scatter plot we can see that the modules are scattered in every quadrant but most of the modules are in third and second quadrant which means that system is more reliable.

From the graphs and figure s we conclude that the Average Cyclomatic complexity and the Average Essential complexities are well within the thresholds. However, these values are lesser for MARF when compared with GIPSY. Hence, we can conclude that MARF has an overall better maintainability and reliability when compared to GIPSY. We can also conclude that GIPSY has a higher percentage of unstructured code which results in reduced maintainability.

*Quality Trends of Classes of the Case Studies*

The Average values of CBO, WMC, RFC, DIT and NOC for MARF and GIPSY are taken from the System Complexity report in McCabe. The thresholds discussed are according to McCabe IQ metrics provided by the McCabe IQ tool.

● Coupling Between the Objects CBO: 2.00

● Number of Weighted Methods per class WMC: 14.00

● Response for Messages RFC: 100.00

● Depth in Inheritance Tree DIT: 7.00

● Number of Children NOC: 3.00

*Coupling Between Objects*

The results for the Coupling between Objects are observed in table 33 and figure 36.

| Case Study | Coupling Between Objects [CBO] | |
|---|---|---|
| | Total | Avg |
| **MARF** | 1 | 0.01 |
| **GIPSY** | 38 | 0.07 |

**Table 33: Coupling Between Objects for the Case Studies**

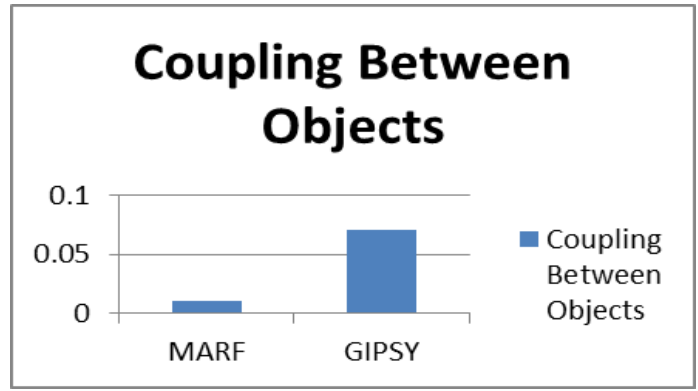

**Figure 36: Graphical Comparison of Coupling Between Objects for the Case Studies**

*KEY OBSERVATIONS*

The figure shows that MARF has high WMC than GIPSY so from this we can conclude that MARF is more application specific and has limited possibility of reuse.

*Weighted Method per Class [WMC]*

The results for the Weighted Method per Class are observed in table 34 and figure 37.

| Case Study | Weighted Method Per Class [WMC] | |
|---|---|---|
| | Total | Avg |
| **MARF** | 2060 | 11.44 |
| **GIPSY** | 6128 | 10.55 |

**Table 34: Coupling Between Objects for the Case Studies**

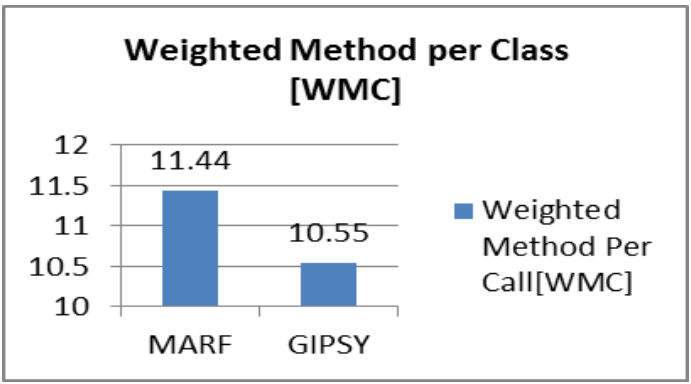

Figure 37: Graphical Comparison of Weighted Methods per class for the Case Studies

*KEY OBSERVATIONS*

The figure shows that MARF has high WMC than GIPSY so from this we can conclude that MARF is more application specific and has limited possibility of reuse.

*Response for Class*

The results for the Coupling between Objects are observed in table 35 and figure 38.

| Case Study | Response For Class[RFC] | |
|---|---|---|
| | Total | Avg |
| **MARF** | 2967 | 16.48 |
| **GIPSY** | 7339 | 12.63 |

Table 35: Response for Class for the Case Studies

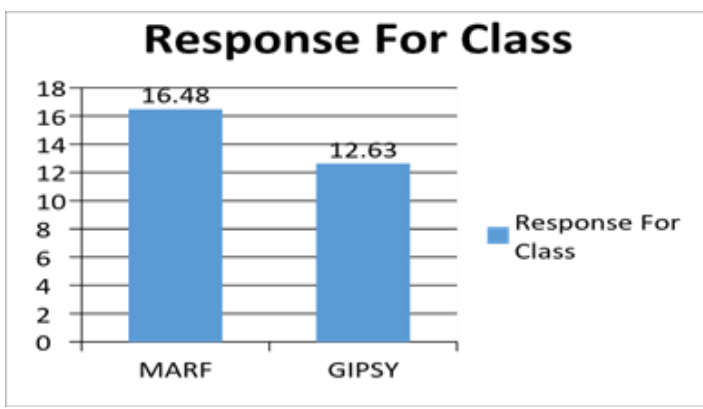

Figure 38: Graphical Comparison of Response for Class for the Case Studies

*KEY OBSERVATIONS*

From the graph it is clear that MARF has high RFC than GIPSY so GIPSY has better understandability than MARF. Furthermore due to low value of RFC in GIPSY, it can be easily tested.

*Depth of Inheritance Tree*

The results for the Coupling between Objects are observed in table 36 and figure 39.

| Case Study | Depth of Inheritance Tree[DIT] | |
|---|---|---|
| | Total | Avg |
| **MARF** | 386 | 2.14 |
| **GIPSY** | 1181 | 2.02 |

Table 36: Depth of Inheritance Tree

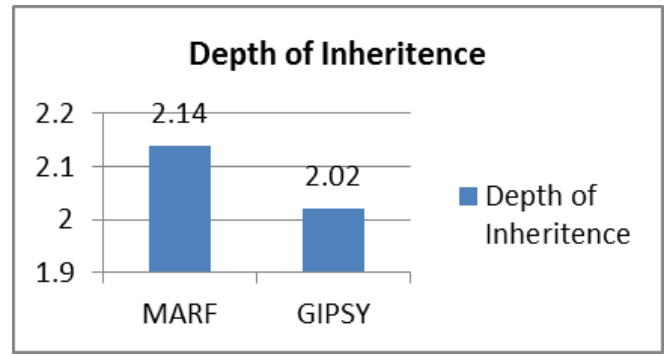

Figure 39: Graphical Comparison of Depth of Inheritance for the Case Studies

*KEY OBSERVATIONS*

From the results it is clear that MARF has slightly high DIT than GIPSY, hence we can say that MARF design is more complex and require more maintenance effort.

*Number of Children [NOC]*

The results for the Coupling between Objects are observed in table 37 and figure 40.

| Case Study | Number of Children[NOC] | |
|---|---|---|
| | Total | Avg |
| **MARF** | 45 | 0.25 |
| **GIPSY** | 120 | 0.21 |

Table 37: Number of Children for the Case Studies

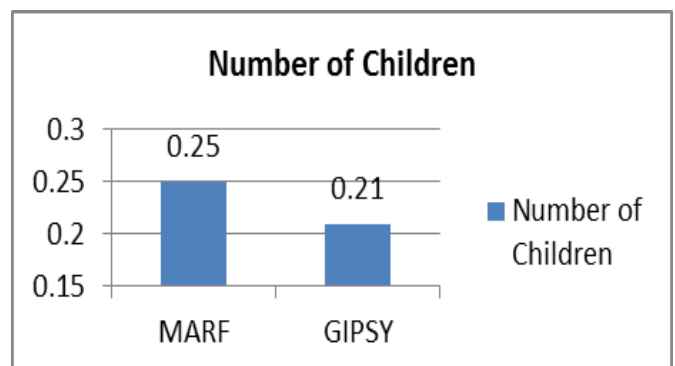

Figure 40: Graphical Comparison of Number of Children for the Case Studies

KEY OBSERVATIONS

MARF has high NOC, so we can say it has high reuse of the base classes.

FOR CLASS METRICS

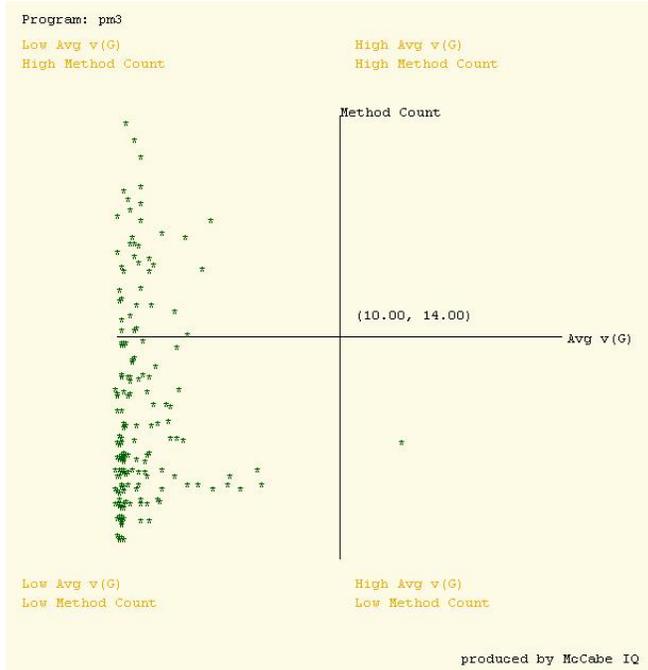

**Figure 41: Scatter Plot for MARF for Method Counts to Complexity**

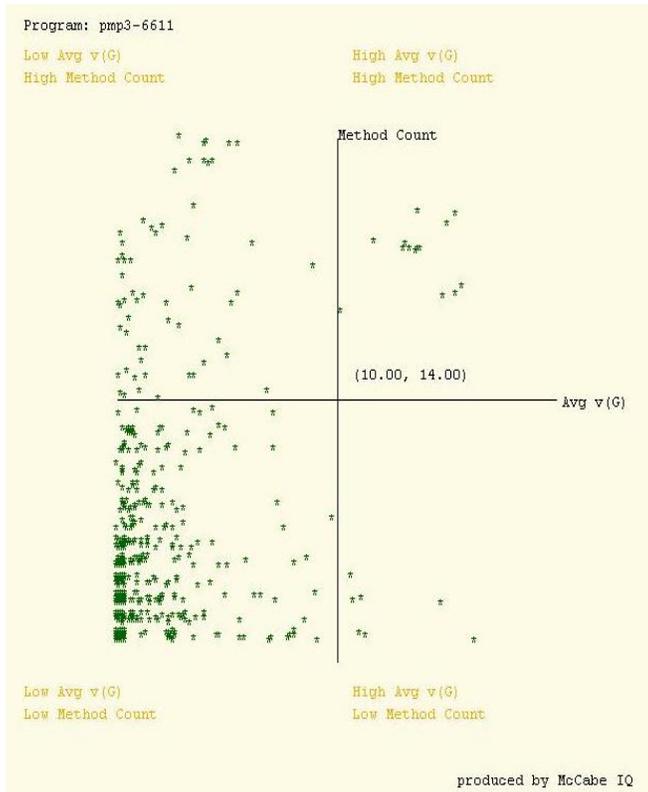

**Figure 42: Scatter Plot for GIPSY for Method Counts to Complexity**

KEY OBSERVATIONS

The Scatter plots show the method counts with respect to Complexity. Most classes for both case studies have less method count. However, MARF has a higher WMC and has a higher complexity when compared to GIPSY.

*HISTOGRAMS FOR THE CLASS METRICS*

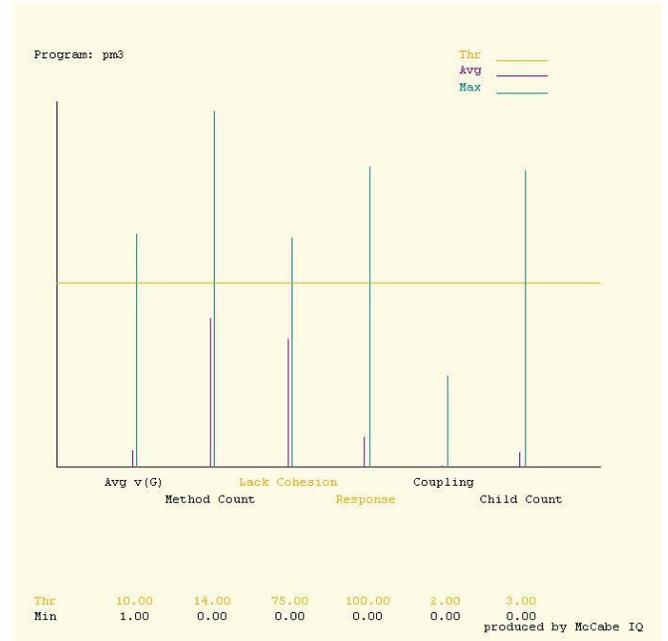

**Figure 43: Histogram for MARF**

KEY OBSERVATIONS

We notice that the classes of MARF have very less coupling. The average child count for all of the classes in MARF is very less and less than the threshold value. Similarly, we can also notice the same trends for the average complexities and cohesion for the case studies.

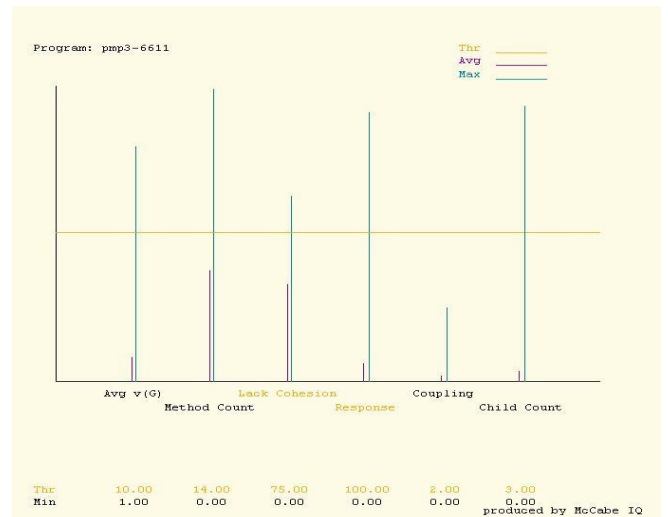

**Figure 44: Histogram for GIPSY**

*KEY OBSERVATIONS*

We notice that the classes of GIPSY have very less coupling. The average child count for all of the classes in MARF is very less and less than the threshold value. Similarly, we can also notice the same trends for the average complexities. The cohesion for the class is fairly closer to the threshold value.

## C. Summary

In this section we summarize the overall understanding of both the tools and list the advantages and disadvantages of LOGISCOPE and McCabe IQ tools.

### I. LOGISCOPE

LOGISCOPE provides metrics and graphical representations of the source code. Although it supports visualizing the results in terms of Call graphs, Control graphs, Use graphs, Inheritance graphs and Kiviat graphs of Metrics, the scope is limited to Kiviat graphs of the metrics and Pie charts in the Quality Report.

*Advantages of LOGISCOPE*

- Creates and provides support for creating Quality Reports with over 190 Procedural and Object-Oriented Metrics for C, C++, Ada & Java.
- Customizable Automated Reporting Facilities (HTML, Word) Integrated with development environments (IDEs)
- Context-free file parsing for C++, Ada and JAVA

*Disadvantages of LOGISCOPE*

- Extremely Limited support towards programming languages.
- Visualization of classes and modules is not supported.

### II. McCabe IQ

McCabe IQ analyzes many different kinds of programming languages running on any platform, enabling organizations to use the same tools and methodologies across all company projects. It lets you view the bigger picture in terms of complexity when dealing with multi language systems or interfaces.

*Advantages of McCabe IQ*

- McCabe tool supports various languages like Ada, ASM86, C, C#, C++.NET, C++, JAVA, JSP, VB, VB.NET, COBOL, FORTRAN, Perl, and PL1.
- It is platform independent.
- It presents detailed result; we can get metrics result for each module and for each class.
- It also provides us with the facility to analyze the result in various graphs.

*Disadvantages of McCabe IQ*

- Metrics calculation is not descriptive.
- Less User Friendly Interface.
- Execution speed is slow.

## 4.2 Design and Implementation with JDeodorant and MARFCAT

This section consists of an overview of JDeodorant and MARFCAT which are used for the implementation of the top ranked metrics along with the analysis of the results of these metrics with respect to the case studies.

### 4.2.1 Overview

#### A. JDeodorant

JDeodorant is an Eclipse plug-in that helps to find bad smells in codes, and solve them by refactoring techniques [74]. God Class is one of the bad smells in code, which is a class that holds many functionalities. Consequently it affect some quality factors like complexity, maintainability, understandability and etc. In order to remove that smell JDeodorant provides a refactoring technique, i.e. extract class [75]. There are other smells like feature envy, long method and type checking, and similarly in order to fix them JDeodorant suggest refactoring solutions respectively move method, extract method and replace conditional with polymorphism or replace type code with state/strategy [74]. In the implementation part of this study, JDeodorant provides a framework to add new metrics suites. We compute all the different metrics which have been ranked earlier.

#### B. MARFCAT

MARFCAT is a MARF-based code analysis tool, which detects, classifies and reports vulnerabilities or bad coding practices found in programming languages [72]. It uses machine learning approach to analyze code statically and also routine testing of any kind of code which further results in the improvement of efficiency in terms of speed, high precision, and robustness. SAMATE data set has been used here for practical validation [72].

There are three main principles used for training MARFCAT:

- Machine learning and dynamic programming
- Spectral and signal processing technique
- NLP n-gram and smoothing techniques

To understand weak code and interpret it in the form of signal, Common Vulnerabilities and exposures (CVEs) and Common Weaknesses and Exposures (CWEs) are used as knowledge in order. Apart from that there are many other different algorithms used, but in terms of accuracy and speed among these algorithms Signal Pipeline and NLP Pipeline are the best. These algorithms are used to test the weak and vulnerable code from our case studies by checking the threshold value.

### 4.2.2 Identification of Problematic Classes in Case Studies

The Logiscope and McCabe tools have ranked the classes under the fair and poor category. The two most problematic classes are selected for each case study and are compared in terms of the results of different metric tools in their respective packages. The top problematic classes identified in the packages are listed as follows –

*MARF*

1. `marf.MARF.java`

2. `marf.storage.StorageManager.java`

*GIPSY*

1. `gipsy.GIPC.intensional.SIPL.JOOIP.JavaParser.java`

2. `gipsy.GIPC.intensional.SIPL.JOOIP.ast.visitor.DumpVisitor.java`

*Interpretation of Problematic Classes*

These problematic classes are selected on the basis of the values generated by McCabe IQ. They have been categorized as the top problematic classes as they have high CBO, LCOM, WMC and RFC values. They also have higher amount of lines of code which increases their complexity and causes problems while parsing the code in the Eclipse environment. High CBO, WMC and RFC values make these classes very complex and they have very bad understandability. They have high LCOM values as well which means that they have very low cohesion when compared to the other classes in their respective packages.

| Class Name | Metrics | | | |
|---|---|---|---|---|
| | CBO | WMC | LOCM | RFC |
| `marf.MARF.java` | 2 | 58 | 97 | 64 |
| `marf.storage.StorageManager.java` | 2 | 39 | 84 | 53 |
| `gipsy.GIPC.intensional.SIPL.JOOIP.JavaParser.java` | 1 | 565 | 99 | 566 |
| `gipsy.GIPC.intensional.SIPL.JOOIP.ast.visitor.DumpVisitor.java` | 1 | 85 | 2 | 163 |

**Table 38: Interpretation of Problematic Classes from McCabe IQ tool**

### 4.2.3 Design and Implementation

The design and implementation of the top ranked metrics are implemented in the JDeodorant plugin and they are tested with the respective test cases by running it as in independent application from the Eclipse environment. The source files of these metrics are then tested on themselves. To test these metrics with respect to the case studies, we identify two problematic classes from each of the case studies and then run the metric implementations on them. The case studies are then run in the MARFCAT tool to check for weak and vulnerable code.

**A. Implementation of metrics in JDeodorant**

We have ranked CK metrics and QMOOD as the top metrics with respect to quality attributes mentioned in the case studies. Every metric is implemented in the metrics package of JDeodorant. The metrics are run on the projects, packages and the class. The test cases used to test these metrics are in a separate project Test. Each package within the test project corresponds to the name of the metric being implemented.

*1. Coupling between Objects CBO Metric*

This metric is implemented in the .java class named CBO.java. The test case for this metric is present in the test project under the `CBO_CF_DACTest` package. The connections between classes is computed by checking if the method invocations, types of attributes, method parameters and the method invocation of the super class.

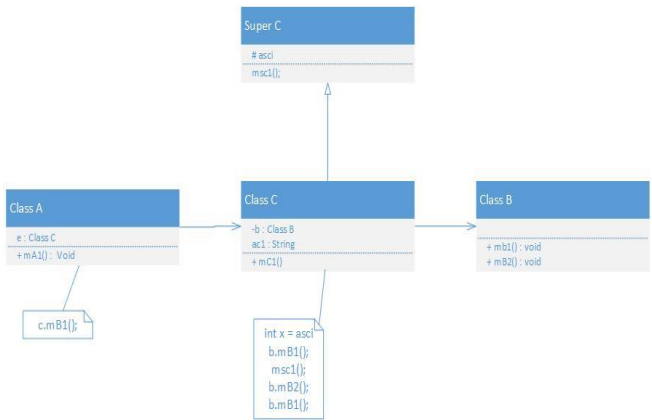

**Figure 45: Class Diagram for Test Cases for CBO**

From the test cases, we can see that a method in class C is invoking the methods of class B and Super Class C. Furthermore, it is also accessing the attribute of Super Class C. Hence, the CBO of class C is 2. The CBO value of Class A is 1 because it is invoking the method of class C. As CBO takes into account import and export coupling, we check if the method or attributes of the class is invoked or accessed by the other class making the CBO values of Class Super C and Class B as 1.

| Class Name | Value |
|---|---|
| Test.CBO_DAC_CF.ClassA.java | 1 |
| Test.CBO_DAC_CF.ClassB.java | 1 |
| Test.CBO_DAC_CF.ClassC.java | 2 |
| Test.CBO_DAC_CF.ClassD.java | 1 |

**Table 39: Test case results for CBO**

*2. Weighted Methods per Class (WMC) metric*

This metric is implemented in the .java class named WMC.java. The test case for this metric is present in the test project under the `WMCTest package`. The Cyclomatic Complexities of the classes are calculated by checking the composite statement of each method in the class. In composite statements we count the number of all if/else conditions, loop constructs and switch case constructs. To calculate CC there is the addition of methods in `CompositeStatementObject` class which are `getSwitchStatements, getForStatements, getWhileStatements and getDoStatements`.

```
public class ClassA
{
    public void m1()
    {
        int i = 0;
        while(i < 10)
        {
            System.out.println(i);
        }
    }

    public void m2()
    {
        int i = 3;
        do
        {
            if(i%3 == 0)
                System.out.println(i);

        }while(i < 10);
    }
}
```

**Figure: 46 Test Class for WMC (Eclipse IDE)**

As seen in figure 45 class from the test case, the number of conditional constructs in method m1() is 1 as there is only one while statement making its CC 1+1 = 2. In method m2() the CC is 3 as there is one if condition and one while condition making it 1+1+1 = 3. Therefore, the WMC of the entire class is 5.

*3. Coupling Factor (CF) Metric*

This metric is implemented in the .java class named WMC.java. The test case for this metric is present in the test project under the `CBO_CF_DACTest` package. The coupling factor is computed for the entire system similarly as in the CBO metric by checking if the method invocations, types of attributes, method parameters and the method invocation of the super class. The values are then computed by adding the coupling values of each class and substituting the values based on the formula described earlier. It considers only the import coupling.

With respect to the test case described in figure 45 the coupling of Class C is computed as 2, Class A as 1. These values are substituted in the formula of CF as 0.25.

*4. Data Abstraction Coupling (DAC) metric*

This metric is implemented in the .java class named DAC.java. The test case for this metric is present in the test project under the `CBO_CF_DACTest` package. The DAC is computed by checking the fields of each class and by checking if their type matches the fields in the other classes.

The test case in figure 45 describes that Class C has two fields where one field is of the type string and the other of type Class B which is one of the system classes. Hence, the DAC of C is 1.

*5. Response for Class (RFC) metric*

This metric is implemented in the .java class named `DAC.java`. The test case for this metric is present in the test project under the `RFCTest` package. In RFC, we check the method objects of each class and count the number of methods, method invocation objects and super method invocation objects and then compute their sums.

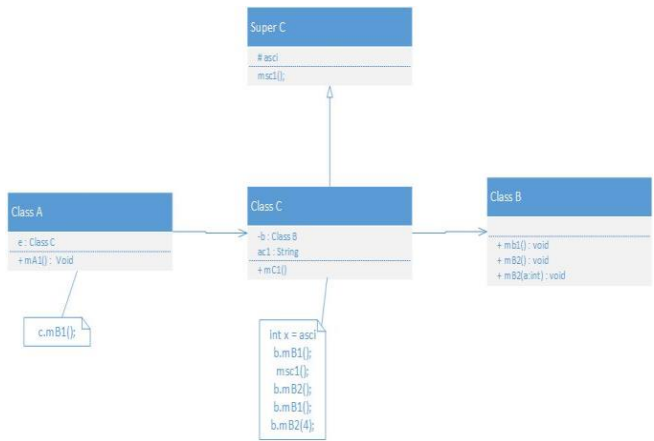

**Figure 47: Class Diagram for Test Cases for RFC**

As seen in the figure for the test case, the RFC for Class C is computed as 4 as it is invoking 1 method of Class Super C and 3 methods of Class B.

| Class Name | Value |
|---|---|
| Test.RFCTest.ClassA.java | 2 |
| Test.RFCTest.ClassB.java | 3 |
| Test.RFCTest.ClassC.java | 5 |
| Test.RFCTest.ClassD.java | 1 |

**Table 40: Test case results for RFC**

*6. Depth Inheritance Tree (DIT) metric*

This metric is implemented in the .java class named `DIT.java`. The test case for this metric is present in the test project under the `DITTEST` package. The metric checks if a class has a super class, and then if that super class has another super class. This process is repeated until the final class has no super class. The result will be the maximum number of inherited classes.

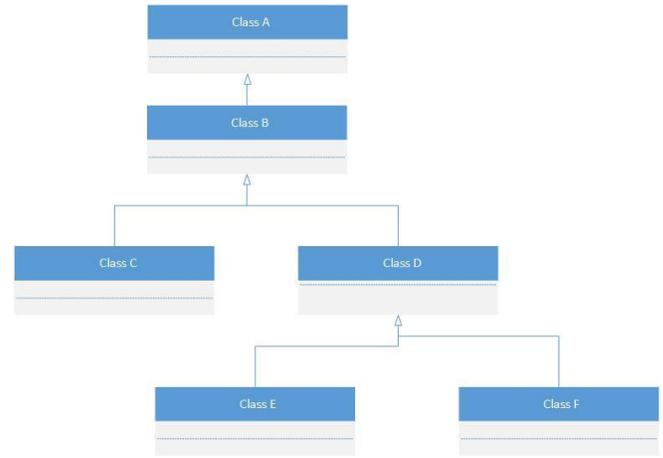

**Figure 48: Class Diagram for Test Cases for DIT**

From the test case, the values of DIT are computed as shown in the table.

| Class Name | Value |
|---|---|
| Test.DITTEST.ClassA.java | 0 |
| Test.DITTEST.ClassB.java | 1 |
| Test.DITTEST.ClassC.java | 2 |
| Test.DITTEST.ClassD.java | 2 |
| Test.DITTEST.ClassE.java | 3 |
| Test.DITTEST.ClassF.java | 3 |

**Table 41: Test case results for DIT**

*7. Class Interface Size (CIS) metric*

This metric is implemented in the .java class named `CIS.java`. The test case for this metric is present in the test project under the `CISTest` package. The access specifiers of each method and constructors are checked. The result is the count of the methods and constructors that are declared as public.

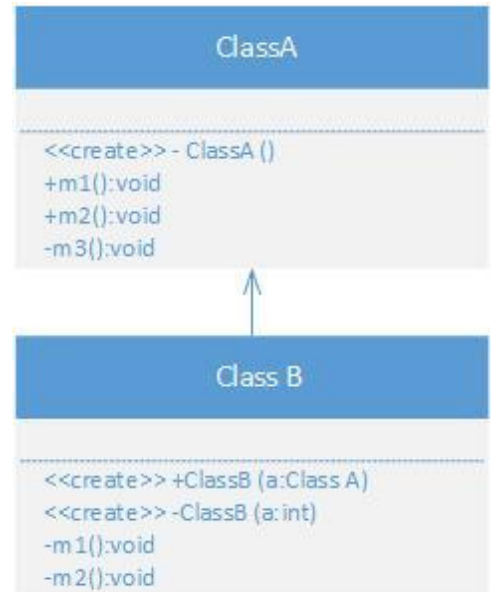

**Figure 49: Class Diagram for Test Cases for CIS**

From the given test case, as seen in the figure the value of CIS of Class A is computed as 2 and for Class B as 1.

*8. Data Access Metric (DAM)*

This metric is implemented in the .java class named DIT.java. The test case for this metric is present in the test project under the `DAMTest` package. The access specifier of each field in a class is checked and the result is computed as the ratio of the number of fields declared as private or protected to the total number of fields in the class.

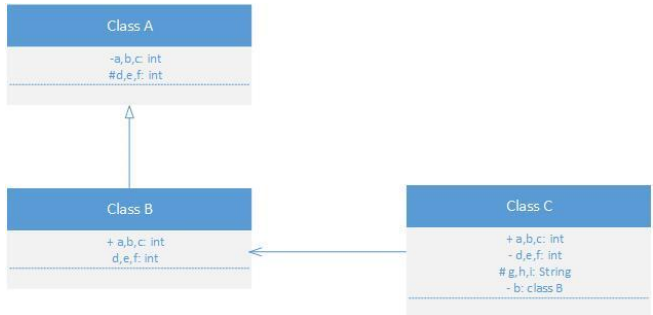

**Figure 50: Class Diagram for Test Cases for DAM**

In the test case described in figure the DAM for Class A is computed as 1 as it has 3 private attributes and three public attributes which makes the sum 6. The ratio is then computed as 1. In class B as there is no specified access specifiers for some attributes the CBO is computed as 0.

| Class Name | Value |
|---|---|
| Test.DAMTest.ClassA.java | 2 |
| Test.DAMTest.ClassB.java | 3 |
| Test.DAMTest.ClassC.java | 5 |

**Table 42: Test case results for DAM**

*9. Direct Class Coupling (DCC) metric*

This metric is implemented in the .java class named DCC.java. The test case for this metric is present in the test project under the `DCCTest` package. The field and parameter types are checked if they belong to the system class.

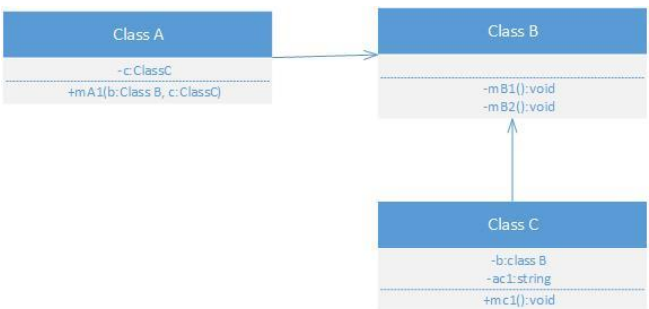

**Figure 51: Class Diagram for Test Cases for DCC**

In the test case described in figure, one of the fields of Class A is of type Class C and parameter type of the method is Class B. Therefore the DCC of Class A is computed as 2. The result set for the test case is given in the table.

*10. Number of Hierarchies (NOH) metric*

This metric is implemented in the `.java class` named `DCC.java`. This metric is computed for an entire package or an entire project and not for each individual class as it is a design level metric. The count of the number of different Super classes are computed here. These super classes do not have any inherited methods.

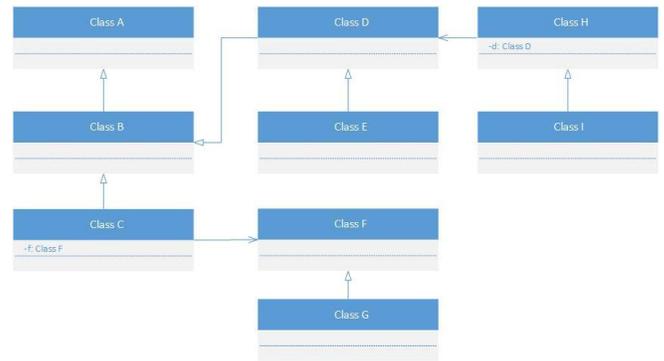

**Figure 52: Class Diagram for Test Cases for NOH**

In the given test case as show in figure, the Super classes are Class A, Class D, Class F and Class H. Therefore, the NOH is computed as 4.

*11. Number of Polymorphic Methods (NOP) metric*

This metric is implemented in the .java class named `NOP.java.` The test case for this metric is present in the test project under the `NOPTest` package. The type of the method objects are checked and the NOP is computed as the count of the number of abstract methods in the class.

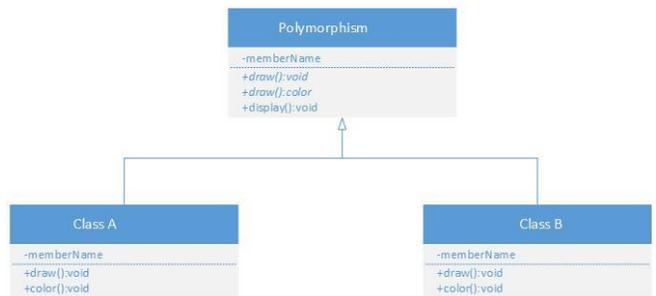

**Figure 53: Class Diagram for Test Cases for NOP**

In the test case described in figure, the class polymorphism has two abstract methods: draw and color. The Class A and B do are overriding these methods. Therefore, its NOP is computed as 2.The Class A and B do are overriding these methods and they're NOP values are computed as 0.

*12. Design Size in Classes (DSC) metric*

This metric is implemented in the .java class named `DSC.java`. This metric is computed for an entire package or an entire project and not for each individual class as it is a design level metric. It is computed as the count of the number of classes in the system.

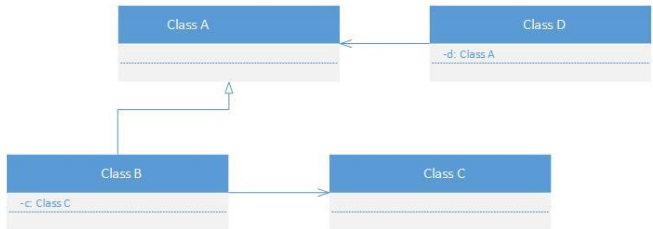

**Figure 54: Class Diagram for Test Cases for DSC**

These metric packages are run as test cases using JDeodorant in an independent Eclipse Application and the results can be seen in table X.

| Test Case/Metrics | METRICS | | | | | | | | | |
|---|---|---|---|---|---|---|---|---|---|---|
| | CBO | CIS | DAC | DCC | DAM | DIT | LCOM | NOP | RFC | WMC |
| CBO | 8 | 3 | 0 | 2 | 1.00 | 0 | 6 | 0 | 43 | 19 |
| CIS | 4 | 3 | 0 | 2 | 0.00 | 0 | 1 | 0 | 12 | 9 |
| DAC | 5 | 3 | 0 | 2 | 0.00 | 0 | 1 | 0 | 16 | 3 |
| DCC | 7 | 3 | 0 | 2 | 1.00 | 0 | 1 | 0 | 21 | 8 |
| DAM | 3 | 3 | 0 | 2 | 0.00 | 0 | 1 | 0 | 12 | 5 |
| DIT | 4 | 3 | 0 | 2 | 0.00 | 0 | 1 | 0 | 10 | 5 |
| LCOM | 3 | 3 | 0 | 2 | 1.00 | 0 | 3 | 0 | 18 | 9 |
| NOH | 4 | 3 | 0 | 2 | 1.00 | 0 | 1 | 0 | 9 | 5 |
| NOP | 3 | 3 | 0 | 2 | 0.00 | 0 | 1 | 0 | 9 | 3 |
| RFC | 4 | 3 | 0 | 2 | 0.00 | 0 | 1 | 0 | 16 | 4 |
| WMC | 5 | 3 | 0 | 3 | 0.00 | 0 | 3 | 0 | 17 | 4 |
| CF | 8 | 2 | 0 | 2 | 1.00 | 0 | 3 | 0 | 39 | 15 |
| DSC | 1 | 3 | 0 | 1 | 1.00 | 0 | ~ | 0 | 4 | 1 |

**Table 43: Result of Metrics as Test Cases**

NOH, DSC and CF being system-level metrics are calculated for the entire package. The NOH = 0 and DSC = 13.

### 4.2.4 Implementation of Metrics on Problematic Classes

To implement the metrics on the problematic classes, first, the packages of the class in question are identified along with the rest of the classes in that package. Then we run, the JDeodorant plugin in eclipse as an independent application and import the MARF and GIPSY projects into the workspace and then run the metrics on the problematic class packages. The results of running the metrics on the problematic classes are shown in the following tables.

*MARF Classes*

1. `MARF.java`

| Test Case/Metrics | METRICS | | | | | | | | | |
|---|---|---|---|---|---|---|---|---|---|---|
| marf package | CBO | CIS | DAC | DCC | DAM | DIT | LCOM | NOP | RFC | WMC |
| MARF.java | 29 | 55 | 6 | 4 | 0.24 | 0 | 1260 | 0 | 86 | 77 |
| Configuration.java | 4 | 39 | 1 | 2 | 1.00 | 0 | 534 | 0 | 59 | 39 |
| Version.java | 3 | 12 | 0 | 0 | 0.20 | 0 | 85 | 0 | 17 | 17 |

**Table 44: Result of Metric Implementations on `MARF.marf` package**

NOH = 0
DSC = 6

2. `StorageManager.java`

| Test Case/Metrics | METRICS | | | | | | | | | |
|---|---|---|---|---|---|---|---|---|---|---|
| Storage Package | CBO | CIS | DAC | DCC | DAM | DIT | LCOM | NOP | RFC | WMC |
| StorageManager.java | 17 | 39 | 0 | 0 | 1.00 | 0 | 278 | 0 | 43 | 49 |
| IDatabaseConnection.java | 1 | 0 | 0 | 1 | 0.00 | 0 | 10 | 0 | 5 | 0 |
| SampleLoader.java | 7 | 21 | 1 | 2 | 1.00 | 0 | 172 | 0 | 24 | 21 |
| ITrainingSample.java | 4 | 0 | 0 | 0 | 0.00 | 0 | 78 | 0 | 13 | 0 |
| Sample.java | 31 | 19 | 1 | 1 | 0.71 | 0 | 6 | 0 | 31 | 23 |
| SampleRecorder.java | 1 | 3 | 1 | 2 | 0.75 | 2 | 0 | 0 | 18 | 5 |
| MARFAudioFileFormat.java | 3 | 7 | 0 | 2 | 0.22 | 1 | 3 | 0 | 4 | 3 |
| FeatureSet.java | 5 | 12 | 0 | 0 | 1.00 | 2 | 3 | 0 | 30 | 21 |
| Idatabase.java | 1 | 0 | 0 | 0 | 0.00 | 0 | 10 | 0 | 0 | 0 |
| ISampleLoader.java | 8 | 0 | 0 | 1 | 0.00 | 0 | 105 | 0 | 15 | 0 |
| ModuleParams.java | 9 | 18 | 0 | 0 | 1.00 | 0 | 28 | 0 | 26 | 37 |
| Result.java | 11 | 16 | 1 | 0 | 1.00 | 0 | 18 | 0 | 13 | 9 |
| StorageException.java | 1 | 5 | 0 | 0 | 1.00 | 2 | 0 | 0 | 1 | 1 |
| ByteArrayFileReader.java | 1 | 12 | 0 | 0 | 0.67 | 0 | 9 | 0 | 10 | 12 |
| Cluster.java | 6 | 12 | 0 | 0 | 1.00 | 1 | 35 | 0 | 24 | 15 |
| Database.java | 3 | 10 | 0 | 0 | 1.00 | 1 | 34 | 0 | 14 | 9 |
| ResultSet.java | 15 | 31 | 0 | 2 | 0.80 | 0 | 68 | 0 | 45 | 41 |
| IStorageManager.java | 7 | 0 | 0 | 0 | 0.00 | 0 | 91 | 0 | 14 | 0 |
| TrainingSet.java | 9 | 17 | 0 | 0 | 0.56 | 1 | 34 | 0 | 39 | 27 |
| SampleLoaderFactory.java | 4 | 3 | 0 | 0 | 0.00 | 0 | 3 | 0 | 7 | 15 |
| TrainingSample.java | 4 | 20 | 0 | 0 | 1.00 | 0 | 81 | 0 | 37 | 24 |

**Table 44: Result of Metric Implementations on `marf.storage` package**

NOH = 0
DSC = 22

*GIPSY Classes*

1. `JavaParser.java`

| Test Case/Metrics | METRICS | | | | | | | | | |
|---|---|---|---|---|---|---|---|---|---|---|
| JOOIP Package.java | CBO | CIS | DAC | DCC | DAM | DIT | LCOM | NOP | RFC | WMC |
| JavaParser.java | 64 | 126 | 6 | 4 | 0.68 | 0 | 111872 | 0 | 596 | 2915 |
| JavaCharStream.java | 2 | 31 | 0 | 0 | 0.72 | 0 | 171 | 0 | 33 | 91 |
| JavaClassSymbolTable.java | 1 | 2 | 0 | 0 | 1 | 0 | NA | 0 | 0 | 0 |
| JavaIdentifierSymbolTable.java | 2 | 2 | 1 | 2 | 0 | 0 | NA | 0 | 0 | 0 |
| JavaParserConstants.java | 3 | 0 | 0 | 0 | 0 | 0 | NA | 0 | 0 | 0 |
| JavaParserTokenManager.java | 6 | 7 | 1 | 3 | 0.04 | 0 | 153 | 0 | 51 | 739 |
| JOOIPCompiler.java | 8 | 3 | 1 | 2 | 0 | 0 | 1 | 0 | 56 | 48 |
| JOOPToJavaTranslationItem.java | 1 | 2 | 1 | 2 | 0 | 0 | NA | 0 | 0 | 0 |
| ParseException.java | 2 | 4 | 1 | 2 | 0.4 | 1 | 1 | 0 | 13 | 24 |
| Token.java | 5 | 2 | 0 | 0 | 0 | 0 | 1 | 0 | 2 | 6 |
| TokenMgrError.java | 1 | 4 | 0 | 0 | 0 | 1 | 3 | 0 | 12 | 15 |

**Table 45: Result of Metric Implementations on `gipsy.GIPC.intensional.SIPL.JOOIP` package**

DSC = 15

2. `DumpVisitor.java`

| Test Case/Metrics | METRICS | | | | | | | | | |
|---|---|---|---|---|---|---|---|---|---|---|
| JOOIP.ast.visitor package | CBO | CIS | DAC | DCC | DAM | DIT | LCOM | NOP | RFC | WMC |
| DumpVisitor.java | 81 | 79 | 1 | 79 | 1 | 0 | 0 | 0 | 149 | 249 |
| GenericVisitor.java | 78 | 78 | 0 | 78 | 0 | 0 | 3003 | 0 | 78 | 0 |
| SourcePrinter.java | 1 | 7 | 0 | 0 | 1 | 0 | 10 | 0 | 10 | 10 |
| VoidVisitor.java | 78 | 78 | 0 | 78 | 0 | 0 | 3003 | 0 | 78 | 0 |

**Table 46: Result of Metric Implementations on `gipsy.GIPC.intensional.SIPL.JOOIP.ast.visitor` package**

### 4.2.5 Analysis of Problematic Classes

This section gives a comparison of the classes with respect to the other classes in their respective packages. The lines of code are computed with the help of the tools used earlier.

*MARF Classes*

1. `MARF.java`

The number of lines of code in the entire package is 2792. The MARF.java class occupies 69% of the package and the less problematic classes occupy the remaining 31% of the packages.

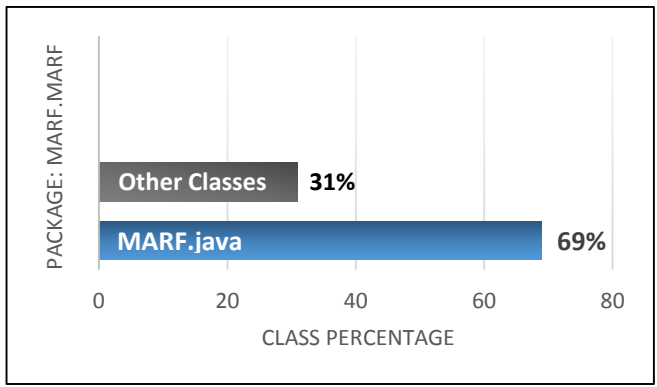

**Figure 55: Size ratio of MARF.java with other classes in package MARF.marf**

2. `StorageManager.java`

The number of lines of code in the entire package is 6028. The StorageManager.java class occupies around 15% of the entire package. This is a fairly high percentage as there are 21 classes present in this package.

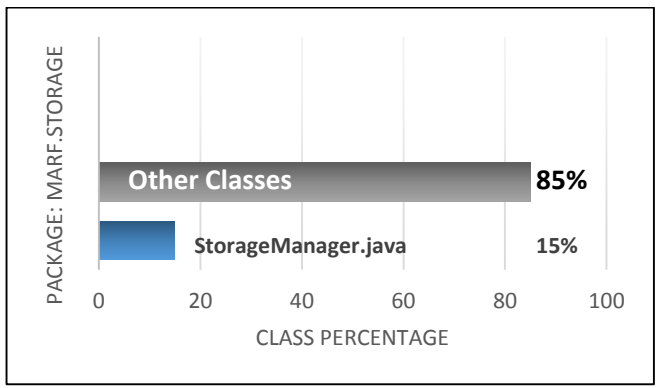

**Figure 56: Size ratio of StorageManager.java with other classes in package MARF.Storage**

GIPSY Classes

1. `JavaParser.java`

The number of lines of code in the entire package is 12112. The `JavaParser.java` class occupies a huge 64% of the package and the less complex classes occupy the remaining 36%.

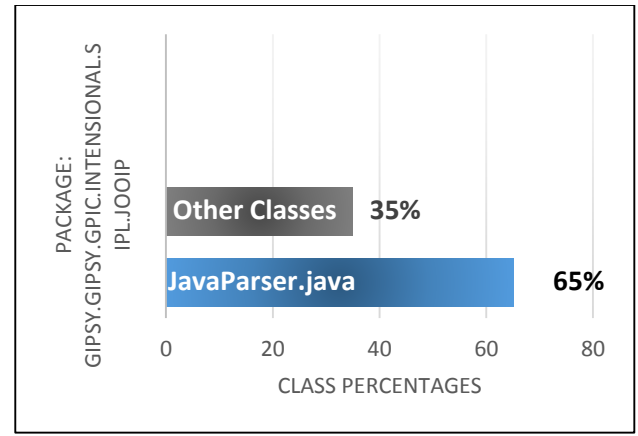

**Figure 57: Size ratio of StorageManager.java with other classes in package GIPSY.GIPSY.GIPC.INTENSIONAL.SIPL.JOOIP**

2. `DumpVisitor.java`

The number of lines of code in the entire package is 1794. The `DumpVisitor.java` class occupies 68% of the class. The remainder is occupied by the less problematic classes.

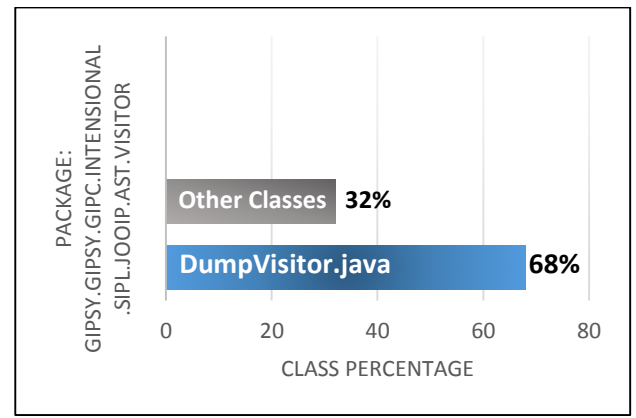

**Figure 58: Size ratio of StorageManager.java with other classes in package GIPSY.GIPSY.GIPC.INTENSIONAL.SIPL.JOOIP.AST. VISITOR**

Comparison of Problematic Classes:

`Marf.Java`, `StorageManager.java`, `Java Parser.java` and `DumpVisitor.java` have high values in their respective packages to the corresponding metrics with respect to the other classes in their packages.

- *Coupling between object (CBO):*
  It means, these classes are more dependent on other classes, and potential error in other classes can cause malfunctionality in these classes. Higher CBO means less Reusability, Flexibility and Understandability.

- *Class Interface Size (CIS):*
  Number of public methods in this class is more than the other classes. Potentially it can be a security issue, because public methods are visible and accessible for everyone.

- *Data Abstraction Coupling (DAC):*
  The number of abstract data type in the class shows number of objects which are from other class type. It means any malfunction or security issue in other class can affect this class. Moreover, higher DAC means less Reusability, Flexibility and Understandability.

- *Lack of Cohesion of Methods (LCOM):*
  The cohesion in classes should be good, which means LCOM should be as less as possible. Larger the LCOM value, lesser cohesion in class which decrease Reusability, Functionality and Understandability.

- *Response For Class (RFC):*
  The response for a class is a set of methods that can potentially be executed in response to a message received by an object of that class. More RFC makes testing more difficult for all possible outcomes.

- *Weighted Method per Class (WMC):*
  A count of methods implemented within a class (rather than all methods accessible within the class hierarchy). Classes with large numbers of methods are likely to be more application specific, limiting the possibility of reuse.

`Marf.java` has lesser value in:

- *Data Access Metric (DAM) :*
  The ratio of the number of private (protected) attributes to the total number of attributes declared in the class. It is an average metrics, less DAM means less private and protected attributes compare to total number of attributes including public attributes. More public attributes can be a security issue.

## 4.2.6 Analysis of Quality Attributes with respect to the metric Implementations

The main motivation behind this study is to derive the quality attributes with respect to the case studies and the metric implementations. This section describes about how each metric influences a quality attribute with respect to the results achieved in the test cases and the case studies. The results of the metric implementations focus on the problematic classes and their respective packages. Only the identified problematic classes give high values of CBO, RFC, WMC, DAC, etc. resulting in higher complexity and less understandability. Also, as seen in the percentages of size occupied are very high indicate higher complexity, less reusability and less flexibility. The high LCOM values show low cohesion among the methods of the class. However, these observations are only limited to the problematic classes. The values shown by the metrics for the other classes are relatively low which indicate that the overall quality of the case studies is good. The Logiscope and McCabe tools gave a comprehensive analysis on these case studies and their results have indicated the same. Therefore, these implementations help understand the quality of the case studies in a greater depth and help provide recommendations to help improve future projects.

## 4.3 MARFCAT

### 4.3.1 Detecting vulnerable Java files with the help of MARFCAT

The case studies of both MARF and GIPSY test files are scanned with the provided default MARFCAT and apache-tomcat training set to detect vulnerable code. After logging into Linux, a folder is created which holds all necessary files such as marfcat.jar, marf.jar, apache-tomcat-5.5.13-src_train.xml, gipsy_test.xml, marf_test.xml, collectfilesmeta.pl, cve.marf.Storage.TrainingSet.709marf.apps.MARFCAT.Storage.AnyToWAVLoader.0.0.107.301.512.gzbin, java.util.ArrayList.ref, javalimited.sh, Makefile are kept along with the test-case folder under which we had our MARF and GIPSY original projects. The scripts are executed by using the commands chmod u+x collect-files-meta.pl \ javalimited.sh marfcat and run using make test-quick-gipsy-cve and make test-quick-marf-cve commands.

After creation of both log files the code is checked and the testing collection size for GIPSY is computed 654 and MARF as 257. The config for both files were –nonpreprep –raw –fft –cheb –flucid. From that it is understood that the data was raw which means that there was no preprocessing. For feature extraction, it has used Fast Fourier transform (FFT) algorithm and for classification, Chebyshev Distance. In both log files, *File* gives us the file location, *Processing time* gives us the time it took to process, *strFileType* tells us what kind of file it is, *Date/time* tells us the date and time of running the test and *Resultset* says that it is suppressed. In both log files for each case study the distance threshold is 0.1, computed raw of P is 0, normalized P is 0.0 and warning to be reported is false. From these values it is understood that no files have surpassed the default threshold value. The log files can be viewed in the implementation folder of this project.

### 4.3.2 Human Generated Code Vs Generated Code

The MARF and GIPSY project java files are scanned to detect generated code as opposed to human written code. The Java files which have highest likelihood of generated code for both MARF and GIPSY are given in the following tables.

| Marf java files that have highest generated code |
|---|
| marf.nlp.Parsing.GrammarCompiler.GrammarTokenType.java |
| marf.nlp.Parsing.TokenSubType.java |
| marf.nlp.Parsing.TokenType.java |
| marf.Classification.ClassificationFactory.java |
| marf.Configuration.java |
| marf.MARF.java |
| marf.Version.java |

**Table 47: MARF Classes with most generated code**

| Gipsy java files that have highest generated code |
|---|
| gipsy.GIPC.intensional.GIPL.GIPLParser.java |
| gipsy.GIPC.intensional.GIPL.GIPLParserTokenManager.java |
| gipsy.GIPC.DFG.DFGAnalyzer.DFGParserTokenManager.java |
| gipsy.GIPC.intensional.SIPL.ForensicLucid.ForensicLucidParserTokenManager.java |
| gipsy.GIPL.intensional.SIPL.IndexicalLucid.IndexicalLucidParserTokenManager.java |
| gipsy.GIPC.Preprocessing.PreprocessorParserTokenManager.java |
| gipsy.GIPC.intensional.SIPL.ObjectiveLucid.ObjectiveGIPLParserTokenManager.java |
| gipsy.GIPC.intensional.SIPL.ObjectiveLucid.ObjectiveIndexicalLucidParserTokenManage.java |
| gipsy.GIPC.intensional.SIPL.Lucx.LucxParserTokenManager.java |
| gipsy.GIPC.intensional.SIPL.JOOIP.JavaParserTokenManager.java |
| gipsy.GIPC.intensional.SIPL.JLucid.JGIPLParserTokenManager.java |
| gipsy.GIPC.intensional.SIPL.JLucid.JIndexicalLucidParserTokenManager.java |

**Table 48: GIPSY Classes with most generated code**

### 4.3.3 Classes with High CBO Values

Both MARF and GIPSY original project are tested as test cases and the java classes which have high CBO values are given in the following table

| Marf Files that have high CBO value | CBO Value |
|---|---|
| Under default package test.java | 158 |
| marf.Storage.Sample.java | 31 |
| marf.Marf.java | 29 |
| marf.nlp.Parsing.Token.java | 18 |
| marf.Preprocessing.Preprocessing.java | 17 |

**Table 49: MARF Classes with most CBO values**

| Gipsy Files that have high CBO value | CBO value |
|---|---|
| gipsy.GIPC.intensional.SIPL.JOOIP.ast.visitor.DumpVisitor.java | 81 |
| gipsy.GIPC.intensional.SIPL.JOOIP.ast.visitor.VoidVisitor.java | 78 |
| gipsy.GIPC.intensional.SIPL.JOOIP.ast.visitor.GenericVisitor.java | 78 |
| gipsy.GIPC.intensional.SIPL.JOOIP.JavaParser.java | 64 |
| gipsy.GIPC.intensional.SimpleNode.java | 39 |
| gipsy.GEE.IDP.demands.Idemand.java | 38 |
| gipsy.Configuration.java | 34 |

**Table 50: GIPSY Classes with most CBO values**

## 5. CONCLUSION

Through a series of comparisons and discussions, the metrics have been ranked with respect to the quality attributes. MARF is primarily used as an audio processing tool but it allows researchers and developers to test new and existing algorithms to provide new and improved projects. GIPSY can be used as a platform to provide superior solutions solve critical problems with the utilization of Intensional Programming. The main motivation of this study has been achieved by identifying the quality attributes from within the case studies. To further understand the relationships between the quality attributes and the case studies, the computation of these quality attributes, their computation techniques and how they help understand the overall quality of a software, quality reports are generated for the MARF and GIPSY case studies with the help of the LOGISCOPE and McCabe IQ metric tools. These tools help in identifying problematic classes which are tested with the design and implementation of 13 metrics in JDeodorant. These implementations have promised to understand the quality attributes that very important role in software comprehension, maintenance and delivery. Finally, MARFCAT was first trained to learn a knowledge base. It was then tested to detect vulnerable code in the case studies.

In the process of generating the quality reports in Logiscope and McCabe IQ tools and during the design and implementations of the top ranked metrics, we have achieved a better way of understanding and measuring the quality of the code. This will help reduce maintainability of a software project in the future and help improve overall quality. This research experience has been invaluable and has helped provide positive evaluation of the quality attributes.

### 5.1 Future Scope

The future and ongoing work within the context of MARFCAT is to integrate the MARFCAT invocation within JDeodorant and to re-train MARFCAT to differentiate between machine code and human generated code. There is also a scope to retrain it to detect classes with high CBO values.

### REFERENCES

[1]. Serguei A. Mokhov. Introducing MARF: a modular audio recognition framework and its applications for scientific and software engineering research. In Advances in Computer and Information Sciences and Engineering, pages 473-478, University of Bridgeport, U.S.A., December 2007. Springer Netherlands. ISBN

## APPENDIX

Classification of classes of MARF and GIPSY ranked under the fair and poor categories.

### A. MARF

| Class Names for MARF with fair Analyzability |
|---|
| marf.Classification.Classification |
| marf.Classification.RandomClassification.RandomClassification |
| marf.Classification.Stochastic.MaxProbabilityClassifier |
| marf.Classification.Stochastic.ZipfLaw |
| marf.FeatureExtraction.FFT.FFT |
| marf.FeatureExtraction.FeatureExtractionAggregator |
| marf.FeatureExtraction.LPC.LPC |
| marf.MARF |
| marf.Preprocessing.CFEFilters.BandStopFilter |
| marf.Preprocessing.CFEFilters.CFEFilter |
| marf.Preprocessing.CFEFilters.LowPassFilter |
| marf.Preprocessing.FFTFilter.FFTFilter |
| marf.Preprocessing.Preprocessing |
| marf.Preprocessing.WaveletFilters.WaveletFilter |
| marf.Stats.ProbabilityTable |
| marf.Stats.StatisticalEstimators.StatisticalEstimator |
| marf.Storage.Cluster |
| marf.Storage.SampleRecorder |
| marf.Storage.StorageManager |
| marf.Storage.TrainingSet |
| marf.math.Matrix |
| marf.nlp.Parsing.GrammarCompiler.Grammar |
| marf.nlp.Parsing.GrammarCompiler.ProbabilisticGrammarCompiler |
| marf.nlp.Parsing.ProbabilisticParser |
| marf.util.Arrays |
| test |

**Table A.1: MARF classes with fair Analyzability**

| Class Names for MARF with poor Analyzability |
|---|
| marf.Classification.NeuralNetwork.NeuralNetwork |
| marf.nlp.Parsing.GrammarCompiler.GrammarCompiler |
| marf.nlp.Parsing.LexicalAnalyzer |

**Table A.2: MARF classes with poor Analyzability**

| Class Names for MARF with fair Changeability |
|---|
| marf.Classification.Stochastic.ZipfLaw |
| marf.Storage.ResultSet |
| marf.Storage.StorageManager |
| marf.math.ComplexMatrix |
| marf.math.Matrix |
| marf.nlp.Parsing.GrammarCompiler.Grammar |
| marf.nlp.Parsing.GrammarCompiler.GrammarCompiler |
| marf.nlp.Parsing.Parser |
| marf.util.Arrays |
| marf.util.OptionProcessor |

**Table A.3: MARF classes with fair Changeability**

| Class Names for MARF with poor Changeability |
|---|
| marf.Classification.NeuralNetwork.NeuralNetwork |
| marf.Configuration |
| marf.MARF |
| marf.nlp.Storage.Corpus |

**Table A.4: MARF classes with poor Changeability**

| Class Names for MARF with fair Stability |
|---|
| marf.Classification.Classification |
| marf.Classification.Distance.Distance |
| marf.Classification.NeuralNetwork.NeuralNetwork |
| marf.Classification.Stochastic.ZipfLaw |
| marf.FeatureExtraction.FeatureExtraction |
| marf.FeatureExtraction.IFeatureExtraction |
| marf.MARF.ENgramModels |
| marf.MARF.NLP |
| marf.Preprocessing.IPreprocessing |
| marf.Storage.Loaders.AudioSampleLoader |
| marf.Storage.MARFAudioFileFormat |

| Class Names for MARF with fair Stability |
|---|
| marf.Storage.ModuleParams |
| marf.Storage.Result |
| marf.Storage.ResultSet |
| marf.Storage.TrainingSet |
| marf.math.ComplexMatrix |
| marf.math.ComplexVector |
| marf.nlp.Parsing.CodeGenerator |
| marf.nlp.Parsing.CompilerError |
| marf.nlp.Parsing.GenericLexicalAnalyzer |
| marf.nlp.Parsing.GrammarCompiler.Grammar |
| marf.nlp.Parsing.GrammarCompiler.GrammarTokenType |
| marf.nlp.Parsing.LexicalError |
| marf.nlp.Parsing.SymTabEntry |
| marf.nlp.Parsing.SymbolTable |
| marf.nlp.Parsing.TokenSubType |
| marf.nlp.Parsing.TokenType |
| marf.nlp.Storage.Corpus |
| marf.util.Arrays |
| marf.util.BaseThread |
| marf.util.Debug |
| marf.util.MARFException |
| marf.util.OptionProcessor |
| marf.util.SortComparator |

**Table A.5: MARF classes with fair Stability**

| Class Names for MARF with poor Stability |
|---|
| marf.MARF |
| marf.Preprocessing.FFTFilter.FFTFilter |
| marf.Stats.StatisticalEstimators.StatisticalEstimator |
| marf.Storage.Sample |
| marf.Storage.StorageManager |
| marf.math.Matrix |
| marf.math.Vector |
| marf.nlp.Parsing.SyntaxError |

**Table A.6: MARF classes with poor Stability**

| Class Names for MARF with fair Testability |
|---|
| marf.Configuration |
| marf.Stats.ProbabilityTable |
| marf.Storage.ResultSet |
| marf.math.ComplexMatrix |
| marf.nlp.Parsing.GrammarCompiler.Grammar |
| marf.nlp.Parsing.GrammarCompiler.GrammarCompiler |
| marf.nlp.Parsing.LexicalAnalyzer |
| marf.nlp.Storage.Corpus |

**Table A.7: MARF classes with fair Testability**

| Class Names for MARF with poor Testability |
|---|
| marf.Classification.NeuralNetwork.NeuralNetwork |
| marf.MARF |
| marf.Storage.StorageManager |
| marf.math.Matrix |
| marf.util.Arrays |

**Table A.8: MARF classes with poor Testability**

### B. GIPSY

| Class Names for GIPSY with fair Analyzability |
|---|
| gipsy.GEE.IDP.DemandGenerator.Interpreter |
| gipsy.GEE.IDP.DemandGenerator.LegacyInterpreter |
| gipsy.GEE.IDP.DemandGenerator.jini.rmi.JINITA |
| gipsy.GEE.IDP.DemandGenerator.jini.rmi.JINITransportAgent.JTABackend |
| gipsy.GEE.IDP.DemandGenerator.jini.rmi.JINITransportAgent.JINITransportAgentProxy |
| gipsy.GEE.IDP.DemandGenerator.jini.rmi.JiniDemandDispatcher |
| gipsy.GEE.IDP.DemandGenerator.jini.rmi.MulticastJiniServiceDiscoverer |
| gipsy.GEE.IDP.DemandGenerator.jms.DemandController |
| gipsy.GEE.IDP.DemandGenerator.jms.JMSDemandDispatcher |
| gipsy.GEE.IDP.DemandGenerator.rmi.IdentifierContextServer |
| gipsy.GEE.IDP.demands.Demand |
| gipsy.GEE.IVW.Warehouse.NetCDFFileManager |
| gipsy.GEE.multitier.DST.jms.JMSDSTWrapper |
| gipsy.GEE.multitier.DWT.DWTFactory |
| gipsy.GEE.multitier.GIPSYNode |
| gipsy.GEE.multitier.GMT.demands.DSTRegistration |
| gipsy.GEE.multitier.GMT.demands.NodeRegistration |
| gipsy.GEE.multitier.GMT.demands.TierAllocationResult |
| gipsy.GEE.multitier.TAExceptionHandler |
| gipsy.GIPC.DFG.DFGAnalyzer.DFGParserTokenManager |
| gipsy.GIPC.DFG.DFGAnalyzer.LucidCodeGenerator |
| gipsy.GIPC.GIPC |
| gipsy.GIPC.Preprocessing.PreprocessorParserTokenManager |
| gipsy.GIPC.imperative.ImperativeCompiler |
| gipsy.GIPC.imperative.Java.JavaCommunicationProcedureGenerator |
| gipsy.GIPC.intensional.GIPL.GIPLParserTokenManager |
| gipsy.GIPC.intensional.GenericTranslator.TranslationLexer |
| gipsy.GIPC.intensional.IntensionalCompiler |
| gipsy.GIPC.intensional.SIPL.ForensicLucid.ForensicLucidParserTokenManager |
| gipsy.GIPC.intensional.SIPL.IndexicalLucid.IndexicalLucidParserTokenManager |
| gipsy.GIPC.intensional.SIPL.JLucid.JGIPLParserTokenManager |
| gipsy.GIPC.intensional.SIPL.JLucid.JIndexicalLucidParserTokenManager |
| gipsy.GIPC.intensional.SIPL.JLucid.JLucidCompiler |
| gipsy.GIPC.intensional.SIPL.JOOIP.JOOIPCompiler |
| gipsy.GIPC.intensional.SIPL.JOOIP.JavaCharStream |
| gipsy.GIPC.intensional.SIPL.JOOIP.ast.expr.CharLiteralExpr |
| gipsy.GIPC.intensional.SIPL.JOOIP.ast.expr.DoubleLiteralExpr |
| gipsy.GIPC.intensional.SIPL.JOOIP.ast.expr.IntegerLiteralMinValueExpr |
| gipsy.GIPC.intensional.SIPL.JOOIP.ast.expr.LongLiteralMinValueExpr |
| gipsy.GIPC.intensional.SIPL.JOOIP.ast.visitor.GenericVisitor |
| gipsy.GIPC.intensional.SIPL.JOOIP.ast.visitor.VoidVisitor |
| gipsy.GIPC.intensional.SIPL.Lucx.LucxParserTokenManager |
| gipsy.GIPC.intensional.SIPL.ObjectiveLucid.ObjectiveGIPLParserTokenManager |
| gipsy.GIPC.intensional.SIPL.ObjectiveLucid.ObjectiveIndexicalLucidParserTokenManager |
| gipsy.GIPC.util.SimpleCharStream |
| gipsy.RIPE.RIPE |
| gipsy.RIPE.editors.RunTimeGraphEditor.core.GraphDataManager |
| gipsy.RIPE.editors.RunTimeGraphEditor.operator.GIPSYGMTController |
| gipsy.RIPE.editors.RunTimeGraphEditor.operator.GIPSYTiersController |
| gipsy.RIPE.editors.RunTimeGraphEditor.ui.dialogs.GIPSYNodeDialog |
| gipsy.RIPE.editors.WebEditor.WebEditor |
| gipsy.apps.marfcat.MARFCATDGT |
| gipsy.apps.marfcat.MARFCATDWT |
| gipsy.apps.marfcat.MARFCATDWT.MARFCATDWTApp |
| gipsy.apps.marfcat.MARFPCATDGT |
| gipsy.apps.marfcat.MARFPCATDWT |
| gipsy.apps.marfcat.MARFPCATDWT.MARFPCATDWTApp |
| gipsy.lang.GIPSYContext |
| gipsy.lang.context.OrderedFinitePeriodicTagSet |
| gipsy.lang.context.OrderedInfiniteNonPeriodicTagSet |
| gipsy.lang.context.OrderedInfinitePeriodicTagSet |
| gipsy.lang.context.UnorderedFiniteNonPeriodicTagSet |
| gipsy.lang.context.UnorderedFinitePeriodicTagSet |
| gipsy.lang.context.UnorderedInfinitePeriodicTagSet |
| gipsy.tests.GEE.IDP.demands.DemandTest |
| gipsy.tests.GEE.multitier.DGT.PseudoDGT |
| gipsy.tests.GEE.multitier.DST.PseudoJiniDSTWrapper |
| gipsy.tests.GEE.multitier.GIPSYNodeTestDriver |
| gipsy.tests.GEE.multitier.GMT.GMTTestConsole |
| gipsy.tests.GEE.multitier.GMT.GMTTestConsole.KeyInputProcessor |
| gipsy.tests.GEE.simulator.DGTDialog |
| gipsy.tests.GEE.simulator.DSTSpaceScalabilityTester |
| gipsy.tests.GEE.simulator.DemandResponseTimeTester |
| gipsy.tests.GEE.simulator.ProfileDialog |
| gipsy.tests.GEE.simulator.ProfileDialog.ProfileToolbar |
| gipsy.tests.GEE.simulator.ResultAnalyst |
| gipsy.tests.GEE.simulator.demands.LightUniqueDemand |
| gipsy.tests.GEE.simulator.demands.SizeAdjustableDemand |
| gipsy.tests.GEE.simulator.demands.WorkResultHD |
| gipsy.tests.GEE.simulator.demands.WorkResultPi |
| gipsy.tests.jooip.CopyOfGIPLtestVerbose |
| gipsy.tests.junit.GEE.multitier.DGT.DGTWrapperTest |
| gipsy.tests.junit.GEE.multitier.DWT.DWTWrapperTest |

**Table A.9: GIPSY classes with fair Analyzability**

| Class Names for GIPSY with poor Analyzability |
|---|
| gipsy.GEE.IDP.DemandGenerator.LegacyEductiveInterpreter |
| gipsy.GEE.IDP.DemandGenerator.jms.JMSTransportAgent |
| gipsy.GEE.multitier.DST.jini.JiniDSTWrapper |
| gipsy.GEE.multitier.DST.jini.JiniERIDSTWrapper |
| gipsy.GEE.multitier.GMT.GMTWrapper |
| gipsy.GIPC.DFG.DFGAnalyzer.DFGParser |
| gipsy.GIPC.DFG.DFGGenerator.DFGCodeGenerator |
| gipsy.GIPC.DFG.DFGGenerator.DFGTranCodeGenerator |
| gipsy.GIPC.Preprocessing.PreprocessorParser |
| gipsy.GIPC.SemanticAnalyzer |
| gipsy.GIPC.imperative.Java.JavaCompiler |
| gipsy.GIPC.intensional.GIPL.GIPLParser |
| gipsy.GIPC.intensional.GenericTranslator.TranslationParser |
| gipsy.GIPC.intensional.SIPL.ForensicLucid.ForensicLucidParser |
| gipsy.GIPC.intensional.SIPL.ForensicLucid.ForensicLucidSemanticAnalyzer |
| gipsy.GIPC.intensional.SIPL.IndexicalLucid.IndexicalLucidParser |
| gipsy.GIPC.intensional.SIPL.JLucid.JGIPLParser |
| gipsy.GIPC.intensional.SIPL.JLucid.JIndexicalLucidParser |
| gipsy.GIPC.intensional.SIPL.JOOIP.JavaParser |
| gipsy.GIPC.intensional.SIPL.JOOIP.JavaParserTokenManager |
| gipsy.GIPC.intensional.SIPL.JOOIP.ast.visitor.DumpVisitor |
| gipsy.GIPC.intensional.SIPL.Lucx.LucxParser |
| gipsy.GIPC.intensional.SIPL.ObjectiveLucid.ObjectiveGIPLParser |
| gipsy.GIPC.intensional.SIPL.ObjectiveLucid.ObjectiveIndexicalLucidParser |
| gipsy.RIPE.editors.RunTimeGraphEditor.ui.GIPSYGMTOperator |
| gipsy.apps.memocode.genome.AlignDGT |
| gipsy.apps.memocode.genome.AlignDWT |
| gipsy.tests.GEE.simulator.DGTSimulator |
| gipsy.tests.GIPC.intensional.SIPL.Lucx.SemanticTest.LucxSemanticAnalyzer |

**Table A.10: GIPSY classes with poor Analyzability**

| Class Names for GIPSY with fair Changeability |
|---|
| gipsy.GEE.GEE |
| gipsy.GEE.IDP.DemandGenerator.jini.rmi.JINITA |
| gipsy.GEE.IDP.DemandGenerator.jini.rmi.JINITransportAgent |
| gipsy.GEE.IDP.DemandGenerator.jini.rmi.JiniDemandDispatcher |
| gipsy.GEE.IDP.DemandGenerator.jms.JMSTransportAgent |
| gipsy.GEE.IDP.demands.Demand |
| gipsy.GEE.IVW.Warehouse.Cache |

| gipsy.GEE.multitier.DST.jini.JiniDSTWrapper |
|---|
| gipsy.GEE.multitier.DST.jini.JiniERIDSTWrapper |
| gipsy.GEE.multitier.GIPSYNode |
| gipsy.GEE.multitier.GMT.GMTWrapper |
| gipsy.GIPC.DFG.DFGAnalyzer.LucidCodeGenerator |
| gipsy.GIPC.DFG.DFGGenerator.DFGCodeGenerator |
| gipsy.GIPC.DFG.DFGGenerator.DFGTranCodeGenerator |
| gipsy.GIPC.GIPC |
| gipsy.GIPC.SemanticAnalyzer |
| gipsy.GIPC.intensional.GenericTranslator.TranslationLexer |
| gipsy.GIPC.intensional.SIPL.ForensicLucid.ForensicLucidSemanticAnalyzer |
| gipsy.GIPC.intensional.SIPL.JOOIP.ast.visitor.DumpVisitor |
| gipsy.RIPE.editors.RunTimeGraphEditor.core.GIPSYTier |
| gipsy.RIPE.editors.RunTimeGraphEditor.core.GraphDataManager |
| gipsy.RIPE.editors.RunTimeGraphEditor.ui.InstancesNodesPanel |
| gipsy.RIPE.editors.RunTimeGraphEditor.ui.MapEditor |
| gipsy.RIPE.editors.RunTimeGraphEditor.ui.dialogs.GIPSYNodeDialog |
| gipsy.RIPE.editors.RunTimeGraphEditor.ui.dialogs.TierPropertyDialog |
| gipsy.RIPE.editors.WebEditor.WebEditor |
| gipsy.lang.GIPSYContext |
| gipsy.tests.GEE.multitier.GMT.GMTTestConsole |
| gipsy.tests.GEE.simulator.DGTDialog |
| gipsy.tests.GEE.simulator.ProfileDialog |
| gipsy.tests.GEE.simulator.jini.WorkerJTA |
| gipsy.tests.GIPC.intensional.SIPL.Lucx.SemanticTest.LucxSemanticAnalyzer |
| gipsy.tests.Regression |

**Table A.11: GIPSY classes with fair Changeability**

| Class Names for GIPSY with poor Changeability |
|---|
| gipsy.GIPC.DFG.DFGAnalyzer.DFGParser |
| gipsy.GIPC.DFG.DFGAnalyzer.DFGParserTokenManager |
| gipsy.GIPC.Preprocessing.PreprocessorParser |
| gipsy.GIPC.Preprocessing.PreprocessorParserTokenManager |
| gipsy.GIPC.intensional.GIPL.GIPLParser |
| gipsy.GIPC.intensional.GIPL.GIPLParserTokenManager |
| gipsy.GIPC.intensional.GenericTranslator.TranslationParser |
| gipsy.GIPC.intensional.SIPL.ForensicLucid.ForensicLucidParser |
| gipsy.GIPC.intensional.SIPL.ForensicLucid.ForensicLucidParserTokenManager |
| gipsy.GIPC.intensional.SIPL.IndexicalLucid.IndexicalLucidParser |
| gipsy.GIPC.intensional.SIPL.IndexicalLucid.IndexicalLucidParserTokenManager |
| gipsy.GIPC.intensional.SIPL.JLucid.JGIPLParser |
| gipsy.GIPC.intensional.SIPL.JLucid.JGIPLParserTokenManager |
| gipsy.GIPC.intensional.SIPL.JLucid.JIndexicalLucidParser |
| gipsy.GIPC.intensional.SIPL.JLucid.JIndexicalLucidParserTokenManager |
| gipsy.GIPC.intensional.SIPL.JOOIP.JavaCharStream |
| gipsy.GIPC.intensional.SIPL.JOOIP.JavaParser |
| gipsy.GIPC.intensional.SIPL.JOOIP.JavaParserTokenManager |
| gipsy.GIPC.intensional.SIPL.Lucx.LucxParser |
| gipsy.GIPC.intensional.SIPL.Lucx.LucxParserTokenManager |
| gipsy.GIPC.intensional.SIPL.ObjectiveLucid.ObjectiveGIPLParser |
| gipsy.GIPC.intensional.SIPL.ObjectiveLucid.ObjectiveGIPLParserTokenManager |
| gipsy.GIPC.intensional.SIPL.ObjectiveLucid.ObjectiveIndexicalLucidParser |
| gipsy.GIPC.intensional.SIPL.ObjectiveLucid.ObjectiveIndexicalLucidParserTokenManager |
| gipsy.GIPC.util.SimpleCharStream |
| gipsy.RIPE.editors.RunTimeGraphEditor.core.GlobalInstance |
| gipsy.RIPE.editors.RunTimeGraphEditor.ui.GIPSYGMTOperator |
| gipsy.tests.junit.lang.GIPSYContextTest |
| gipsy.tests.junit.lang.context.GIPSYContextTest |

**Table A.12: GIPSY classes with poor Changeability**

| Class Names for GIPSY with fair Stability |
|---|
| gipsy.GEE.IDP.DMSException |
| gipsy.GEE.IDP.DemandDispatcher.DemandDispatcherException |
| gipsy.GEE.IDP.DemandGenerator.DemandGenerator |
| gipsy.GEE.IDP.DemandGenerator.jini.rmi.JINITA |
| gipsy.GEE.IDP.DemandGenerator.jini.rmi.JiniDemandDispatcher |
| gipsy.GEE.IDP.DemandGenerator.jms.JMSTransportAgent |
| gipsy.GEE.IDP.DemandGenerator.threaded.IdentifierContext |
| gipsy.GEE.IDP.demands.DemandSignature |
| gipsy.GEE.IDP.demands.DemandState |
| gipsy.GEE.IDP.demands.DemandType |
| gipsy.GEE.IDP.demands.IDemand |
| gipsy.GEE.IDP.demands.ProceduralDemand |
| gipsy.GEE.IDP.demands.SystemDemand |
| gipsy.GEE.IDP.demands.TimeLine |
| gipsy.GEE.multitier.GIPSYNode |

Concordia
UNIVERSITY

| |
|---|
| gipsy.GEE.multitier.GMT.GMTWrapper |
| gipsy.GEE.multitier.GenericTierWrapper |
| gipsy.GEE.multitier.IMultiTierWrapper |
| gipsy.GIPC.DFG.DFGAnalyzer.DFGParser |
| gipsy.GIPC.GIPCException |
| gipsy.GIPC.Preprocessing.PreprocessorParser |
| gipsy.GIPC.imperative.CommunicationProcedureGenerator.CommunicationProcedure |
| gipsy.GIPC.imperative.ImperativeCompiler |
| gipsy.GIPC.intensional.GIPL.GIPLParser |
| gipsy.GIPC.intensional.IntensionalCompiler |
| gipsy.GIPC.intensional.SIPL.ForensicLucid.ForensicLucidParser |
| gipsy.GIPC.intensional.SIPL.IndexicalLucid.IndexicalLucidParser |
| gipsy.GIPC.intensional.SIPL.IndexicalLucid.IndexicalLucidParserTreeConstants |
| gipsy.GIPC.intensional.SIPL.JLucid.JGIPLParser |
| gipsy.GIPC.intensional.SIPL.JLucid.JIndexicalLucidParser |
| gipsy.GIPC.intensional.SIPL.JOOIP.JavaCharStream |
| gipsy.GIPC.intensional.SIPL.JOOIP.JavaParser |
| gipsy.GIPC.intensional.SIPL.JOOIP.Token |
| gipsy.GIPC.intensional.SIPL.JOOIP.ast.body.BodyDeclaration |
| gipsy.GIPC.intensional.SIPL.JOOIP.ast.body.VariableDeclaratorId |
| gipsy.GIPC.intensional.SIPL.JOOIP.ast.expr.Expression |
| gipsy.GIPC.intensional.SIPL.JOOIP.ast.expr.NameExpr |
| gipsy.GIPC.intensional.SIPL.JOOIP.ast.stmt.BlockStmt |
| gipsy.GIPC.intensional.SIPL.JOOIP.ast.stmt.Statement |
| gipsy.GIPC.intensional.SIPL.JOOIP.ast.type.Type |
| gipsy.GIPC.intensional.SIPL.Lucx.LucxParser |
| gipsy.GIPC.intensional.SIPL.ObjectiveLucid.ObjectiveGIPLParser |
| gipsy.GIPC.intensional.SIPL.ObjectiveLucid.ObjectiveIndexicalLucidParser |
| gipsy.GIPC.util.ParseException |
| gipsy.GIPC.util.Token |
| gipsy.GIPC.util.TokenMgrError |
| gipsy.RIPE.editors.RunTimeGraphEditor.core.AppConstants |
| gipsy.RIPE.editors.RunTimeGraphEditor.core.GIPSYPhysicalNode |
| gipsy.RIPE.editors.RunTimeGraphEditor.core.GIPSYTier |
| gipsy.RIPE.editors.RunTimeGraphEditor.core.GlobalInstance |
| gipsy.RIPE.editors.RunTimeGraphEditor.ui.AppLogger |
| gipsy.RIPE.editors.RunTimeGraphEditor.ui.GIPSYGMTOperator |
| gipsy.interfaces.GIPSYProgram |
| gipsy.lang.GIPSYInteger |
| gipsy.storage.DictionaryItem |
| gipsy.tests.GEE.simulator.GlobalDef |
| gipsy.util.GIPSYException |
| gipsy.util.NetUtils |

**Table A.13: GIPSY classes with fair Stability**

| Class Names for GIPSY with poor Stability |
|---|
| gipsy.Configuration |
| gipsy.GEE.CON |
| gipsy.GEE.IDP.ITransportAgent |
| gipsy.GEE.IDP.demands.Demand |
| gipsy.GEE.multitier.DGT.DGTWrapper |
| gipsy.GEE.multitier.DST.DSTWrapper |
| gipsy.GEE.multitier.DWT.DWTWrapper |
| gipsy.GIPC.GIPC |
| gipsy.GIPC.intensional.GIPL.GIPLParserTreeConstants |
| gipsy.GIPC.intensional.SIPL.JOOIP.ast.Node |
| gipsy.GIPC.intensional.SIPL.JOOIP.ast.body.TypeDeclaration |
| gipsy.GIPC.intensional.SIPL.JOOIP.ast.expr.StringLiteralExpr |
| gipsy.GIPC.intensional.SimpleNode |
| gipsy.GIPC.util.SimpleCharStream |
| gipsy.lang.GIPSYContext |
| gipsy.lang.GIPSYType |
| gipsy.lang.context.TagSet |

**Table A.14: GIPSY classes with poor Stability**

| Class Names for GIPSY with fair Testability |
|---|
| gipsy.Configuration |
| gipsy.GEE.IDP.DemandGenerator.jini.rmi.JINITA |
| gipsy.GEE.IDP.DemandGenerator.jini.rmi.JiniDemandDispatcher |
| gipsy.GEE.IDP.DemandGenerator.jms.JMSTransportAgent |
| gipsy.GEE.IDP.demands.Demand |
| gipsy.GEE.multitier.GMT.GMTWrapper |
| gipsy.GIPC.DFG.DFGAnalyzer.DFGParserTokenManager |
| gipsy.GIPC.DFG.DFGGenerator.DFGCodeGenerator |
| gipsy.GIPC.DFG.DFGGenerator.DFGTranCodeGenerator |
| gipsy.GIPC.Preprocessing.PreprocessorParserTokenManager |
| gipsy.GIPC.SemanticAnalyzer |
| gipsy.GIPC.intensional.GIPL.GIPLParserTokenManager |
| gipsy.GIPC.intensional.SIPL.ForensicLucid.ForensicLucidParserTokenManager |
| gipsy.GIPC.intensional.SIPL.ForensicLucid.ForensicLucidSemanticAnalyzer |
| gipsy.GIPC.intensional.SIPL.IndexicalLucid.IndexicalLucidParserTokenManager |
| gipsy.GIPC.intensional.SIPL.JLucid.JGIPLParserTokenManager |
| gipsy.GIPC.intensional.SIPL.JLucid.JIndexicalLucidParserTokenManager |
| gipsy.GIPC.intensional.SIPL.JOOIP.JavaCharStream |
| gipsy.GIPC.intensional.SIPL.JOOIP.ast.visitor.GenericVisitor |
| gipsy.GIPC.intensional.SIPL.JOOIP.ast.visitor.VoidVisitor |
| gipsy.GIPC.intensional.SIPL.Lucx.LucxParserTokenManager |
| gipsy.GIPC.intensional.SIPL.ObjectiveLucid.ObjectiveGIPLParserTokenManager |
| gipsy.GIPC.intensional.SIPL.ObjectiveLucid.ObjectiveIndexicalLucidParserTokenManager |
| gipsy.GIPC.util.SimpleCharStream |
| gipsy.RIPE.editors.RunTimeGraphEditor.core.GlobalInstance |
| gipsy.lang.GIPSYContext |
| gipsy.lang.GIPSYInteger |
| gipsy.tests.GIPC.intensional.SIPL.Lucx.SemanticTest.LucxSemanticAnalyzer |

**Table A.15: GIPSY classes with fair Testability**

| Class Names for GIPSY with poor Testability |
|---|
| gipsy.GIPC.DFG.DFGAnalyzer.DFGParser |
| gipsy.GIPC.Preprocessing.PreprocessorParser |
| gipsy.GIPC.intensional.GIPL.GIPLParser |
| gipsy.GIPC.intensional.GenericTranslator.TranslationParser |
| gipsy.GIPC.intensional.SIPL.ForensicLucid.ForensicLucidParser |
| gipsy.GIPC.intensional.SIPL.IndexicalLucid.IndexicalLucidParser |
| gipsy.GIPC.intensional.SIPL.JLucid.JGIPLParser |
| gipsy.GIPC.intensional.SIPL.JLucid.JIndexicalLucidParser |
| gipsy.GIPC.intensional.SIPL.JOOIP.JavaParser |
| gipsy.GIPC.intensional.SIPL.JOOIP.JavaParserTokenManager |
| gipsy.GIPC.intensional.SIPL.JOOIP.ast.visitor.DumpVisitor |
| gipsy.GIPC.intensional.SIPL.Lucx.LucxParser |
| gipsy.GIPC.intensional.SIPL.ObjectiveLucid.ObjectiveGIPLParser |
| gipsy.GIPC.intensional.SIPL.ObjectiveLucid.ObjectiveIndexicalLucidParser |
| gipsy.RIPE.editors.RunTimeGraphEditor.ui.GIPSYGMTOperator |

**Table A.16: GIPSY classes with poor Testability**